\documentclass[apj,twocolumn]{openjournal}
\usepackage{graphicx}
\usepackage{xcolor}
\usepackage{amsmath}
\usepackage{amsfonts}
\usepackage{amssymb}
\usepackage{multirow}
\usepackage{upgreek}
\usepackage[pdfborder={0    0 0}]{hyperref}
\usepackage{txfonts}
\usepackage{lipsum} 
\usepackage{booktabs}
\usepackage{array} 

\newcommand{\orcidauthor}[3]{\author{{#2}$^{#3}$}}

\usepackage{mathrsfs}

\begin{document}
\title{The SKAO Pulsar Timing Array}

\orcidauthor{}{Ryan M. Shannon}{1,2}
\orcidauthor{}{N. D. Ramesh Bhat}{3}
\orcidauthor{}{Aur\'elien Chalumeau}{4,5}
\orcidauthor{}{Siyuan Chen}{6}
\orcidauthor{}{H. Thankful Cromartie}{7}
\orcidauthor{}{A. Gopukumar}{8}
\orcidauthor{0009-0000-2963-6641}{Kathrin Grunthal}{9}
\orcidauthor{}{Jeffrey S. Hazboun}{10}
\orcidauthor{}{Francesco Iraci}{11,12}
\orcidauthor{}{Bhal Chandra Joshi}{13,14}
\orcidauthor{}{Ryo Kato}{15}
\orcidauthor{}{Michael J. Keith}{16}
\orcidauthor{}{Kejia Lee}{17}
\orcidauthor{}{Kuo Liu}{6,18}
\orcidauthor{}{Hannah Middleton}{19}
\orcidauthor{}{Matthew T. Miles}{20,2}
\orcidauthor{}{Chiara M. F. Mingarelli}{21}
\orcidauthor{}{Aditya Parthasarathy}{22,23,9}
\orcidauthor{}{Daniel J. Reardon}{1,2}
\orcidauthor{}{Golam M. Shaifullah}{24,25,12}
\orcidauthor{}{Keitaro Takahashi}{26}
\orcidauthor{}{Caterina Tiburzi}{11}
\orcidauthor{}{Riccardo J. Truant}{1,2}
\orcidauthor{}{Xiao Xue}{27}
\orcidauthor{}{Andrew Zic}{28}
\orcidauthor{}{The SKAO pulsar science working group}{}

\affiliation{$^{1}$Centre for Astrophysics and Supercomputing, Swinburne University of Technology, Hawthorn, VIC, 3122, Australia}
\email{rshannon@swin.edu.au}
\affiliation{$^{2}$Australian Research Council Centre of Excellence for Gravitational Wave Astronomy (OzGrav)}

\affiliation{$^{3}$International Centre for Radio Astronomy Research, Curtin University, Bentley, WA 6102, Australia}

\affiliation{$^{4}$Laboratoire de Physique et Chimie de l’Environnement et de l’Espace, Universit\'e d’Orl\'eans / CNRS, 45071 Orl\'eans Cedex 02, France}
\affiliation{$^{5}$Observatoire Radioastronomique de Nan\c{c}ay, Observatoire de Paris, Universit\'e PSL, Universit\'e d’Orl\'eans, CNRS, 18330 Nan\c{c}ay, France}

\affiliation{$^{6}$Shanghai Astronomical Observatory, Chinese Academy of Sciences, 80 Nandan Road, Shanghai 200030, P. R. China}

\affiliation{$^{7}$National Research Council Research Associate, National Academy of Sciences, Washington, DC 20001, USA resident at Naval Research Laboratory, Washington, DC 20375, USA}

\affiliation{$^{8}$Data Institute of Fundamental Research, II Homi Bhabha Road, Mumbai 400005, India}

\affiliation{$^{9}$Max-Planck-Institut für Radioastronomie, Auf dem Hügel 69, D-53121 Bonn, Germany}

\affiliation{$^{10}$Department of Physics, Oregon State University, Corvallis, OR 97331, USA}

\affiliation{$^{11}$Dipartimento di Fisica, Università di Cagliari, Cittadella Universitaria, I-09042 Monserrato (CA), Italy}

\affiliation{$^{12}$INAF - Osservatorio Astronomico di Cagliari, via della Scienza 5, 09047 Selargius (CA), Italy}

\affiliation{$^{13}$National Centre for Radio Astrophysics, SP Pune University Campus, Pune 411007, Maharashtra, India}
\affiliation{$^{14}$Department of Physics, Indian Institute of Technology Roorkee, Roorkee 247667, Uttarakhand,}

\affiliation{$^{15}$Mizusawa VLBI Observatory, National Astronomical Observatory of Japan, 2-21-1 Osawa, Mitaka, Tokyo 181-8588, Japan}

\affiliation{$^{16}$Jodrell Bank Centre for Astrophysics, Department of Physics and Astronomy, University of Manchester, Manchester M13 9PL, UK}

\affiliation{$^{17}$Department of Astronomy, School of Physics, Peking University, Beijing 100871, P. R. China}

\affiliation{$^{18}$State Key Laboratory of Radio Astronomy and Technology, A20 Datun Road, Chaoyang District, Beijing, 100101, P. R. China}

\affiliation{$^{19}$Institute for Gravitational Wave Astronomy \& School of Physics and Astronomy, University of Birmingham, Birmingham, United Kingdom}

\affiliation{$^{20}$Department of Physics and Astronomy, Vanderbilt University, 2301 Vanderbilt Place, Nashville, TN 37235, USA}

\affiliation{$^{21}$Department of Physics, Yale University, New Haven, CT USA, 06520}

\affiliation{$^{22}$ASTRON, Netherlands Institute for Radio Astronomy, Oude Hoogeveensedijk 4, 7991 PD Dwingeloo, The Netherlands}
\affiliation{$^{23}$Anton Pannekoek Institute for Astronomy, University of Amsterdam, Science Park 904, 1098 XH Amsterdam, The Netherlands}

\affiliation{$^{24}$Dipartimento di Fisica “G. Occhialini”, Universit\'a degli Studi di Milano-Bicocca, Piazza della Scienza 3, I-20126 Milano, Italy.}
\affiliation{$^{25}$INFN, Sezione di Milano-Bicocca, Piazza della Scienza 3, 20126 Milano, Italy}

\affiliation{$^{26}$Faculty of Advanced Science and Technology, Kumamoto University, Japan}

\affiliation{$^{27}$Institut de F\'{i}sica d’Altes Energies (IFAE), The Barcelona Institute of Science and Technology, Campus UAB, 08193 Bellaterra (Barcelona), Spain}

\affiliation{$^{28}$CSIRO, Space \& Astronomy, Marsfield, NSW, 1710, Australia}

\begin{abstract}
Pulsar timing arrays (PTAs) are ensembles of millisecond pulsars observed for years to decades. The primary goal of PTAs is to study gravitational-wave astronomy at nanohertz frequencies, with secondary goals of undertaking other fundamental tests of physics and astronomy. Recently, compelling evidence has emerged in established PTA experiments for the presence of a gravitational-wave background. To accelerate a confident detection of such a signal and then study gravitational-wave emitting sources, it is necessary to observe a larger number of millisecond pulsars to greater timing precision. The SKAO telescopes, which will be a factor of three to four greater in sensitivity compared to any other southern hemisphere facility, are poised to make such an impact. In this chapter, we motivate an SKAO  pulsar timing array (SKAO PTA) experiment.  We discuss the classes of gravitational waves present in PTA observations and how an SKAO PTA can detect and study them. We then describe the sources that can produce these signals. We discuss the astrophysical noise sources that must be mitigated to undertake the most sensitive searches. We then describe a realistic PTA experiment implemented with the SKA and place it in context alongside other PTA experiments likely ongoing in the 2030s.  We describe the techniques necessary to search for gravitational waves in the SKAO PTA and motivate how very long baseline interferometry can improve the sensitivity of an SKAO PTA. The SKAO PTA will provide a view of the Universe complementary to those of the other large facilities of the 2030s. 
\end{abstract}


\maketitle

\section{Introduction}

The last few decades have ushered in an era of multi-messenger astronomy: the study of the
Universe through combining electromagnetic and  non-electromagnetic signals \citep{SN1987ASNnu,2017PhRvL.119p1101A,TXS18}.
One such non-electromagnetic signal is gravitational radiation: waves in the space-time metric.  
Like the electromagnetic spectrum, the gravitational-wave (GW) spectrum spans decades in frequencies.

GWs are produced by time-changing, non-symmetric mass distributions, which occur in systems like merging black holes or neutron stars.
While the first indirect evidence for GWs came from the damping of the orbit of the Hulse-Taylor binary pulsar~\cite[][]{1982ApJ...253..908T,1975ApJ...195L..51H}, the first detection of GWs was by the LIGO~\cite[][]{AdLIGO:2015} and Virgo~\cite[][]{AdVirgo:2015} Collaborations in 2015, with the ground-based laser interferometric detection of decahertz GWs from the inspiral of two $\sim30\,M_\odot$ black holes \cite[][]{2016PhRvL.116f1102A}.
This was soon followed by the discovery of GWs from the merger of a binary neutron star system~\cite[][]{2017PhRvL.119p1101A}, for which electromagnetic radiation was also observed from a gamma-ray burst and kilonova associated with the merger~\cite[][]{2017ApJ...848L..12A}, ushering in the era of multi-messenger GW astronomy.
To date, the LIGO, Virgo and KAGRA~\cite[][]{KAGRA:2021} collaborations have made $\sim 200$ probable detections~\cite[][]{GWTC4}, with a fourth observation run ongoing.
Ground-based laser interferometers are sensitive to GWs in the hertz to kilohertz regime, a band where signals originate from sources such as stellar mass black holes and neutron stars.
While the observations  directly proved Einstein's prediction for the existence of GWs, they have also facilitated other breakthroughs across physics and astrophysics by providing a new way to observe the Universe with objects impossible to study otherwise \cite[e.g.,][]{2025PhRvL.135k1403A} . 

Another band exists at much lower GW frequency that utilizes pulsars: rotating neutron stars that emit beams of radiation that span the electromagnetic spectrum. When the beams (misaligned with the rotation axis) cross our line of sight, we observe a pulse of radiation, which can be timed to exquisite precision. When observed in the radio band, the time-of-arrival of the pulses observed at Earth can be measured to precisions of better than  hundreds of nanoseconds \cite[][]{2023ApJ...951L...9A,2023PASA...40...49Z,2025MNRAS.536.1467M} for the most precise millisecond pulsars (MSPs): a subclass of rapidly rotating pulsars spun up due to accretion from a companion \cite[][]{1982Natur.300..728A,1982Natur.300..615B}.

It has long been known that GWs passing across the pulsar-Earth line of sight will alter the light propagation time from the pulsar to the Earth~\cite[][]{1978SvA....22...36S,1979ApJ...234.1100D}. Through the study of an ensemble of pulsars, called a pulsar timing array~\cite[PTA, ][]{1990ApJ...361..300F}, it is possible to distinguish the presence of GW from other processes that alter pulse arrival times, as GWs impart distinctive angular correlations~\cite[][]{1983ApJ...265L..39H} due to their quadrupolar nature, whereas other processes (often termed noise) are either uncorrelated between pulsars or show different spatial correlations~\cite[][]{2016MNRAS.455.4339T}. 

Pulsar timing observations are sensitive to GWs with frequencies between the reciprocals of the observation baseline ($\sim$ years to decades) and twice the observing cadence ($\sim$days to months), which is roughly 1-100\,nHz. In this band, the most likely source of GWs is binary supermassive black holes~\cite[SMBHs][]{1980Natur.287..307B} in sub milliparsec orbits. After galaxies merge, the central SMBHs of the progenitors are dragged to the center of the merged system. Through dynamical and viscous friction, the black holes form a gravitationally bound system that continues to harden. When the black holes are sufficiently close, GW emission can dominate the inspiral. The superposition of GW from all supermassive black hole binaries (SMBHBs) in the Universe produces a stochastic GW background~\cite[GWB, ][]{1983ApJ...265L..39H,1995ApJ...446..543R}.  
This is thought to be the first signal that a PTA experiment would detect \cite[][]{2003ApJ...590..691W}
The GWB induces red-noise or temporal correlations in pulse arrival times. This can be modeled as a power-law power spectrum with an amplitude and a known spectral index in the case of SMBHBs inspiralling due to GW emission alone~\cite[][]{2001astro.ph..8028P}.
A GWB could first emerge as red noise with statistically consistent properties, known as common noise \cite[][]{2013CQGra..30v4015S}.
However, the presence of Hellings-Downs angular correlations \cite[][]{1983ApJ...265L..39H} is necessary for a confident  detection of the GWB~\cite[][]{2023arXiv230404767A}, because it is possible to detect common noise in PTA data sets when no GWB is present \cite[][]{2021ApJ...917L..19G,2022MNRAS.516..410Z}.
In addition to a GWB from the population of SMBHBs, other GW signals could exist in PTA data sets.
This includes individual resolvable (spectrally and spatially) SMBHBs \cite[][]{2010CQGra..27h4016S, 2025ApJ...978...31A}, 
bursts from SMBH flybys, and signals originating from exotic physical processes \cite[][]{2015MNRAS.446.1657W,SD23hyp,2023ApJ...951L..11A}.
Detection and study of any of these signals not only allows us to understand the nature of the laws of physics, but also offers glimpses into regions of the Universe otherwise invisible, either because the objects themselves are electromagnetically dim or embedded in centers of galaxies and obscured. Recent advances are summarized in \cite{2025NatAs...9..183M}.

PTA experiments have been ongoing for decades in Australia \cite[The Parkes Pulsar Timing Array, PPTA,][]{2013PASA...30...17M,2023PASA...40...49Z}, Europe \cite[The European Pulsar Timing Array, EPTA,][]{2013CQGra..30v4009K,EPTA+2023a}, and North America \cite[The North American Nanohertz-frequency Gravitational Wave Observatory, NANOGrav,][]{2013ApJ...762...94D,2023ApJ...951L...9A}.
Observations have been taken with the most sensitive radio telescopes at metre-decimetre wavelengths, choosing this band as receivers are sensitive, pulsars are relatively bright, and the highest timing precision can in general be achieved. More recently PTA experiments have started in China \cite[The Chinese Pulsar Timing Array, CPTA][]{2016ASPC..502...19L}, using the Five Hundred Metre Aperture Telescope (FAST);  in India \cite[The Indian Pulsar Timing Array, InPTA][]{2022JApA...43...98J}, using the upgraded Giant Metrewave Radio Telescope (uGMRT) and the Ootucmund Radio Telescope (ORT); and in South Africa \cite[The MeerKAT Pulsar Timing Array, MPTA][]{2023MNRAS.519.3976M}, using the MeerKAT radio telescope. 
There is also now  PTA analysis undertaken using observations of gamma-ray bright millisecond pulsars observed with the Fermi Space Telescope \cite[The Gamma Ray Pulsar Timing Array, GPTA,][]{2022Sci...376..521F}.
The groups collaborate and share data as part of the International Pulsar Timing Array \cite[][]{2010CQGra..27h4013H,2016MNRAS.458.1267V,2019MNRAS.490.4666P}.
More than $160$ MSPs are observed as part of these efforts.

\begin{figure*}
\centering
\begin{tabular}{cc}
\includegraphics[width=0.43\textwidth]{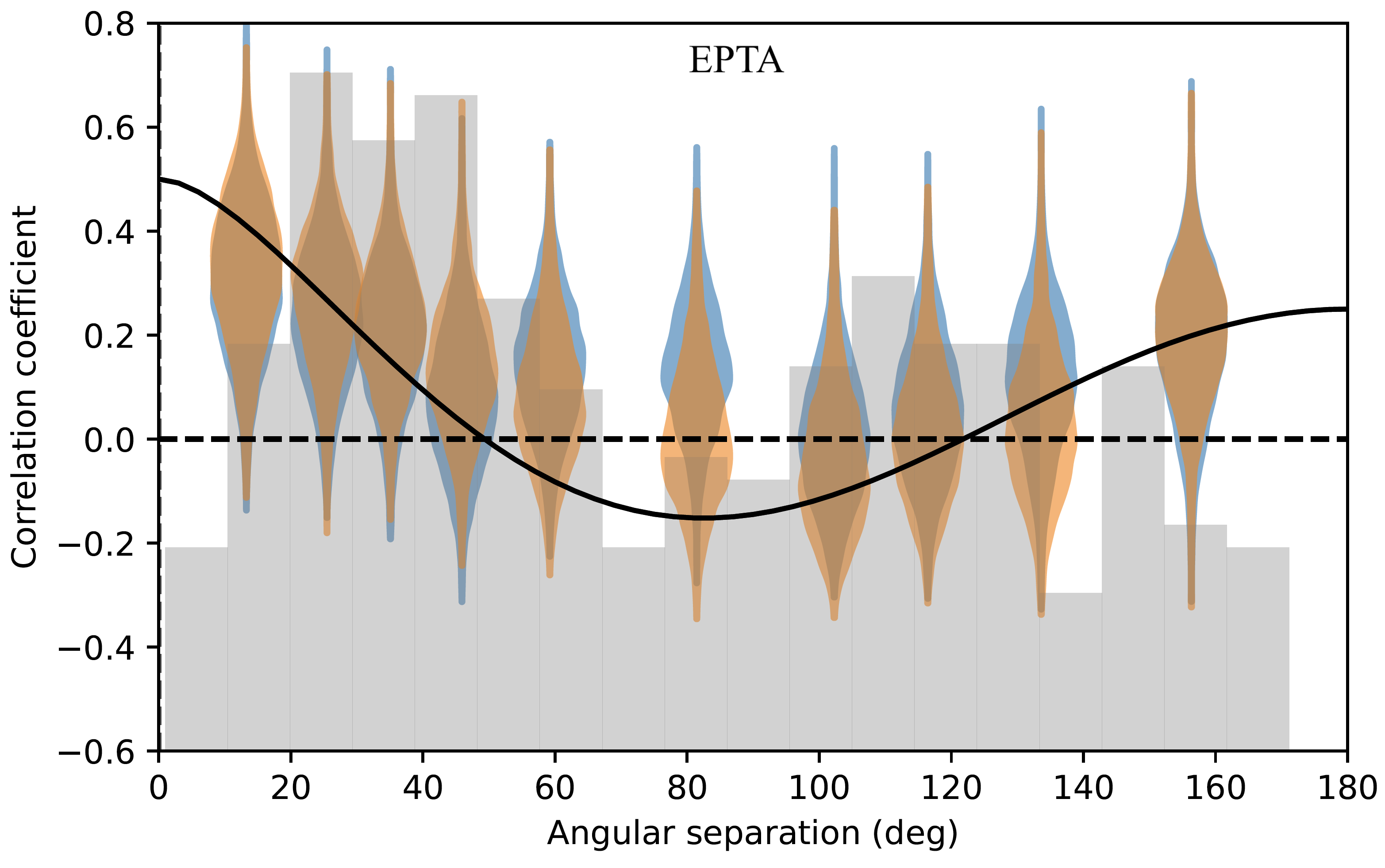} &
\includegraphics[width=0.4\textwidth]{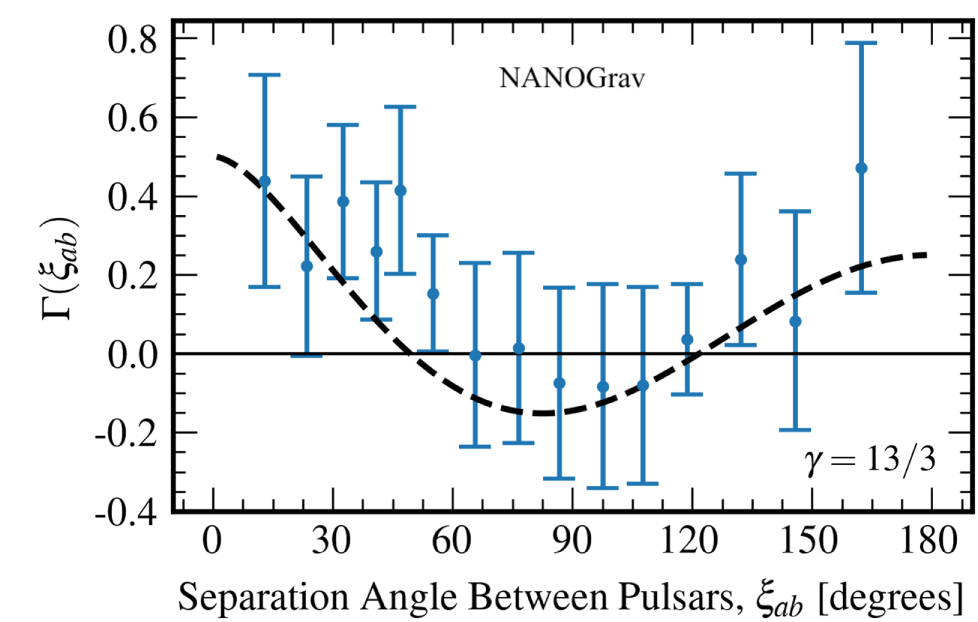} \\ \includegraphics[width=0.43\textwidth]{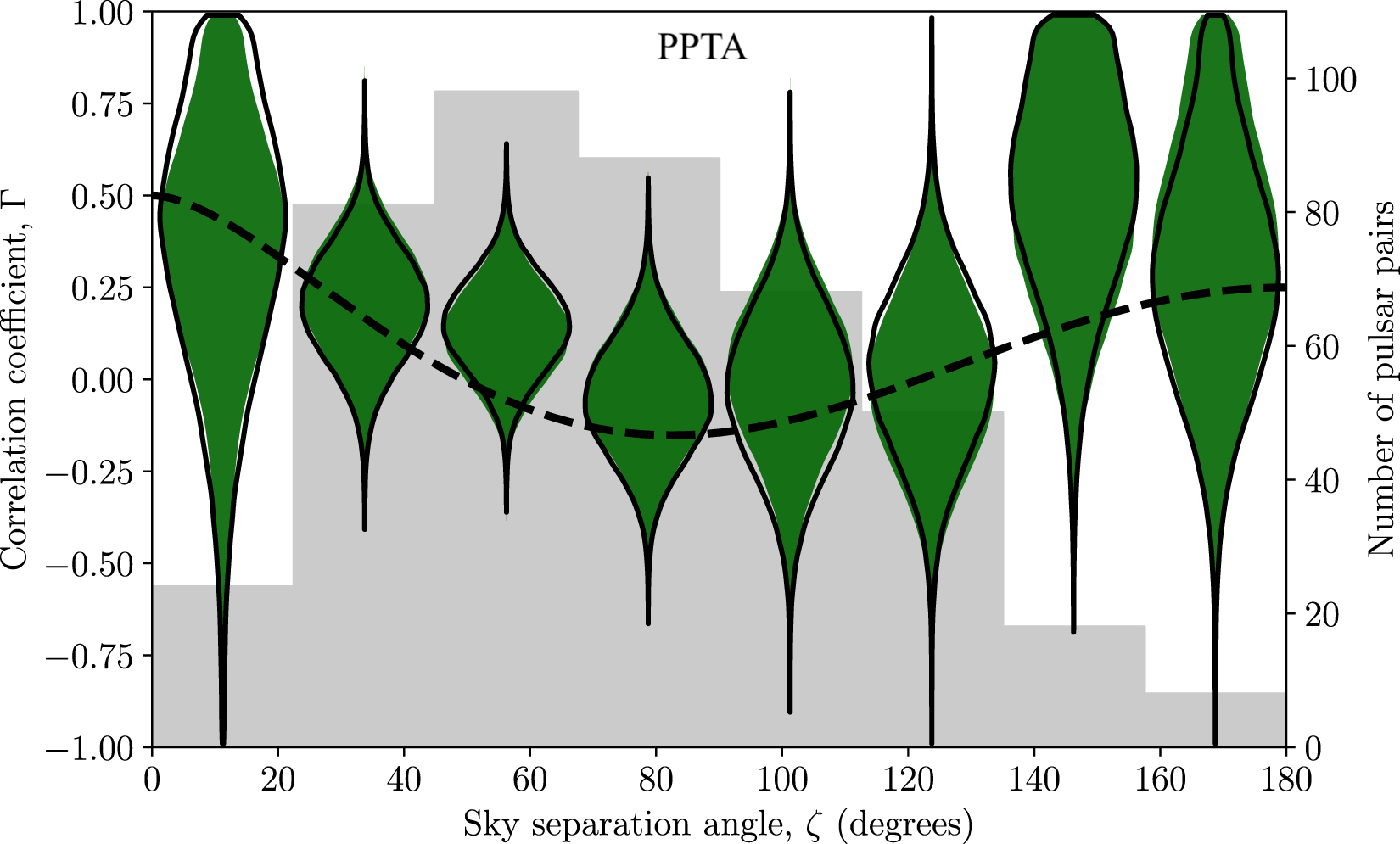} &
\includegraphics[width=0.4\textwidth]{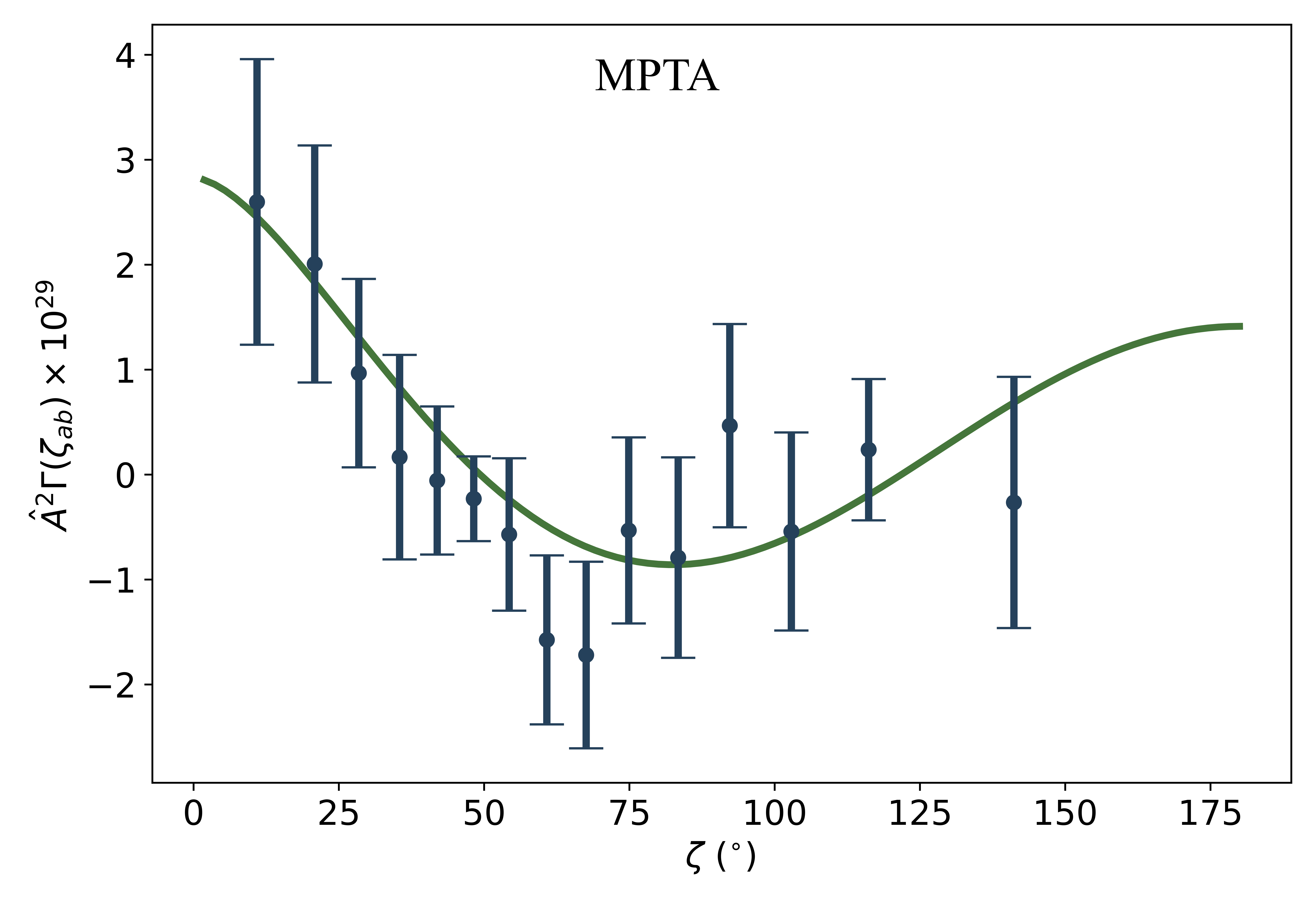}  \\
\includegraphics[width=0.5\textwidth]{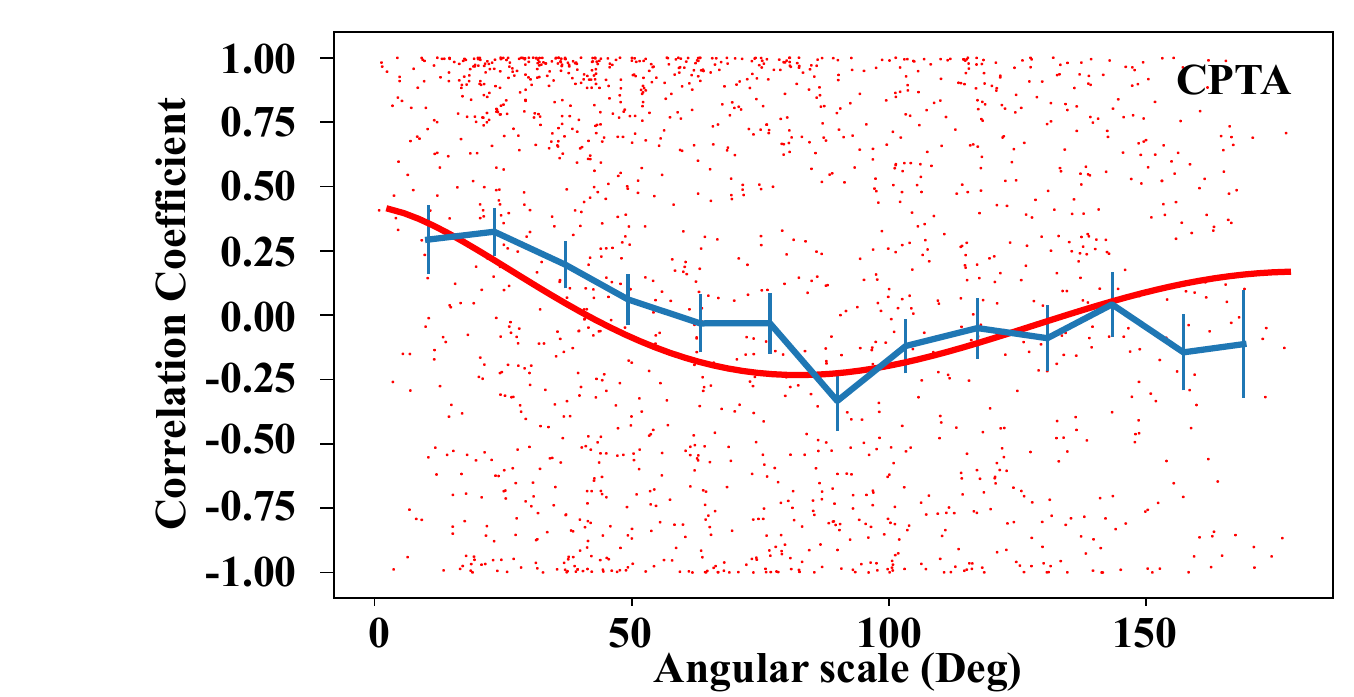}
\end{tabular}
\caption{\em \label{fig:pta_hd}  
Inter-pulsar correlations from recent pulsar timing array gravitational-wave searches.  We show the correlations derived from the 
the European Pulsar Timing Array and Indian Pulsar Timing Array \citep{2023A&A...678A..50E}, the North American Nanohertz Observatory for Gravitational Waves  \citep{2023ApJ...951L...8A}, the MeerKAT Pulsar Timing Array 
\citep{2025MNRAS.536.1489M}, the Parkes Pulsar Timing Array \citep{2023ApJ...951L...6R}, and the Chinese Pulsar Timing Array \citep{2023RAA....23g5024X}.
}
\end{figure*}

These projects have made great progress towards the definitive detection of nanohertz-frequency GWs through searching the data sets over the past decade.
At the beginning of the 2020s, common uncorrelated red noise was detected in NANOGrav~\cite[][]{2020ApJ...905L..34A},  EPTA~\cite[][]{2021MNRAS.508.4970C}, and PPTA~\cite[][]{2021ApJ...917L..19G} data analyses. 
More recently, evidence has emerged for the presence of a GWB in PTA data sets \cite[][]{2023ApJ...951L...8A,2023A&A...678A..50E,2023ApJ...951L...6R,2023RAA....23g5024X,2025MNRAS.536.1489M}.
Hellings and Downs spatial correlations were identified in these searches with $\approx 2-4\sigma$ statistical significance;  visualizations of these correlations can be found in Figure~\ref{fig:pta_hd}.
While these analyses are consistent at the $1-2\sigma$ level \cite[][]{2024ApJ...966..105A}, the amplitude of the signal is louder than found in some previous searches \cite[][]{2015Sci...349.1522S,2023ApJ...951L...6R}.
An increased sensitivity to GWs can be achieved by timing larger numbers of pulsars to higher precision \cite[][]{2013CQGra..30v4015S}. This has been demonstrated through GW searches undertaken by the CPTA and MPTA, which have shown high sensitivity despite timing baselines more than a factor of three shorter than those of other PTA experiments.

With a significant increase in sensitivity relative to MeerKAT and the Murchison Widefield Array, and access to the same sky, the SKA Observatory (SKAO) telescopes (SKAO-Low and SKAO-Mid) promise to be important instruments for PTA science in the 2030s. 
There are two baseline deployments for the telescopes: AA$^{*}$ and AA4.
The SKAO-Low AA$^{*}$ deployment comprises  307 stations and AA4 comprises 512 stations. 
The SKAO-Mid AA$^{*}$ deployment comprises 144 antennas, and AA4 comprises 197 antennas.
Indeed, PTA GW searches have been long identified as a key project for the SKAO \cite[][]{2004NewAR..48.1413C}.  
Here we update the previous PTA science case for the SKAO published nearly a decade ago \cite[][]{2015aska.confE..37J}. 
We describe the types of GW signals that can be observed with the SKAO in Section \ref{sec:ska_signals}.
We then describe the sources that could emit these signals in Sections~\ref{sec:ska_sources_smbhbs} and~\ref{sec:ska_sources_early_universe}.
In Section \ref{sec:noise}, we describe sources of contaminating noise in PTA data sets and how to best remove or mitigate them using the SKAO.
In Section~\ref{sec:ska_obs}, we motivate a potential PTA programme for the SKAO, including requirements for observations, data products, and data analysis.  
We highlight the important roles that can be made by SKAO's Low telescope and very long baseline interferometry in Sections~\ref{sec:ska_long_baseline} and~\ref{sec:ska_low}, respectively. 
In Section \ref{sec:gptas} we highlight the complementarity of  of high energy pulsar timing to the SKAO PTA.
In Section~\ref{sec:ska_requirements} we outline some of requirements for the SKAO PTA. 
We connect the SKAO PTA to other science outcomes that will be achieved by the SKAO and other 2030s facilities in Section~\ref{sec:other_science} and conclude the paper in Section~\ref{sec:conclusions}. 

\section{Pulsar Timing Array Signals}
\label{sec:ska_signals}

There are a plethora of different source classes and signals that we might expect to detect with the SKAO PTA.
Before we describe the sources of GWs we expect to observe, we describe the signals that can be measured, and  motivate how the SKAO PTA has the potential to play a transformational role.


\subsection{Gravitational-wave backgrounds}

A  GWB is the superposition of GWs from all of  GW emitting sources. For an individual gravitational wave source, the strain $h_{ij}$ at a distance $d_L$ from a source with mass quadrupole moment $Q_{ij}$ is \cite[][]{2017AnP...52900209A}
\begin{equation}
    h_{ij} = \frac{2}{c^4}\frac{G}{d_L}\frac{d^2}{dt^2} Q_{ij} 
\end{equation}
In the case of compact objects in a circular binary orbit, this can be approximated as \cite[][]{1963PhRv..131..435P,2003ApJ...583..616J}
\begin{eqnarray}
h  &&=  4 \sqrt{\frac{2}{5}} \left( \frac{G \mathscr{M}}{c^3} \right)^{5/3} \left( \frac{2 \pi}{P_b} \right)^{2/3} \frac{c}{d_L}
\label{eqn:strain_period} \\
&&  \approx 1.2\times 10^{-15} \left(\frac{\mathscr{M}}{10^9 M_\odot} \right)^{5/3} \left(\frac{P_b}{\rm yr} \right)^{-2/3}  \left(\frac{d_L}{\rm Gpc}\right)^{-1}
\end{eqnarray}
where $P_b$ is the binary orbital period, $\mathscr{M}=(m_1 m_2)^{3/5}/(m_1+m_2)^{1/5}$ is the chirp mass, and $m_1$ and $m_2$ are the component masses. 
We note that sensitivity to GW strain scales $\propto d_L^{-1}$, unlike electromagnetic observations which scale  $\propto d_L^{-2}$.  

A GWB can be characterized by a strain spectrum, $h_{\rm c}(f)$, which quantifies the gravitational wave strain amplitude emitted at different frequencies.
The GWB is often assumed to follow a power law spectrum for astrophysically motivated reasons \cite[e.g.,][]{2001astro.ph..8028P}.  In this case, the strain spectrum is defined as 
\begin{equation}\label{eq:hc}
h_{\rm c}(f) = A_{\rm yr} \left( \frac{f}{f_{\rm yr}}\right)^\alpha\,,
\end{equation}
where $A_{\rm yr}$ is the characteristic amplitude of the GWB at a frequency of $f_{\rm yr}=1$\,yr$^{-1}$, and $\alpha$ is the strain spectral index.  The most recent PTA analyses find  $A_{\rm yr} \approx 2\times 10^{-15}$ \cite[][]{2024ApJ...966..105A}.

The strain spectrum can be related to the closure energy density in GWs following
\begin{equation}
\label{eqn:gwdensity}
\Omega_{\rm gw} (f) = \frac{2 \pi^2}{3 H_0^2} f^2 h_c^2(f)\,, 
\end{equation}
where $H_0$ is the Hubble constant.
For most expected sources in the nanohertz-frequency band, the strain amplitude of GWs is expected to be greater at lower frequencies (i.e., $\alpha<0$).  For a GWB arising from SMBHBs, in circular orbits, hardening solely due to gravitational radiation, the exponent is expected to be $\alpha=-2/3$ \cite[][]{2001astro.ph..8028P}. Other sources, for example of cosmological origin, may produce a GWB with a different spectral index. This can be used to distinguish GWB from different sources.

To compare the GWB signal to other sources of noise in PTA data sets, the effect of the GWB is often expressed in terms of its power spectral density in the pulsar timing residuals, which is \cite[][]{2013CQGra..30v4015S}
\begin{equation}
  P_{\rm r}(f) = \frac{A_{\rm yr}^2}{12 \pi^2} \left( \frac{f}{f_{\rm yr}} \right)^{2\alpha -3}\,.
    \end{equation}

A GWB is manifested in two parts.
The first part is a common red noise: stochastic processes that are statistically independent in different pulsars but are drawn from the same underlying distribution. This is the combined effect of gravitational waves passing through the Earth and all of the pulsars. The second part is the characteristic correlations between different pulsars. This effect is related to the perturbations that GWs passing through the Earth induce on the times of arrival.  
The GWB induces angular correlations between pulsar pairs that follow the Hellings-Downs function \cite[][]{1983ApJ...265L..39H}, which deviates from a pure quadrupole due to averaging over a large number of GW sources that are uniformly distributed across the sky with random GW polarization angles.

For PTAs with a small number of pulsars, the common red noise is expected to be the first signal detected. However, it is possible to mistakenly detect the presence of such a signal \cite[][]{2021ApJ...917L..19G,2022MNRAS.516..410Z}, due to the presence of other sources of noise (including pulsar spin noise, as discussed below).
Thus, in order to confidently detect a GWB and study it robustly, it is necessary to also find the common correlations it induces across an ensemble of pulsars.

The sensitivity to a GWB can be improved through longer observing campaigns, observing with higher precision (or equivalently higher cadence), and observing a larger number of pulsars \cite[][]{2013CQGra..30v4015S}. However, once a PTA is dominated by a common red noise with unclear correlations, the improvement to sensitivity only scales weakly with the respect to the first two. Thus, the best way to improve the PTA sensitivity is to observe a larger sample of pulsars. With increased sensitivity and access to the southern sky, an SKAO PTA can include the largest sample of pulsars ever. With the ability to form multiple sub-arrays with the SKAO telescopes, the pulsars in the SKAO PTA can be observed efficiently.

\subsection{Individual sources of gravitational waves}

Continuous GW sources (CGWs) are the most considered individual sources of GWs \cite[][]{2023ApJ...951L..50A,2024A&A...690A.118E,2025arXiv250813944Z}.  
These are typically modeled as sinusoidal signals in pulsar timing residuals, as the sources thought to emit them (binary supermassive black holes, discussed below) are expected to be in circular orbits in the pulsar timing band, and can be decomposed into two parts: one representing the passage of the gravitational wave by the Earth (the Earth term) and the second  passing the pulsar (the pulsar term). 
The amplitude of the signal depends on the GW strain and polarization, the relative positions of the  source and the pulsar, and the pulsar distance.
The distance of the pulsar is treated as an unknown if a pulsar distance is known to a precision greater than the GW wavelength (the case, currently, for most MSPs). 
GW searches also need to consider whether the source is evolving, in which case the frequency in the pulsar term might be different than that in the Earth term.
While most searches consider these models,  efforts are ongoing to model PTA responses to individual massive black holes 
spiraling along general relativistic eccentric orbits and understanding their implications for PTAs
\cite[][]{AS20,AS23,N15ecc}.

In addition to periodic sources, it is possible to search for transient GW bursts as well.
Bursts with memory are semipermanent distortions in space-time caused by the merger of compact objects \cite[][]{2010CQGra..27h4036F}.  They will manifest as sudden changes in the spin frequency of the pulsar.
They could manifest as changes that are contemporaneous between pulsars and show the expected quadrupolar correlation \cite[][]{2010MNRAS.402..417P,2010MNRAS.401.2372V}.  Alternatively, pulsar term memory bursts would manifest in an individual pulsar \cite[][]{2012ApJ...752...54C}.
Bursts produced by other sources may have different wave forms.   
Hyperbolic passages of SMBHBs in general relativistic hyperbolic 
orbits could provide bursts with linear memory sources in the  PTA frequency window 
\cite[][]{SD23hyp,SDN125hyp}.

\subsection{Beyond backgrounds: gravitational wave maps}

If the nanohertz GWB is primarily generated by SMBHBs in massive galaxies, its angular anisotropies should mirror the cosmic large-scale structure (LSS) \citep{SemenzatoEtAl2025} at low GW frequencies, or be dominated by loud individual binaries at high frequencies \citep{Mingarellietal2013}. Otherwise, a GWB  can also originate from a wide range of early-Universe processes, such as red-shifted primordial GWs \citep{2016PhRvX...6a1035L,Domenech:2021ztg}, phase transitions \citep{Caprini:2019egz, Hindmarsh_2021} or cosmic string collapse \citep{Hiramatsu:2013qaa}, discussed further below.
The majority of these cosmological models predict an isotropic GWB. 

Hence, a pillar stone of identifying the origin of the GWB is searching for variations in GW across the sky. Most commonly, the GW power is expanded in spherical harmonics~\citep{Mingarellietal2013}, where the corresponding coefficients are determined from the PTA data. The resulting sky distribution is visualized in terms of sky maps, the so-called anisotropy maps.

Recent simulation-based studies coupling halo catalogs, semi-analytic SMBHB population models, and PTA map-making pipelines demonstrate that cross-correlations between GWB anisotropy maps and galaxy clustering can recover this imprint at high significance once PTAs achieve multipole sensitivities of $\ell_{\max} \gtrsim 40$–$70$. In this regime, loud nearby binaries behave as Poisson-like contaminants, and become visible as stand-out hot-spots on the sky map  that can be modeled or excised, allowing the residual anisotropy to act as a novel cosmological tracer~\citep{SemenzatoEtAl2025}. The local angular resolution of a PTA data set is intimately linked to the number of pulsars it comprises, and their sky distribution \citep{Boyle_Pen_2012,alihaimoud2020,alihaimoud2021,Grunthal_inprep}.

The broad sky coverage and timing precision of the SKAO PTA will be decisive in realizing this measurement, enabling anisotropy maps of sufficient resolution to probe the spatial distribution of SMBHB hosts and, ultimately, to connect the nanohertz GWB to the cosmic web itself \citep{SemenzatoEtAl2025}.

\subsection{Ultra-low frequency gravitational waves}

The SKA timing era also opens a complementary ${\rm pHz}$ window ($10^{-16}\!$–$10^{-9}$\,Hz) by exploiting slow, GW-induced drifts in pulsar timing parameters rather than residuals. A Bayesian analysis of pulsar orbital period derivatives $\dot{P}_b$ and spin second derivatives $\ddot{P}$ across many millisecond pulsars sets competitive limits today and forecasts SKAO-enabled sensitivity to continuous ${\rm pHz}$ signals from early-stage SMBHBs (and some cosmological scenarios) \citep{ZhengEtAl2025}. This approach bridges the gap between cosmic microwave background and PTA frequencies, leverages long baselines without overfitting away ultra-low-frequency content, and adds orthogonal evidence to residual-based searches like PTAs. With  improved timing precision of SKAO PTA and new pulsar discoveries, parameter-drift searches become a realistic avenue for first pHz-band detections \citep{ZhengEtAl2025}.

\section{Gravitational waves from supermassive black hole binaries}
\label{sec:ska_sources_smbhbs}

We now summarize how supermassive black hole binaries manifest in the pulsar timing band and the astrophysics that can be gleaned through their observation.

\subsection{Gravitational-wave backgrounds}

SMBHs are expected to reside at the center of all massive galaxies \cite[][]{2013ARA&A..51..511K}. As two galaxies merge, the two central SMBHs will come closer to each other and eventually form a  gravitationally bound binary \cite[][]{1976ApJ...204L...1T}. The evolution of the binary is driven by several mechanisms: dynamical friction, stellar hardening, and interactions with gaseous disks are the main contributors. The question of which mechanism is the dominant one that drives the binary to shrink down to sub-pc scale has been termed the final parsec problem~\cite[][]{2003AIPC..686..201M}. Eventually, when the distance is below $\sim$pc, the binary will merge as the orbital energy diminishes through the emission of GWs \cite[][]{1980Natur.287..307B}. This hierarchical process produces SMBHBs of increasingly greater masses \cite[][]{2003ApJ...582..559V}. As the SMBHBs slowly spiral into each other they emit GWs related to their orbital period (see equation \ref{eqn:strain_period}). These GWs can be measured by PTAs and are thus a major candidate for the possible source of the PTA signal.

The energy emitted in GWs depends on the properties of the SMBHBs and the physics driving the evolution. For an individual binary, this manifests in the frequencies at which GWs are emitted. For a circular SMBHB purely driven by GW emission the GW frequency is twice the orbital frequency. 
Eccentric binaries and environmental evolution will cause  deviation from the power law. The most prominent effect is a shallower spectrum at low frequencies, corresponding to long orbital periods, as this is where environmental evolution dominates over the GW emission.

Assuming that the GWB is composed of signals from circular SMBHBs, the resulting characteristic strain follows a power-law described by Eq.~\eqref{eq:hc} with slope $\alpha=-2/3$ 
and amplitude $A$ determined by the underlying population model~\citep{2001astro.ph..8028P}. If one takes into account the environmental effect and non-zero eccentricity, the characteristic strain manifests a broken power-law pattern with a spectral turnover, which could occur at frequencies near $\sim 3\times 10^{-9}{\rm Hz}$. A single broken power-law model can be modeled as \cite[][]{Sampson2015, 2020ApJ...905L..34A}:
\begin{equation}
    \begin{aligned}
        h_c(f) = A_{\rm yr} \left(\frac{f}{{\rm yr}^{-1}}\right)^{\alpha} \left[1 + \left(\frac{f}{f_{\rm break}}\right)^{1/\ell}\right]^{\ell(\beta-\alpha)}\ ,
    \end{aligned}
\end{equation}
where $f_{\rm break}$ is the transition frequency, $\ell$ affects the smoothness of the transition and $\beta$ is the spectral index describing the power law at high frequencies compared to $\alpha$ at lower frequencies. The phenomenological model can then be compared against astrophysically motivated models to determine the driving mechanism of the SMBHB evolution. The SKAO PTA will initially contribute by measuring the properties of a GWB at high frequency, with lower frequency properties inferred from the combination of SKAO PTA  and legacy IPTA data sets.
 
Finally, a finite-number-of-sources effect challenges the Gaussian ensemble assumption, which would also cause a spectral turn-over at  frequencies $>10^{-8}{\rm Hz}$ \citep{2008MNRAS.390..192S,ng15-discreteness}. This effect can also be accounted for by introducing a model based prior distribution for $h_c(f)$. With the SKAO PTA, we may be able to measure this effect and use it to infer the properties of the population.

\subsection{Continuous gravitational-wave signals from individual supermassive black hole binaries}

The SKAO PTA will also enable exquisite study of individual supermassive black hole binaries. 
With SKAO-enabled PTAs, it will become possible to probe the relativistic dynamics of SMBHBs in a regime inaccessible to any other experiment.
Eccentric binaries or binaries whose frequency evolution are influenced by environmental effects, will emit GWs at multiple frequencies with different powers \cite[][]{2011MNRAS.411.1467K,2013CQGra..30v4014S,2014MNRAS.442...56R,2017MNRAS.470.1738C}. If measured by PTAs, this can be a hint as to the exact mechanism driving the evolution of the SMBHB. 
For bright CGW sources where both the Earth and pulsar terms can be coherently recovered, the $\sim10^3$\,yr phase separation between these signals encodes post-Newtonian corrections to the orbital evolution, including measurable signatures of the binary mass ratio and spin–orbit coupling~\citep{Mingarelli2012}. The timing precision and parallax-based distance determinations (discussed further below), enhanced by joint Gaia and pulsar-timing analyses \citep{MoranEtAl2023}, make recovery of the pulsar term feasible for a subset of high-mass, higher-frequency systems, allowing direct inference of component masses and spin parameters. This capability transforms PTA detections from statistical evidence of a GWB into precision probes of weak-field general relativity and the astrophysical evolution of massive binaries over millennia \citep{Mingarelli2012}.

It is also possible to use electromagnetic observations to enhance the sensitivity to sources of individual GWs.
Targeted searches benefit by having fewer trial factors as positions and binary periods can be well constrained and can increase Bayes factors by a few to ten \cite[][]{2021ApJ...921..178L,Agarwaletal2025}.
Optical periodicity catalogs provide sky position, frequency priors, and luminosity distances that boost PTA sensitivity to individual binaries. Using a vetted sample of periodic AGN, it is possible to forecast CGW strains and produce all-sky detection maps for IPTA and SKAO phases, showing that SKA should realistically reach several candidates within a decade of observation. Injections targeted by known periods can double the effective sensitivity relative to blind all-sky searches, and even non-detections constrain supermassive black hole binary masses. This turns the SKAO PTA into an active experiment directly hunting for supermassive black hole binaries in their host galaxies, linking radio pulsar timing with optical variability, possible VLBI imaging of hosts, and future IR spectroscopy to test the binary hypothesis for binary candidates, source by source \citep{Arzoumanianetal2020,XinMingarelliHazboun2021,Agarwaletal2025,caseyclyde2025}.

These considerations open up the exciting prospect of pursuing persistent, multi-messenger nanohertz GW astronomy with promising inspiraling supermassive black-hole binaries \citep{2022JApA...43...98J,2025NatAs...9..183M}. 
The strong evidence for the presence of 
 nHz GW emitting SMBHBs in blazars OJ~287 and 
PKS 2131-021, thanks to decade-long multi-wavelength EM observations, adds real substance to this possibility
\citep{LD18,Spitzer20,RadAstOJ287,PKS25}.

At the sensitivities and redshift reach of the SKAO PTA, strong gravitational lensing will become an important factor in the detection and interpretation of CGWs. Lensing magnification of $\mu \sim 2$--$100$ can raise intrinsically faint binaries above the SKAO detection threshold, substantially increasing the number of resolvable systems and extending the PTA horizon to $z \gtrsim 2$ \citep{KhusidEtAl2023}. The same phenomenon can produce multiple gravitationally lensed images with measurable time delays, offering an unprecedented opportunity to test the coherence of GW and electromagnetic signals across cosmological distances. Because lensed hosts trace overdense environments, lensing will also modulate the angular distribution of individually resolvable binaries, a feature that SKAO’s expanded sky coverage and pulsar yield will uniquely allow us to characterize. In this way, the SKAO PTA will open a new window onto the population of supermassive black hole binaries in the high-redshift Universe \citep{KhusidEtAl2023}.

\section{Gravitational-wave and related signals from the early universe}
\label{sec:ska_sources_early_universe}

In addition to GWs from supermassive black holes, an SKAO PTA has the possibility to detect GWs predicted by fundamental physical processes.
Several early-Universe processes—quantum fluctuations during inflation, first-order phase transitions (FOPTs), and topological defects—could have generated GWs  \citep{Allen:1996vm,Caprini:2018mtu}. Their spectra are characterized by $\Omega_{\rm gw}(f)$, directly linked to the observable strain $h_c(f)$ \citep{PlanckCosmo2018} as shown in Equation \ref{eqn:gwdensity}.

\subsection{Cosmic inflation}

Inflation explains the observed flatness and isotropy while predicting nearly scale-invariant perturbations \citep{1974ZhETF..67..825G,1981PhRvD..23..347G,1982PhLB..108..389L,Starobinsky:1979ty}. Quantum fluctuations produce scalar and tensor modes, the latter forming a stochastic GWB \citep{Grishchuk:1977ky}. The tensor-to-scalar ratio $r<0.036$ (95\% C.L.) \citep{Campeti2022,Tristram2022} constrains the amplitude, while the spectral index $n_t=4/(1+3w)+2$ is linked to the effective equation of state $w$. Slow-roll models ($w>-1$) give red-tilted spectra ($n_t<0$) that fall steeply at PTA frequencies; stiff or phantom phases ($w<-1$) can yield blue tilts enhancing nanohertz power \citep{Kuroyanagi:2014nba,Lasky_2016,Cai:2021uup,Ferreira:2024prl}. The inflationary GWB \citep{Zhao:2013bba}
\begin{equation}
\Omega_{\rm gw}^{\rm IGW}(f)\simeq1.1\,r\times10^{10n_t-15}\!\left(\frac{f}{\rm yr^{-1}}\right)^{n_t},
\end{equation}
implies $\Omega_{\rm gw}\!<\!10^{-16}$ for $r\!\sim\!10^{-2}$, $n_t\!\sim\!-0.01$, though blue or stiff reheating scenarios may reach $\sim10^{-10}$–$10^{-9}$ \citep{Liu:2023ymk}.

\textit{Scalar-induced GWs} arise from large curvature perturbations generating second-order tensor modes during radiation domination \citep{Ananda:2006af,2007PhRvD..76h4019B,Saito:2008jc}. For curvature power $P_\mathcal{R}(k)$,
\begin{equation}
\Omega_{\rm gw}^{\rm SIGW}(f)\!\propto\!\!\int\!du\,dv\,\mathcal{K}(u,v)P_\mathcal{R}(uk)P_\mathcal{R}(vk)
\end{equation}
\citep{Espinosa:2018eve,Kohri:2018awv}, peaking at $f\!\sim\!10^{-9}$–$10^{-8}$\,Hz for $A_\zeta\!\sim\!10^{-2}$ at $k\!\sim\!10^{6\!-\!7}\,{\rm Mpc^{-1}}$ \citep{EPTA:2024DR2-IV,2023ApJ...951L..11A}.

\subsection{First-order phase transitions}

Strong FOPTs driven by vacuum bubble nucleation can produce GWs through collisions, sound waves, and turbulence \citep{Witten:1984rs,Hogan:1986qda,Kamionkowski:1993fg,Caprini:2009yp,Schwaller:2015tja,Caprini:2018mtu,Guo:2020grp}. The dominant sound-wave term \citep{Hindmarsh:2013xza}
\begin{equation}
\Omega_{\rm gw}^{\rm sw}\!\propto\!v_w\!\left(\!\frac{H_*}{\beta}\!\right)\!\!\left(\!\frac{\kappa\alpha_*}{1+\alpha_*}\!\right)^{\!2}\!\!\!S(f/f_{\rm sw}),
\end{equation}
depends on bubble speed $v_w$, transition strength $\alpha_*$, inverse duration $\beta/H_*$. The peak frequency $f_{\rm sw}\!\simeq\!2\times10^{-5}{\rm Hz}(\beta/H_*)(T_*/100\,{\rm GeV})/v_w$ shows that $T_*\!\sim\!100$–$300\,{\rm MeV}$ transitions fall in the PTA range \citep{Xue:2021gyq,NANOGrav:2023NewPhysics,EPTA:2024DR2-IV,Athron:2023hgm}. A smaller turbulent component scales as $(\epsilon_{\rm turb}\alpha_*)^{3/2}$ with $\epsilon_{\rm turb}\!\lesssim\!0.05$ \citep{Caprini:2009yp}.

\subsection{Topological defects}

In Beyond Standard Model particle physics, the early Universe may undergo symmetry breaking as it cools down, leaving relic defects whose dynamics radiate GWs. Discrete breaking forms \textit{domain walls} decaying as the Universe expands \citep{Vilenkin:1981zs,Hiramatsu:2013qaa}, producing
\begin{equation}
\Omega_{\rm gw}^{\rm dw}\!\propto\!\tilde{\epsilon}\!\left(\!\frac{\alpha_*}{0.01}\!\right)^{\!2}\!S(f/f_{\rm dw}), \ \ 
f_{\rm dw}\!\sim\!10^{-9}{\rm Hz}\!\left(\!\frac{T_*}{10\,{\rm MeV}}\!\right),
\end{equation}
with efficiency $\tilde{\epsilon}$, energy fraction $\alpha_*$, and annihilation temperature $T_*$ \citep{Ferreira:2022zzo}. Their spectra at higher frequencies are flatter than FOPT signals. 
\textit{Cosmic strings}, 1D defects from $U(1)$ breaking, form loops radiating via oscillation and cusps \citep{Kibble:1976sj,Vachaspati:1984gt,Allen:1991bk}, yielding
\begin{equation}
\Omega_{\rm gw}^{\rm cs}\!\sim\!\sum_k (G\mu)^2P_k\mathcal{I}_k(f),
\end{equation}
where $G\mu$ is string tension and $P_k\!\propto\!k^{-q}$ ($q\!\sim\!1$) \citep{Lorenz:2010sm,Gouttenoire:2019kij}. PTAs now probe $G\mu\!\sim\!10^{-11}$–$10^{-8}$ \citep{Yonemaru:2020bmr,EuropeanPulsarTimingArray:2023lqe,NANOGrav:2023NewPhysics}. 

\subsection{Non-GW and exotic correlated signals}

PTAs also respond to coherent non-GW phenomena that mimic or add to the GWB. 
This includes signals caused by dark matter. Some models predict dark matter is produced by low mass particles, which can interact with normal matter through gravitation, and potentially through scalar or vector fields. 

\textit{Ultralight dark matter} (ULDM) with $m\!\lesssim\!10^{-18}$\,eV produces coherent oscillations \citep{Khmelnitsky:2013lxt,Porayko:2014rfa,Porayko:2018sfa,Xia:2023hov,PPTA:2022eul}, yielding a common signal $|\delta t|\!\simeq\!0.02\,{\rm ns}(m/10^{-22}{\rm eV})^{-2}(\rho_{\rm DM}/0.4\,{\rm GeV\,cm^{-3}})$ at $f_0\!\simeq\!4.8\times10^{-8}{\rm Hz}(m/10^{-22}{\rm eV})$. PTA analyses exclude gravitational ULDM down to $m\!\lesssim\!10^{-21}$\,eV and place competitive limits on conformal couplings \cite[][]{EPTA:2023uldm,Smarra:2024ConformalULDM,EPTA:2024DR2-IV,NANOGrav:2023NewPhysics}. 

If a dark matter scalar field  couples to photons via a Chern–Simons term \cite[][]{d648afea-bdd6-33d6-bac5-32fb94e6cccf}, it rotates the linear polarization of pulsar emission, i.e., cosmic birefringence \citep{Sikivie:1983ip,Carroll:1989vb}. The induced angle difference $\Delta\theta(t)=g[a(t-r_{\rm psr}/c,\mathbf{x}_{\rm psr})-a(t,0)]$ depends on coupling $g$ and field $a$, producing correlated sky patterns \citep{Liu:2019brz,Caputo:2019tms,Castillo:2022zfl,2023PhRvL.130l1401L,Li:2025xlr}. Polarimetry limits from EPTA DR2 and the PPTA DR3 have been used to set complementary bounds on axion-like dark matter \citep{Porayko:2025EPTA-Polarimetry-ALDM,Xue:2024PPA-ALDM}. 

Vector fields kinetically mixing with photons or gravitons can produce narrow-band strain backgrounds with $\Omega_{\rm gw}\!\sim\!10^{-13}$–$10^{-11}$ near $f\!\sim\!10^{-8}$–$10^{-6}$\,Hz \citep{Co:2022bqq}. If vector fields couple with Standard Model particles, their effect can be directly observed in PTA \citep{PPTA:2021uzb}. Scalar–tensor couplings to baryonic mass or $\alpha$ variations induce time-dependent potentials \citep{Arvanitaki:2015iga,Blas:2016ddr}; PTA data constrain  $g_\phi\!\lesssim\!10^{-9}$ for $m_\phi\!\sim\!10^{-23}$–$10^{-21}$\,eV \citep{Nomura:2020tpc}. PTA are also capable of detecting dark matter substructures \citep{Siegel:2007fz,Lee:2020wfn,Kim:2023kyy} through gravitational effects.

\subsection{Other early-Universe contributions}

Additional mechanisms may add subdominant low-frequency power: An ``Audible axion'' is an axion or axion-like particle that couples with dark photons \citep{Machado:2018nqk,Ellis:2024AandA};
primordial gravitational collapses can also introduce a signal \citep{Zeng:2025law}; 
relic-neutrino damping altering the tensor slope \citep{Weinberg:2004kr,Boyle:2005se}; and modulated reheating or curvature variations imprinting secondary tensor modes \citep{Domenech:2021ztg,Inomata:2021zel}.

\noindent
PTAs thus probe physics from QCD and dark-sector transitions to axion-like and vector dark matter, providing a direct window onto pre-recombination dynamics.
The SKA Observatory will be able to contribute by accurately measuring the properties of the signals it detects.  The large number of pulsars and high timing precision will enable the spatial and temporal correlations of all signals measured to be precisely characterized.

\section{Noise sources in pulsar timing array experiments}
\label{sec:noise}

In order to study the sources of nanohertz frequency GWs and make best use of a PTA, it is essential to identify, model, and, where possible, eliminate noise from the observations.  

In addition to thermal noise innate to a telescope (which can be reduced through lower system temperature and mitigated with larger gain, longer integration times, and wider bandwidths), pulsar timing observations include additional sources of noise that are largely astrophysical in origin \cite[][]{2010arXiv1010.3785C,2018CQGra..35m3001V,ng15detchar}. 

The noise processes are usually divided into phenomenological classes, based on their degree of temporal correlation and chromaticity, i.e., the dependence on radio frequency.  
Processes (assumed to be) uncorrelated between individual observing epochs are termed ``white noise'' sources.  
Red noise processes exhibit long time correlations between observations. 
They are often characterized in the Fourier domain by a red power-law spectrum.
We note that the terms red and  white do not imply the degree of chromaticity.

Noise processes can be achromatic, affecting all frequencies of radio emission identically. 
A GWB comprises one such red noise source.
However, many processes show a high degree of chromaticity \cite[][]{2010arXiv1010.3785C,2017MNRAS.464.2075S}, and these are often among the strongest sources of noise in MSP observations.
While many noise sources are expected to be spatially uncorrelated, there are stochastic processes associated with an observatory, the Earth, and the Solar system that can be common to all pulsars \cite{2016MNRAS.455.4339T}. 
The nature of the noise, and the covariance of the noise with GW signals of interest, dictate how it is best mitigated.

\subsection{Red noise sources}

\subsubsection{Spin noise}

It is well known that slowly spinning canonical pulsars exhibit rotational irregularities that manifest as spin noise \cite[][]{1975ApJS...29..453G}.
While millisecond seconds pulsars are much more stable rotators,   they are also intrinsically unstable at some level \cite[][]{2010ApJ...725.1607S}.
The noise can potentially induce fluctuations in residuals with similar redness to that of a GWB, potentially causing false detections of a common noise \cite[][]{2022MNRAS.516..410Z}.
It has also been proposed to use  population level models for spin noise \cite[][]{2022ApJ...932L..22G,2024ApJS..273...23V} to derive unbiased properties of the GWB.
Such models could be developed using MSP observations observed as part of the SKAO-PTA and the IPTA, in combination with studies of spin noise in young (non-recycled) pulsar timing data sets.

\subsubsection{Secular profile variability}

Pulsar timing typically assumes that the observed profile converges to an identical pulse profile at each epoch when calculating pulse times of arrival (TOAs). If this assumption is broken there will be biases in measured pulse arrival times.
While secular changes in pulse shape have been known about non-recycled pulsars for half a century \cite[][]{1970Natur.228.1297B}, only recently have shape variations been observed in millisecond pulsars.
The most dramatic event was the sudden and significant distortion in the pulse profile of (at the time) one of the most important millisecond pulsars for PTA work, PSR~J1713$+$0747 \cite[][]{2021MNRAS.507L..57S,2024ApJ...964..179J,2025arXiv250918972M}. 
However, it has also been observed in other millisecond pulsars \cite[][]{lkl+15,2016ApJ...828L...1S,2021MNRAS.502..478G}, and is relatively prevalent in slower spinning pulsars \cite[][]{2025arXiv250103500L}.
Some MSPs also show evidence for mode changing \cite[][]{2022MNRAS.510.5908M,2023MNRAS.523.4405N}, in which the pulse profile appears to converge to one of two (or a few) distinct states \cite[][]{1970Natur.228.1297B}.

GWs do not cause pulse shape variability. Pulse timing techniques that both simultaneously model pulse shape variability and search for signal of interest can break covariances between shape variations and GWs in timing measurements.  Such techniques have been  developed \cite[][]{2017MNRAS.466.3706L,2022MNRAS.510.5908M,2023MNRAS.523.4405N, 2024ApJ...972...49L} and will become important with the high fidelity pulse profiles SKAO PTA will deliver.

\subsection{Noise from the interstellar medium}

As radio waves propagate through the interstellar medium (ISM), they are affected by the tenuous ionized interstellar medium (IISM).  The medium disperses the radio emission due to a frequency-dependent refractive index, resulting in radio emission at lower frequencies arriving later than that at higher frequencies. The degree of refraction and hence dispersion depends on the density of the interstellar medium, with the dispersion dependent on the total column density, which is measured in units of dispersion measure (DM, see \citealt{Tiburzi2025_SKA_SKAPTA}).

As the pulsar, Earth, and IISM move, the pulsar-Earth line of sight probes different columns of plasma.
The largest stochastic process in the arrival times arises from variations in dispersion measure  \cite[][]{2007MNRAS.378..493Y,2016ApJ...821...66L,Donner:2020AandA644A153}.

As the IISM is inhomogeneous \cite[][]{1995ApJ...443..209A}, the variable refractive index  affects the propagation direction of the radio emission.  This results in multi-path propagation effects, where radio waves travelling along many lines of sight are received at the telescope.  The effects can be bifurcated into diffractive and refractive effects which cover the results of propagation effects on small and larger scales respectively.  In pulsar timing observations  diffractive effects  manifest as diffractive scintillation and pulse broadening \cite[][]{1990ARA&A..28..561R}. 
This multi-path propagation produces frequency-dependent timing delays that typically scale close to $\nu^{-4}$ to $\nu^{-4.4}$. Those stochastic scattering variations introduce additional chromatic noise in the timing data, which typically modeled assuming   a power law. Crucially, scattering noise is distinguishable from DM noise through wide-band observations. Recent PTA works have identified significant scattering noise in multiple pulsars \citep[][]{2023ApJ...951L...7R,2023A&A...678A..49E,2024MNRAS.535.1184S}. The wider frequency coverage and enhanced sensitivity of SKAO telescopes will render scattering noise increasingly prominent, necessitating careful  modeling of scattering noise. 

To mitigate propagation effects, it is necessary to observe pulsars at a range of radio frequencies. For PTA experiments aiming at GW detection, timing precision at `infinite frequency' -- i.e., the timing precision after correcting the dispersion measure fluctuations -- is crucial.  Optimal observing strategies have been studied \citep{2012MNRAS.423.2642L,2018ApJ...861...12L,Iraci:2024AandA,2025arXiv251103185K}, and both the waveform driven \citep{2015ApJ...813...65N} and the stochastic model-driven method \citep{2014MNRAS.441.2831L} to correct dispersion measure fluctuations have been developed. Concurrent observations spanning wide ranges of radio frequency, particularly using multiple sub-arrays and radio frequencies between 30 MHz to 2 GHz, have shown the potential to measure DM variations up to sixth decimal places, significantly mitigating chromatic noise associated with DM variations, scattering and solar wind \citep{2020A&A...644A.153D,2022JApA...43...98J,2022PASA...39...53T,2025PASA...42..108R,Susarla:2024AandA}. The SKAO telescopes  provide this wide frequency coverage, with SKAO-Mid telescope's band 1, 2 and 5, together with SKAO-Low  telescope providing nearly continuous frequency coverage from 50 MHz to 15 GHz. Such wide frequency coverage and high sensitivity at frequencies above 3 GHz provide a unique opportunity to decouple the achromatic and chromatic noise processes from e.g. dispersion measure and scattering variations \citep{2020A&A...644A.153D,2022JApA...43...98J,2012MNRAS.423.2642L}.
This will be especially important in the SKAO era, where more distant pulsars (which are more strongly affected by propagation effects) will be included in PTAs.

\subsection{White noise sources} \label{sec:white_noise}

\subsubsection{Jitter Noise}
Pulsar radio emission exhibits clear pulse-to-pulse variability, a phenomenon long observed in canonical pulsars. In recent years, this variability has also been revealed in a number of MSPs \citep[e.g.,][]{ovb+14,lkl+15,lbj+16,2021MNRAS.502..407P,lab+22,2025arXiv251003139G}, thanks to the construction of highly sensitive telescopes and the development of high-time-resolution data recording systems. The variation can be either systematic, such as sub-pulse drifting and mode change \citep[e.g.,][]{wes06,lbj+16,2022MNRAS.510.5908M}, or stochastic. The stochastic process introduces random phase and amplitude variations in integrated pulse profiles, and thus in their arrival time measurements, which can exceed the level expected from radiometer noise.
This process, termed jitter noise \cite[][]{2010arXiv1010.3785C,lkl+12} is present in all pulsar observations but dominates when the signal-to-noise ratio of individual pulses exceeds unity.  In this regime the pulse shape variations are expected to be larger than the radiometer noise \cite[][]{2010arXiv1010.3785C}.  In the case of MSPs, this is the case for the brightest pulsars, or with sensitive telescopes such as the SKAO  \citep{lvk+11,2014MNRAS.443.1463S}.

On narrow frequency scales, the jitter noise is fully correlated across the band and of the same amplitude \citep[e.g.,][]{lkl+12}. However, it is likely that jitter noise is not fully broad band and de-correlates over fractional bandwidths close to unity \cite[e.g.,][]{2014MNRAS.443.1463S,2024MNRAS.528.3658K}.
The magnitude of jitter noise also depends on observing frequency.
Several studies have reported that the magnitude of jitter noise  in millisecond pulsars such as PSR~J0437$-$4715 tends to increase toward lower frequencies~\citep{2014MNRAS.443.1463S, 2021MNRAS.502..407P}.
However, recent InPTA measurements of PSR~J0437$-$4715 with the uGMRT have provided jitter amplitudes in the low-frequency range ($300-500$\,MHz), finding values comparable to those at higher frequencies around $1.4$\,GHz. This result may challenge previous studies which suggested that jitter amplitude increases at lower frequencies, and implies that low-frequency observations in future SKAO surveys may not always be severely limited by jitter noise~\citep{2024PASA...41...36K}. These studies further demonstrate that for timing analysis using wide-band data, jitter noise may need to be modeled independently in multiple parts of the band \citep[e.g.,][]{2021MNRAS.502..478G,2023ApJ...951L...7R,2024MNRAS.528.3658K}.

Jitter noise is expected to decrease as the square-root of the number of pulses averaged, and thus integration time as well. In addition, some frameworks have been developed to mitigate jitter noise from the outset. This can be accomplished using either integrated pulse profiles \citep[e.g.,][]{ovh+11}, or directly single pulses \citep[][]{ker15,2023MNRAS.523.4405N}. So far, none of the schemes has been successfully implemented in a long-term PTA dataset. The SKAO sensitivity will enable detailed study of single pulses for a large number of PTA MSPs, providing the best opportunity to facilitate jitter noise mitigation in the PTA timing dataset.

The flexibility of SKAO telescopes also allows for a more efficient observing strategy, when strong jitter noise is present. For bright pulsars limited by jitter noise, SKAO telescopes  can be divided into sub arrays  \citep{2022JApA...43...98J,2025arXiv251003139G}, where individual sub-arrays are observing different pulsars. This effectively provides more observing time, significantly aiding the jitter-dominated situation \citep{2012MNRAS.423.2642L}.   In analysis of MSPs observed as part of the MPTA, \cite{2025arXiv251003139G}  found that for $10$ of the $83$ pulsars, more than half of the observations were jitter dominated. However this increased to over half of the sample when extrapolating to SKAO  sensitivity, assuming the AA4 configuration. 

\subsubsection{Other white noise sources}

There are several other types of white-noise sources that can be present in a PTA dataset \citep{lvk+11}. For instance, stochastic variations in the pulse broadening can also introduce short term variations in the pulse profile. The stochasticity arises from stochasticity in the scatter broadened image of the pulsar and is termed the finite-scintle effect \cite[][]{2010arXiv1010.3785C}. Instrument-related effects caused by radio frequency interference, imperfect polarization calibration, insufficient bit sampling can also distort the shape of integrated profiles and introduce systematic bias in their arrival times. Additionally, errors in the cross-correlation to measure the arrival times can also introduce white noise in the timing data \citep{lvk+11}. This can be caused by significant mis-match of the template shape to the intrinsic pulsar signal. 

\subsection{Spatially correlated noise sources}

It is of utmost importance to mitigate the effects of processes that can impart signals that are correlated between pulsars.   
\cite{2016MNRAS.455.4339T} investigated many of the noise sources that can introduce such spatial correlations.  
The most common spatial correlations are monopolar and dipolar correlations, which reflect processes related to the spacetime reference frame.
Pulsar timing is necessarily undertaken in an inertial reference frame \citep{lk2004}.  
The arrival time measurements are referred to local atomic clocks, which need to be transferred to TT(BIPM), the most stable terrestrial time (TT) standard.  
The arrival time at the observatory should also be transferred to the solar system barycentre, using a solar system ephemeris (SSE) which provides the positions of the major objects in the solar system \cite[e.g., DE440,][]{2021AJ....161..105P}. 
Errors in the terrestrial time standard  affect all pulsars in the same way and would induce a monopolar correlation.
In contrast, errors in the solar system ephemeris would induce a dipolar correlation. 

Such spatial correlations are noise processes in the context of GW detection, but serve as important means to construct a pulsar-based spacetime reference frame,
The long-term stability of some MSPs is comparable to that of an atomic clock, so it is potentially possible to improve the performance of terrestrial time scales using PTA data \citep{2012MNRAS.427.2780H,Hobbs2020}.
PTA observations can be used to measure the mass of major planets and massive asteroids in the solar system and constrain possible unmodeled objects in the solar system \citep{Champion2010,Caballero2018,Guo2018}.
Despite the precision achievable using current timing data, the special meaning of the pulsar timing method is to provide a completely independent way to examine the space-time standard.

In terms of GW searches, such spatially correlated noise needs to be carefully checked and mitigated or modeled. 
PTAs provide a good way to check the transfer from the observatory time standard to TT. If the time transfer is performed correctly, the error in TT(BIPM) is less concerning for the current timing precision.
Errors in solar system ephemeris were thought to be an important source of noise budgets for PTA data, and SSE errors may induce measurable biases on the GW detection statistics obtained from PTA
data sets \citep{NG11}. However, the
more recent searches have been found to be insensitive to the choice of SSE \citep{2020ApJ...905L..34A,2021ApJ...917L..19G,2021MNRAS.508.4970C}. The covariance between SSE errors and the GWB signal could depend on the specifics of the data set, such as the timing precision, the total time span, and the number of pulsars. 
For SKAO PTA with high timing precision and (initially) short data span, the role of SSE error still needs to be examined.

Several approaches have been developed to model the SSE uncertainties. One way relies on the fact that an SSE error has dipolar correlation, such as searching for dipolar-correlated red noise \citep{2016MNRAS.455.4339T}, or using spherical harmonics to subtract the dipolar mode \citep{Roebber2019}. Another approach is to directly correct the timing residuals based on the physical model of the solar system, including a mass-perturbation method \citep{Champion2010}, quasi-Keplerian approximation \citep{2020ApJ...893..112V}, or numerical dynamical model \citep{Guo2019,Guo2024}. 

\subsection{The solar wind}

The solar wind (SW) introduces time-variable dispersion delays in pulsar signals that must be modeled as part of the PTA noise budget. The SW plasma is a medium that the line-of-sight of all of the PTA pulsars cross -- therefore, if not correctly mitigated, it induces spatial correlations that can mimic a false GWB detection \citep{2016MNRAS.455.4339T}. Until 2022, PTA analyses corrected for a simple $1/r^2$ SW model (with $r$ being the distance from the Sun) using a constant electron density at 1~AU, as implemented in standard pulsar timing software \citep[][]{2006MNRAS.372.1549E, Madison2019}. However, static models are insufficient: residual DM fluctuations of $\sim 10^{-4}$–$10^{-3}$~pc~cm$^{-3}$ remain \citep{Tiburzi2019}, especially near solar conjunctions, and can bias other noise and timing parameters or mimic low-frequency signals \citep{Lam2016,Madison2019,Liu2025}. The dynamic state of the SW and its detectability in pulsar-based measurements has been known for decades \citep{Counselman1972,Bird1980}, but only recently have PTA analyses begun integrating these effects rigorously into the noise model.

Recently,  linear Gaussian process (GP) models have been developed  \cite[][]{2022ApJ...929...39H,Nitu2024,Susarla:2024AandA} that treat the SW amplitude as a time-variable stochastic process, enabling the electron density at 1~AU to vary smoothly in time. This approach improves over the standard static correction and has been implemented in PTA pipelines such as \texttt{enterprise} \cite[][]{2020zndo...4059815E}. The current PTA approach to SW modeling was presented in \citet{Susarla:2024AandA}, where the authors finalized the ``SW GP'' algorithm developed in \citet{2022ApJ...929...39H} and tested it over a decade of LOFAR data. The authors also demonstrated strong correlations between SW variability and average SW electron density, and pulsar ecliptic latitude.
In particular, this confirms the trend presented in \citet{Tiburzi2021} with high-latitude pulsars showing SW-induced DM trends following the Solar cycle over years, and where low-latitude pulsars displaying instead elevated and more stochastic DM progresses, consistent with the slow solar wind.

Such data-driven modeling enables the separation of heliospheric and interstellar dispersion and improves timing precision near solar conjunction. Transient events—such as coronal mass ejections (CMEs)—require separate detection \citep{Shaifullah2020} and treatment. \citet{2021A&A...651A...5K} reported a CME-induced DM spike in PSR~J2145$-$0750, and similar use of pulsars for CME detection has been demonstrated by \citet{Wood_2020}.

\section{A realistic pulsar timing array for the SKA Observatory}
\label{sec:ska_obs}

Here we provide a forecast for what one type of PTA observing campaign with the SKAO telescopes could look like, to highlight the key role SKAO can play in future PTA experiments.   We emphasize that this campaign is not fully optimized.  We expect full optimization of the SKAO PTA to be developed over the coming years.  We focus our efforts on precision timing with SKAO-Mid, and consider both the AA$^*$ and AA4 deployments.   
We can base our forecast on observations of a census of MSPs \cite[][]{2022PASA...39...27S} undertaken with MeerKAT as part of the MeerTime Large Survey project \cite[][]{2020PASA...37...28B}.
$189$ recycled pulsars visible to the SKAO-Mid telescope were observed at more than $8$ epochs using the MeerKAT L-band observing system.  A smaller subset of $>80$ have been observed more regularly as part of the MeerTime PTA~\cite[][]{2023MNRAS.519.3976M,2025MNRAS.536.1467M}.

A formalism for calculating PTA sensitivity to both backgrounds and individual sources has been developed \cite[][]{2019PhRvD.100j4028H}, extending on previous sensitivity forecasting methodologies \cite[][]{2013CQGra..30v4015S}. The formalism can be used both to estimate the signal to noise ratio of a GWB (of specified amplitude and spectral index) or individual source (of specified amplitude and emitting frequency). It can also be used to generate sensitivity curves commonly used to characterize GW detectors.  


We assume that AA$^*$ is  a factor of three more sensitive than MeerKAT and that AA4 is a factor of four more sensitive than MeerKAT.  We base estimates of arrival time precisions obtained from the MeerTime MSP census \cite[][]{2022PASA...39...27S} and the MPTA \cite[][]{2023MNRAS.519.3976M}.  We can use spectral index measurements from the MSP census to extrapolate achievable timing precision to different observing bands.

We base our strategy on what has been implemented with the MPTA. 
We calculate the amount of time it takes to achieve $1\,\mu$s timing precision for a pulsar using either SKAO or a MeerKAT sized sub-array. 
It is observed with a MeerKAT-size subarray if such timing precision can be achieved in   $< 256$\,s.
If a pulsar can be observed with 1\,$\mu$s timing precision with AA$^*$ or AA4 in less than $256$\,s, then it is observed for that duration.
For pulsars that require $>256$\,s to achieve this timing precision we observe for that duration. 
Pulsars that require $\gtrsim 2000$\,s of integration time are not considered for inclusion in the array. 
We assume an observing cadence of two weeks.

We find that such an array can observe $174$ MSPs with $8.9$\,hr of integration time with AA4. For AA$^*$ the array can achieve similar timing precision  with $12.4$\,hr integration time per epoch. 
This highlights the importance of sub-arrays in enhancing observing efficiency. 
A similar strategy, chosen for use by the MPTA, can achieve the same goals with $12$\,hr integration time per epoch on $83$~MSPs.

\subsection{Sensitivity to a GWB}

\begin{figure}
\centering
\begin{tabular}{c}
\includegraphics[width=0.43\textwidth]{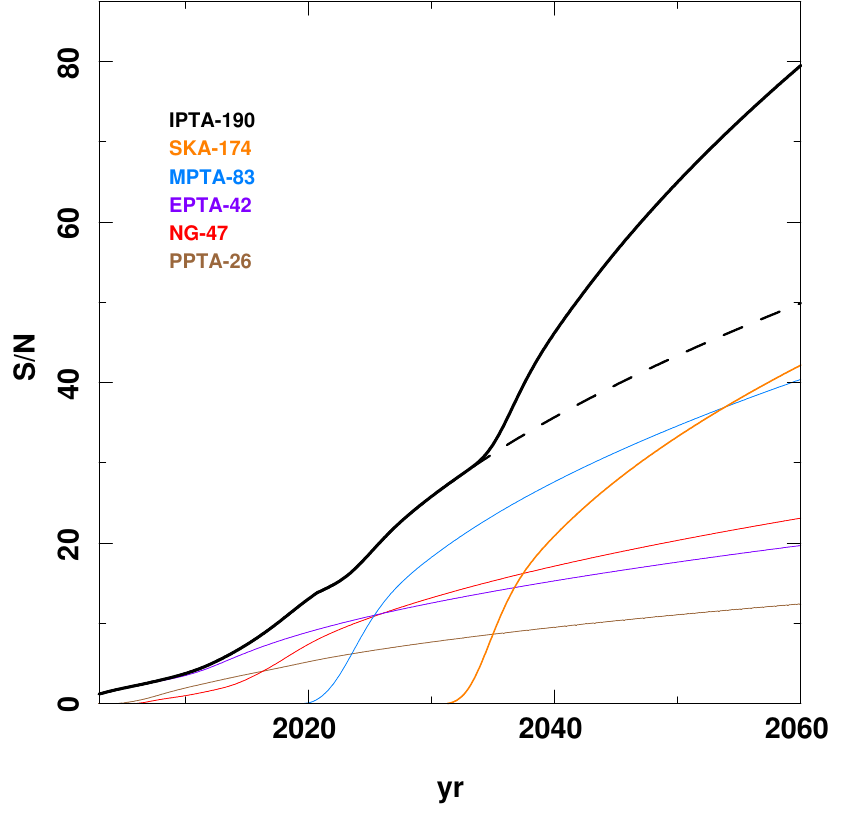} 
\end{tabular}
\caption{\em \label{fig:pta_compare}  
A comparison of PTA sensitivities.  Curves were obtained using methodology discussed in \citep{2022PASA...39...27S}.
}
\end{figure}

Forecasts for PTA sensitivity to a GWB based on scaling relations of \cite{2013CQGra..30v4015S} can be seen in Figure \ref{fig:pta_compare}.  
We compare the forecast sensitivity of this proposed SKAO timing program with ongoing PTA experiments as they were operational in 2020, assuming a GWB of amplitude  of $A_{\rm yr} = 2\times 10^{-15}$. 
The sensitivity uses the median timing precision and cadence observed for each pulsar as part of a given timing array.
When an array has a smaller number of pulsars, signal to noise ratio soon scales with time   $\propto T^{1/2}$ (where $T$ is the observing span), as the GWB  becomes self noise for itself. 
With a large number of pulsars timed to high precision, the SKAO-PTA dominates the international sensitivity within four years of operation. The SKAO-PTA continues to increase in sensitivity quickly as the array remains in the intermediate signal limit for longer owing to the larger number of pulsar pairs.

\begin{figure*}
\centering
\begin{tabular}{cc}
\includegraphics[width=0.43\textwidth]{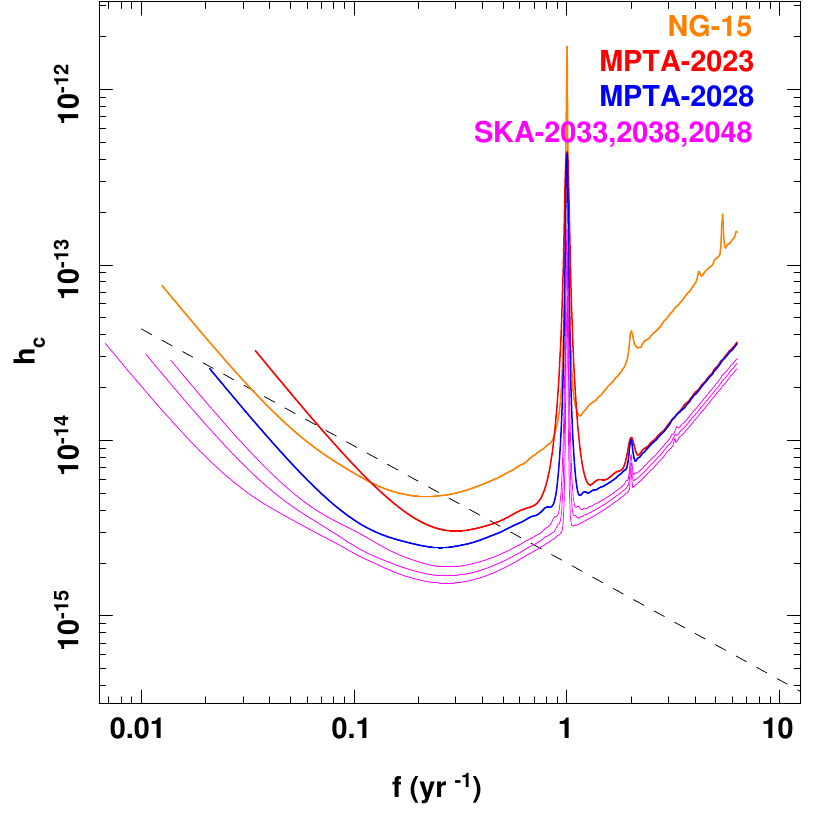} & \includegraphics[width=0.43\textwidth]{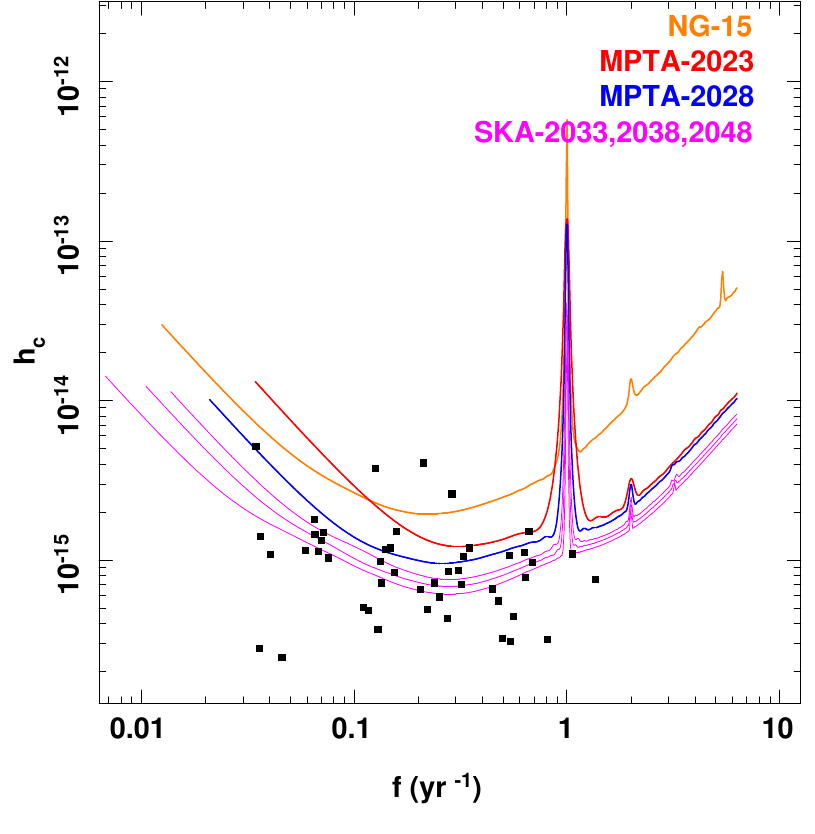} \\
\end{tabular}
\caption{\em \label{fig:strain_compare}  
Strain sensitivity curves for the MeerKAT Pulsar Timing Array and a conceptual SKAO Pulsar Timing Array after observing for $5$, $10$, and $15$ years.  The left panel shows the sensitivity to a GWB. The dashed line shows the spectrum of a GWB at an amplitude of $A_{\rm yr}=2\times 10^{-15}$. The right panel shows the sky-averaged sensitivity to single source. The black points highlight individual sources detectable by the PTA.  The curves highlight where pulsar timing array experiments have the greatest sensitivity.
}
\end{figure*}

Array sensitivity can be also visualized and assessed through sensitivity curves. These curves show array sensitivity to either a GWB or sky averaged sensitivity to an individual source.
In Figure \ref{fig:strain_compare}, we show sensitivity curves for the SKAO PTA after $5$, $10$, and $20$ years of observing, and compare them to that from the MPTA 4.5 year data release \cite[][]{2025MNRAS.536.1467M} and NANOGrav 15-year data set \cite[][]{2023ApJ...951L...9A}.
We find that compared to other PTA experiments that the SKAO-PTA will have greater sensitivity across the GW spectrum. 
These sensitivity curves can also be used to forecast a signal to noise ratio. We find results that are in general agreement with those calculated using scaling relations as presented above.

We emphasize that  the sensitivity calculations presented above are likely to be significantly refined 
considering the wide range of frequencies available to SKAO. In a strategy similar to that employed by InPTA using the SKA pathfinder telescope 
uGMRT, concurrent observations using different sub-arrays observing with  Band 1 and 2 
at the SKAO-Mid telescope simultaneously with the SKAO-Low telescope will provide 
more accurate instantaneous DM, scatter broadening, and solar wind measurements. Currently, 
the near simultaneous measurements by most PTAs implies that such characterization of 
chromatic noise is limited by the day-to-day fluctuations in the IISM biasing both the white 
and red noise posteriors. The resultant reduction in the sensitivity can therefore be 
avoided by adopting such observing strategy. The two telescope nature of SKAO, the 
wide primary beam of the SKAO-Low telescope, the flexibility of using multiple 
sub-arrays with 16 pulsar timing beams in both the SKAO-Low and SKAO-Mid telescope allows 
such flexible, efficient and more effective observing strategies  
\cite[See][for possible configurations of the two SKAO telescopes that can be used]{2022JApA...43...98J}. 

Furthermore, additional MSPs suitable for inclusion in an SKAO PTA have been found since the MeerTime MSP census was undertaken, and will continue to be discovered into the SKA era. This will allow for further expansion  and optimization of the SKAO PTA.

\subsection{Prospects for detecting individual sources}

To assess the detectability of individual continuous gravitational wave (CGW) sources, we use the signal-to-noise ($\rm S/N$) computation introduced in \citep{2025A&A...694A.282T}, applied to a 200 SMBHB populations generated with the \texttt{L-Galaxies} semi-analytical model\citep{2022MNRAS.509.3488I}, applied on the \texttt{Millennium} simulation \cite[][]{2005Natur.435..629S}. We consider the same SKAO PTA observing strategy used to search for and study a GWB described in the previous sections, in particular, an array comprising 174 known pulsars. 

The total noise power spectral density of each pulsar is given by
\begin{equation}
    S_k(f) = S_{\rm w} + S_{\rm GW}(f),
\end{equation}
were $S_{\rm w} = 2\Delta t_{\rm cad} \sigma_{\rm w}^2$ quantifies the stochastic uncertainty in pulse arrival times, $\Delta t_{\rm cad}$ is the cadence time, which will be typically two weeks.  $\sigma_{\rm w}$ encodes all the temporal uncorrelated noise processes related to the telescope sensitivity and pulse instability (see Section \ref{sec:white_noise} ). The term $S_{\rm GW}(f)$ represents the GWB noise power spectral density, produced by an inspiralling SMBHB population, evaluated at a given frequency 
 bin of the array (i.e., $\Delta{f_i}=[i/T_{\rm obs},(i+1)/T_{\rm obs}]$, with $i=1,...,N$).
 
 As such, the strain-squared ($h_c^2$)  associated with each frequency  bin can be written as 
\begin{equation}
    h_c^2(f_i) \,{=}\, \sum_{j=1}^{N_S} \sum_{n=1}^{\infty}  h_{c,n,j}^2(nf_k) \, \delta(\Delta{f_i}\,{-}\,nf_k),
\end{equation}
where the sum is over all sources, $N_S$, and $\delta(\Delta{f_i}\,{-}\,nf_k)$ is a delta function that selects only SMBHBs emitting within the considered frequency bin. The index $n$ accounts for the GW emission of eccentric binaries, which is distributed across harmonics of the orbital frequency. $h_{c,n,j}^2(f)$ is the squared characteristic strain of the  source $j$. The value of $h^2_{c,n}$ is given by \cite{2010MNRAS.402.2308A}. 

\begin{figure} 
    \includegraphics[width=1\columnwidth]{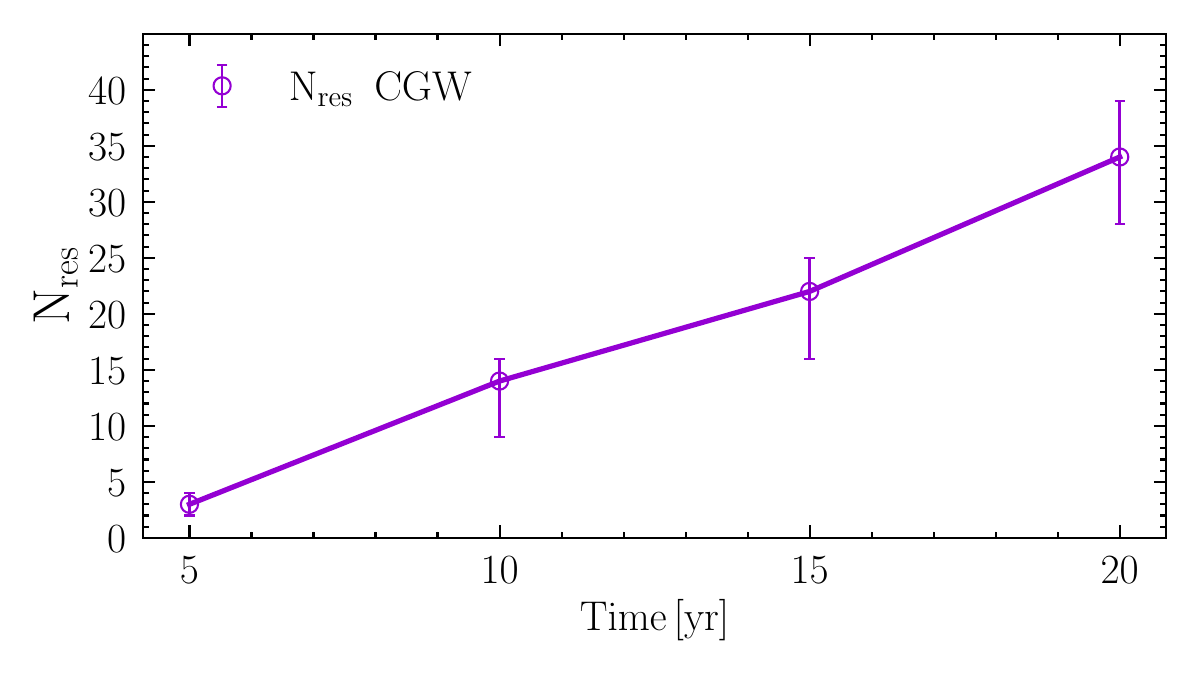}
    \caption{Predicted number of detectable single sources with SKAO-PTA. The open dots represent the median number of resolved continuous gravitational waves  resolved from the 200 SMBHB populations, while error bars represent the 64 and 32 percentiles of the distribution.}
    \label{fig:N_median_CGW}
\end{figure}

We compute the $\rm S/N$, and a CGW emitted by an SMBHB is considered resolved if its $S/N$ exceeds 3. Once resolved, its contribution is subtracted from the SGWB; consequently, the noise budget in the pulsar array is reduced. This lower noise enhances the detectability of fainter CGWs. We repeat this process, reassessing the detectability of the remaining sources with the updated GWB, until no further sources are resolvable. In Figure \ref{fig:N_median_CGW}, we show the median number of resolvable sources as a function of the observing time. The result, though idealized, highlight the capability of the SKAO PTA to resolve multiple CGW in the following years. 

\begin{figure} 
    \includegraphics[width=1\columnwidth]{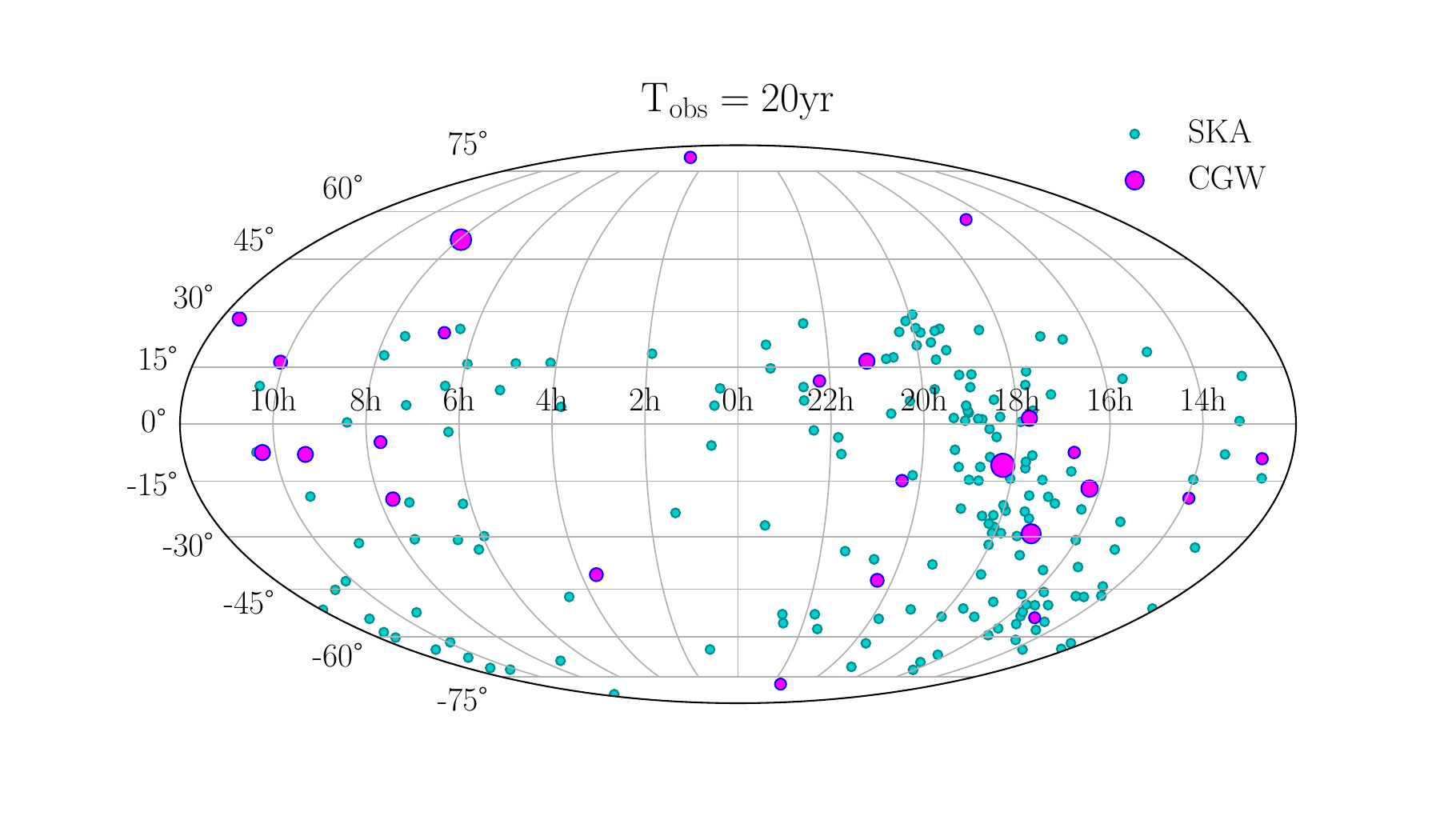}
    \caption{Sky distribution of the SKAO PTA pulsars (light-blue)
    and the SMBHBs detected as continuous gravitational waves (dark violet) in a single realization SMBHB population. The size of the dark violet  points is weighted by the CGW $\rm S/N$ }
    \label{fig:pulsars_CGW_sky_loc}
\end{figure}

The pulsars sky position and location of resolved CGW for a single SMBHB population are shown in Figure \ref{fig:pulsars_CGW_sky_loc}. 

\subsection{Merit of the SKAO PTA for GW sky maps}

The SKAO PTA can  not only improve over current PTAs in terms of the sensitivity, but also in terms of the sky resolution. The long-term timing with the SKAO telescopes will significantly improve the accuracy of the pulsar distance, enabling the arc-min level resolution of the GW sky \citep{2011MNRAS.414.3251L}. In the short run, due to the limited pulsar distance precision, GW strain at the pulsar can not be coherently modeled. As demonstrated in \cite{Grunthal_inprep}, decreasing the angular scale of the GWB fluctuations that a PTA can resolve is a key ingredient to increasing the detectability of small angular scale deviations from isotropy, such as caused by a single, stand-out SMBHB. There are of course challenges. For example, \cite{semenzato2025-leakage} showed that truncating the angular power distribution at any $\ell_\mathrm{max}$ causes power from the higher order multipoles to leak down into the lower order ones, inducing a bias. This will also need to be resolved when interpreting maps of the GW sky.

To demonstrate the improvement that an SKAO PTA can yield compared to the currently operated MPTA, we assume that a potential GWB anisotropy analysis uses similar methods as the MPTA 4.5-year data set analysis \citep{2025MNRAS.536.1501G}. 

As outlined by \cite{Pol_2022,2025MNRAS.536.1501G}, the maximum degree of spherical harmonics constrainable by a PTA of $N_\mathrm{PSR}$ pulsars can be estimated as $\ell_\mathrm{max} = \mathrm{floor}\left( \sqrt{N_\mathrm{PSR}} -1\right)$. Assuming the SKAO PTA containing 174 MSPs as described above, this means $\ell_\mathrm{max}^\mathrm{SKA} = 12$. This conservative, global resolution estimate indicates a significant improvement in angular resolution of the proposed SKAO PTA compared to the MPTA (83 pulsars, $\ell_\mathrm{max}^\mathrm{MPTA} = 8$).

\begin{figure}
    \centering
    \includegraphics[width=\linewidth]{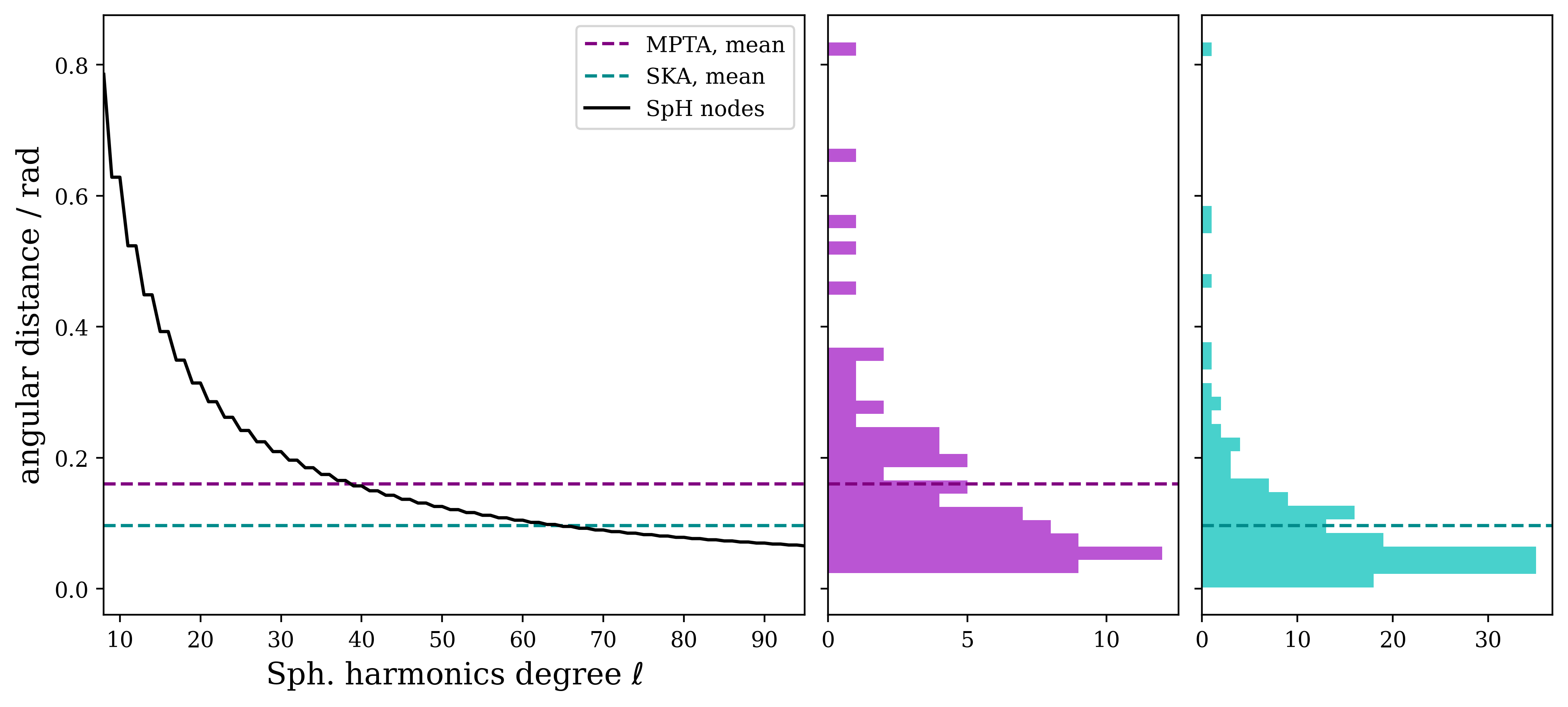}
    \caption{Maximum spherical harmonics degree locally constrainable by the MPTA and proposed SKAO PTA data sets, derived from the angular separations between the pulsars in the respective data sets. Left: Comparison between the minimum angular scale corresponding to the spherical harmonics degree $\ell$  (black curve) and the mean next-neighbor distance in the MPTA (purple line) and the SKAO PTA simulated from the MeerTime MSP census (teal line). Middle: Distribution of next-neighbor distances in the MPTA. Right: Distribution of next-neighbor distances in the proposed SKAO PTA.}
    \label{fig:MPTA-SKAPTA-resolution}
\end{figure}

The resolution potential of the SKAO PTA becomes even more visible using the more realistic estimate strategy developed by \cite{Grunthal_inprep}, which simultaneously respects the diffraction limit of a PTA (i.e., the finite number of pulsars), while accounting for the changing density of pulsars across the sky. Fig.~\ref{fig:MPTA-SKAPTA-resolution} shows that the pulsar-pair separation distribution of the SKAO PTA (right) allows for a local resolution corresponding to a spherical harmonics expansion up to $\ell_\mathrm{max}= 65$\footnote{As outlined in \cite{Grunthal_inprep}, this higher $\ell_\mathrm{max}$ implies regularizing the Fisher matrix inversion following the scheme described in \cite{2025MNRAS.536.1501G}, including not more than its first 174 singular values.}. This corresponds to a mean local angular resolution of 2.7$^\circ$, a factor of $\sim 2$ better than the local angular resolution of the MPTA  (5.8$^{\circ}$) \citep{Grunthal_inprep}.

\begin{figure*}
    \centering
\begin{tabular}{cc}
    \includegraphics[width=0.5\linewidth]{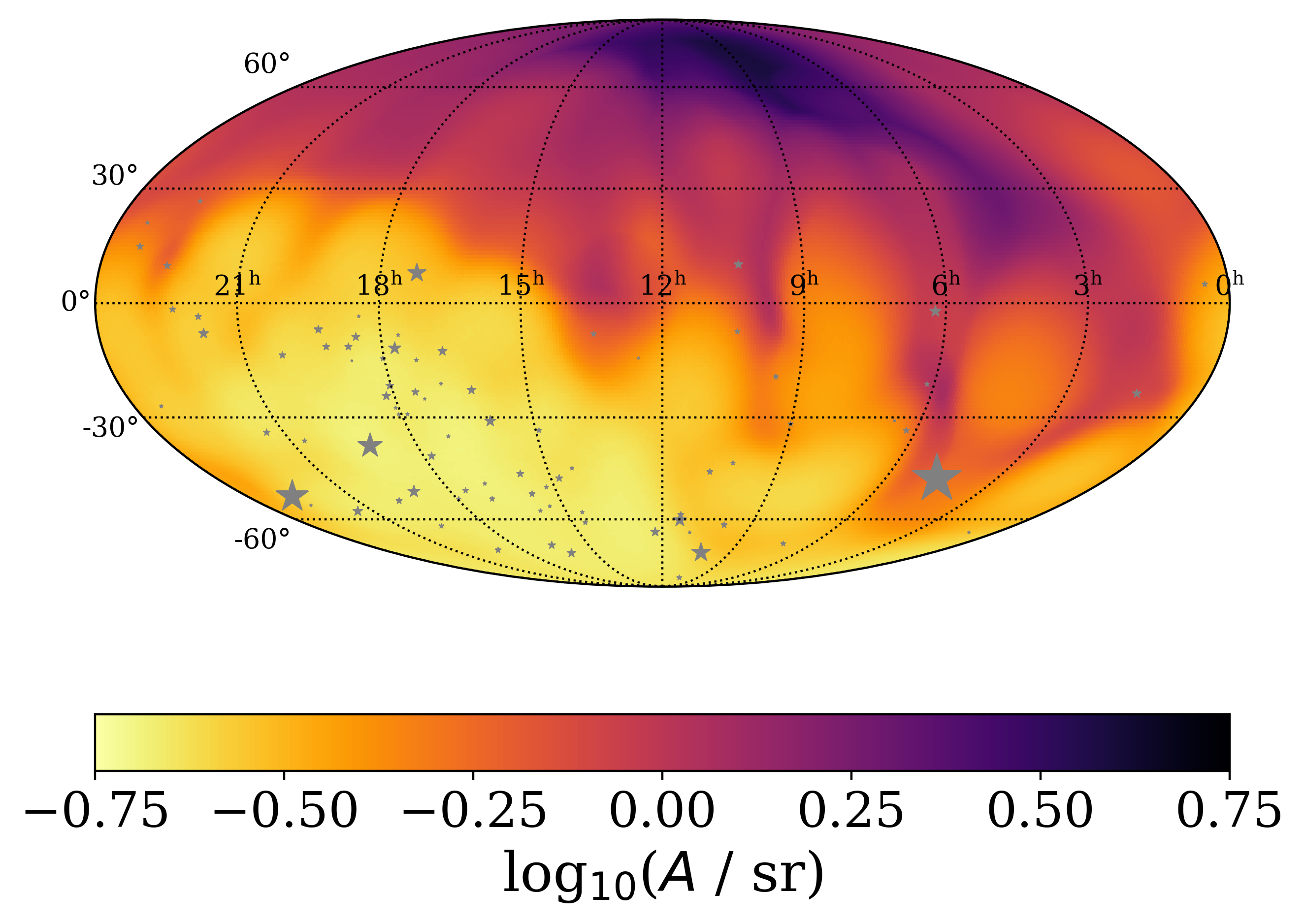} &
    \includegraphics[width=0.5\linewidth]{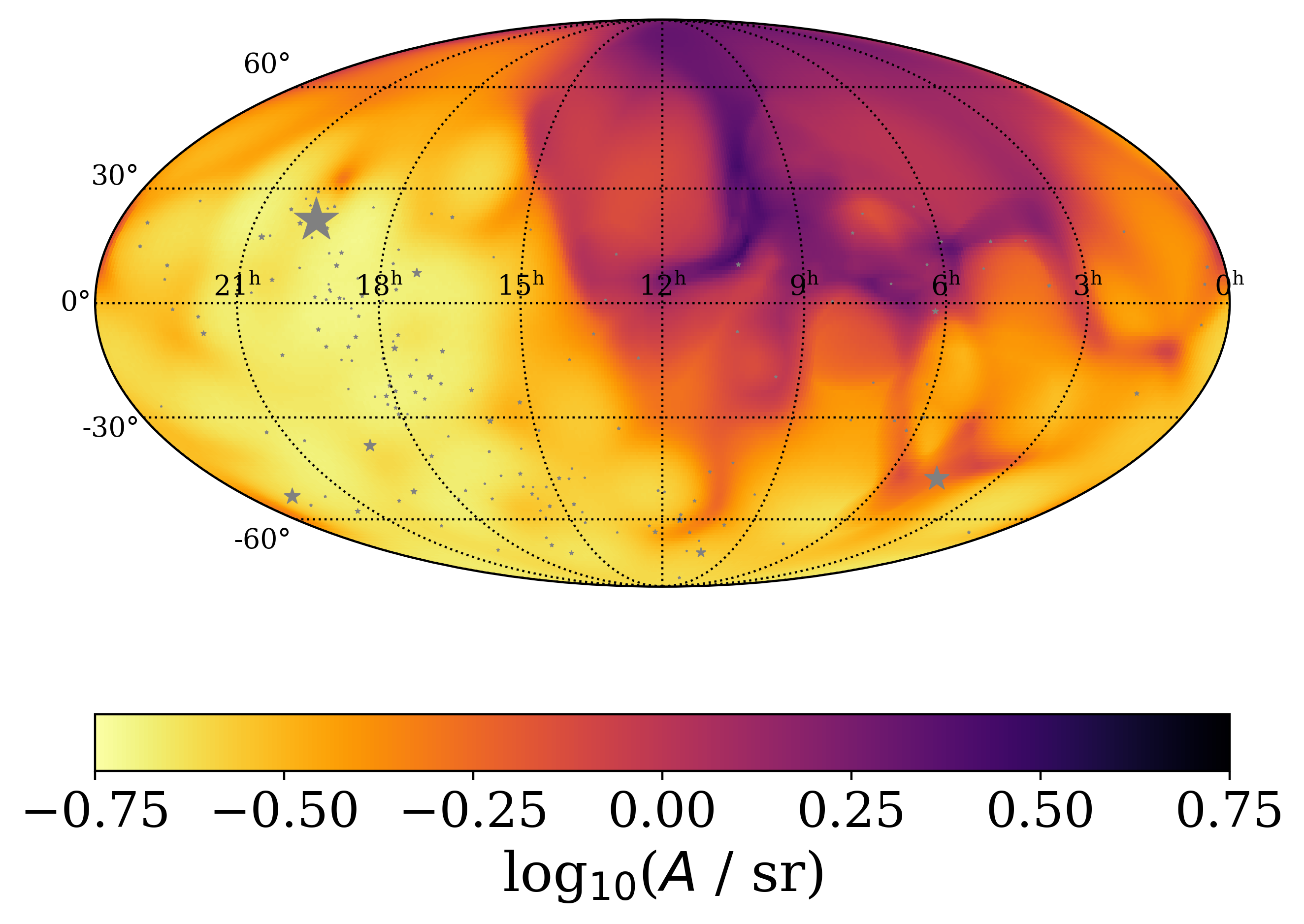}
    \end{tabular}
    \caption{Spread of a point-source (point-spread-area) function of the sky position. Left: A simulated 4.5-year MeerKAT PTA data set. Right: A simulated 4.5-year SKAO PTA data set, with timing precisions based on the MeerTime MSP census. Both simulations assume white noise is the only noise source present in the observations.}
    \label{fig:PSAmaps}
\end{figure*}

Finally, the sky distribution of the point-spread-areas (PSA) of the SKAO PTA compared to the MPTA underline the significant step ahead an SKAO PTA would pose for the calculation of GW sky maps in the nanohertz regime. Fig.~\ref{fig:PSAmaps} shows the PSA distribution for white-noise-only simulations of both the MPTA and the above described SKAO PTA. Each data set spans 4.5 years, and both maps were calculated with $\ell_\mathrm{max}=8$ and a singular value cutoff of $n_\mathrm{SV}=32$, these values were inherited form the MPTA 4.5-year anisotropy analysis \citep{2025MNRAS.536.1501G}. Although the previous investigation has established that this is likely not the best parameter combination for calculating the best-scenario capabilities of the SKAO PTA, it enables a more direct comparison to the MPTA data set. It impressively shows the improvement the SKAO PTA can provide mapping the GW sky compared to the MPTA: The area with comparably small PSAs has grown significantly, especially in the Northern Hemisphere, and the darker shaded area (least sensitivity) is smaller and more sensitive than for the MPTA. Simultaneously, the right plot in Fig.~\ref{fig:PSAmaps} underlines the inevitably outstanding role an SKAO PTA will pose for mapping the GW sky using the combined data sets of all regional PTAs. With no regional PTA providing a comparably sensitive data set for the Southern Hemisphere, the SKAO PTA will serve as an indispensable pillar of any meaningful global\footnote{meaning balanced sensitivities on the Northern and Southern Hemisphere.} nanohertz GW sky map.

\section{The role of SKAO-Low telescope}
\label{sec:ska_low}

While current PTA observing campaigns emphasize decimeter-wavelength timing for maximal raw precision, the low-frequency domain (50–350 MHz) accessible to the SKA-Low telescope will be indispensable for understanding the propagation effects and systematics that limit timing stability. While the SKAO-Low telescope will enable an extensive array of science, as discussed in \citet[][this collection.]{Tiburzi2025_SKA_SKAPTA}, incorporating the SKAO-Low telescope into PTA programs will bridge the gap between precision timing and propagation monitoring, providing direct leverage on chromatic delays, pulse-shape evolution, and DM variability—dominant contributors to the single-pulsar noise budget.

Low-frequency pulsar studies with LOFAR and NenuFAR at, respectively, 100 to 200 MHz and $<100$ MHz have already demonstrated this capability \citep{Porayko:2019MNRASsty3324,Donner:2019AandA624A22,Tiburzi2019,Donner:2020AandA639A77,2021A&A...652A..34B,Tiburzi2021,Wu:2022AandA657A98,Porayko:2023JGeodesy,Susarla:2024AandA,Iraci2025}. These efforts show that high-cadence campaigns at low radio-frequencies can reconstruct the DM and scattering structure of the IISM with sub-$10^{-5}$ pc cm$^{-3}$ precision, constraining stochastic plasma variations that would otherwise contaminate red-noise processes or mimic a GWB signal. Observational studies of pulsar sightlines and local IISM structures have likewise emphasized the need to characterize turbulence, bow shocks, and discrete ionized regions along the line of sight \citep[][]{Ocker2020ApJ, Ocker:2024ApJ_HII, Ocker2024MNRAS}.

The Murchison Widefield Array (MWA) has demonstrated precise DM measurements and scintillation studies of bright southern MSPs \citep{2016ApJ...818...86B,2018ApJS..238....1B,2019ApJ...882..133K,2022ApJ...930L..27K}. Until recently, the limited real-time capability of the array restricted its monitoring cadence, but an ongoing upgrade \citep{2023PASA...40...19M} will soon enable continuous beamformed observations. 
Targeted campaigns on the bright MSP PSR~J2241$-$5236 have already achieved DM precisions of $\sim10^{-6}$ pc,cm$^{-3}$ by exploiting a flexible system design \citep{2019ApJ...882..133K}.  
Contemporaneous observations with the uGMRT and Parkes, which span 70 MHz to 4 GHz, have demonstrated the importance of disentangling profile evolution from chromatic DM effects \citep{2022ApJ...930L..27K}. 
Recent work \citep{2025PASA...42..117L} indicates that a large fraction of southern-sky PTA pulsars are detectable with MWA, offering excellent prospects for high-cadence, low-frequency monitoring that will complement SKAO-Low and provide legacy datasets for future PTA analyses. 
In the full AA4 configuration, the SKAO-Low telescope will push these capabilities to their ultimate limit, reaching DM precisions near $10^{-8}$ pc cm$^{-3}$ \citep{Tiburzi2025_SKA_SKAPTA}, a benchmark that will redefine low-frequency timing and propagation monitoring within PTA experiments.

Complementary progress at higher frequencies demonstrates that the DM precision can be improved even further through instantaneous dual-band (300–500 MHz and 1260–1460 MHz) observations, as shown with the SKA pathfinder uGMRT \citep{2022PASA...39...53T,2025PASA...42..108R}. Joint analysis across this wide frequency range—using both narrow-band \citep{2021A&A...651A...5K} and wideband \citep{2024MNRAS.paladi.527..213P} DM estimation techniques—achieves order-of-magnitude gains in precision when the SKAO-Low and the SKAO-Mid telescopes observe simultaneously. 

Parallel efforts within PTA collaborations are already advancing this integration. LOFAR and NenuFAR data are now incorporated into EPTA-InPTA analyses, with a dedicated LOFAR–EPTA program included in the second EPTA data release \citep{Iraci2025}, while CHIME provides complementary coverage \citep{2025arXiv251016668A} to the NANOGrav collaobration. The third data release of the IPTA 
will present results from the first combination of data from LOFAR, NENUFAR, uGMRT, and CHIME. The synergy among SKA-Low and current high- and low-frequency instruments will deliver a contiguous frequency baseline spanning more than an order of magnitude, enabling detailed modeling of pulse-profile evolution and DM transfer functions. This information will be essential for precise chromatic calibration and for disentangling GW and propagation-induced noise.

However, realizing the full scientific potential of low-frequency timing requires accurate modeling of frequency-dependent pulse-shape evolution and chromatic timing residuals by \citet{Donner:2019AandA624A22} and \cite{Donner:2020AandA639A77}. Their analyses showed that LOFAR’s large fractional bandwidth can isolate intrinsic pulse evolution from propagation-induced effects, while complementary studies \citep[e.g.,][]{Lam2016} have quantified DM variability and scattering-induced distortions at low frequencies. The SKA-Low telescope, with its greater sensitivity and continuous bandwidth, will extend this capability to a far larger pulsar sample, enabling direct, high-cadence tracking of chromatic delays within each PTA epoch.

In addition, the SKA-Low telescope’s large instantaneous bandwidth and sensitivity will permit direct measurement of scattering transfer functions  at the same cadence as PTA timing, allowing correction of stochastic, frequency-dependent effects that otherwise masquerade as excess timing noise. These capabilities—repeatedly highlighted in multi-band calibration studies \citep{Lam2016,Donner:2020AandA639A77,Tiburzi2021,2024MNRAS.535.1184S,Susarla:2024AandA}—will be further enhanced by coordinated observations across SKA-Low, uGMRT, CHIME, and SKA-Mid, enabling direct tests of DM structure functions on astronomical-unit scales and linking PTA noise budgets to measurable IISM dynamics \citep{wcv+23}.

Beyond IISM diagnostics, SKA-Low will enhance PTA science by extending timing baselines for bright southern pulsars that saturate higher-frequency receivers, improving spectral separation between chromatic and achromatic noise components. In concert with SKA-Mid, SKA-Low will thus provide the full frequency lever arm needed to isolate the nanohertz GWB from propagation-induced systematics \citep{Verbiest:2024RINP}. This precision will be crucial for advanced searches for individual SMBHBs \citep{Ferranti:2025AandA} and for detecting anisotropy in the GWB \citep{Pol_2022,gersbach2025fopts,Gersbach2025fopt,2025MNRAS.536.1501G,Moreschi2025aniso}.

\section{The role of very long baseline interferometry and use of the pulsar term}
\label{sec:ska_long_baseline}

The SKAO is expected to measure accurate pulsar distances through pulsar timing, and the pulsar term can be incorporated in data analysis to improve the localization accuracy of single GW sources \citep{2011MNRAS.414.3251L}.
Recent theoretical and simulation studies have further demonstrated that incorporating precise pulsar distance information from external measurements, such as very long baseline interferometry (VLBI), can dramatically enhance the localization accuracy of individual CGWs in PTA experiments.
For example, \cite{2023PhRvD.108l3535K} showed that, even if precise distance information with an accuracy of $\sim$1 pc, comparable to the GW wavelength, is available for only a small subset of pulsars, rather than all pulsars, the sky localization of a single GW source can improve by more than an order of magnitude.
This effect arises because precise distance information which is independent of timing observations allows the phase of the pulsar term in the GW signal to be tightly constrained, thereby breaking degeneracies between the source position and other parameters.

\begin{figure}
\centering
\includegraphics[width=0.45\textwidth]{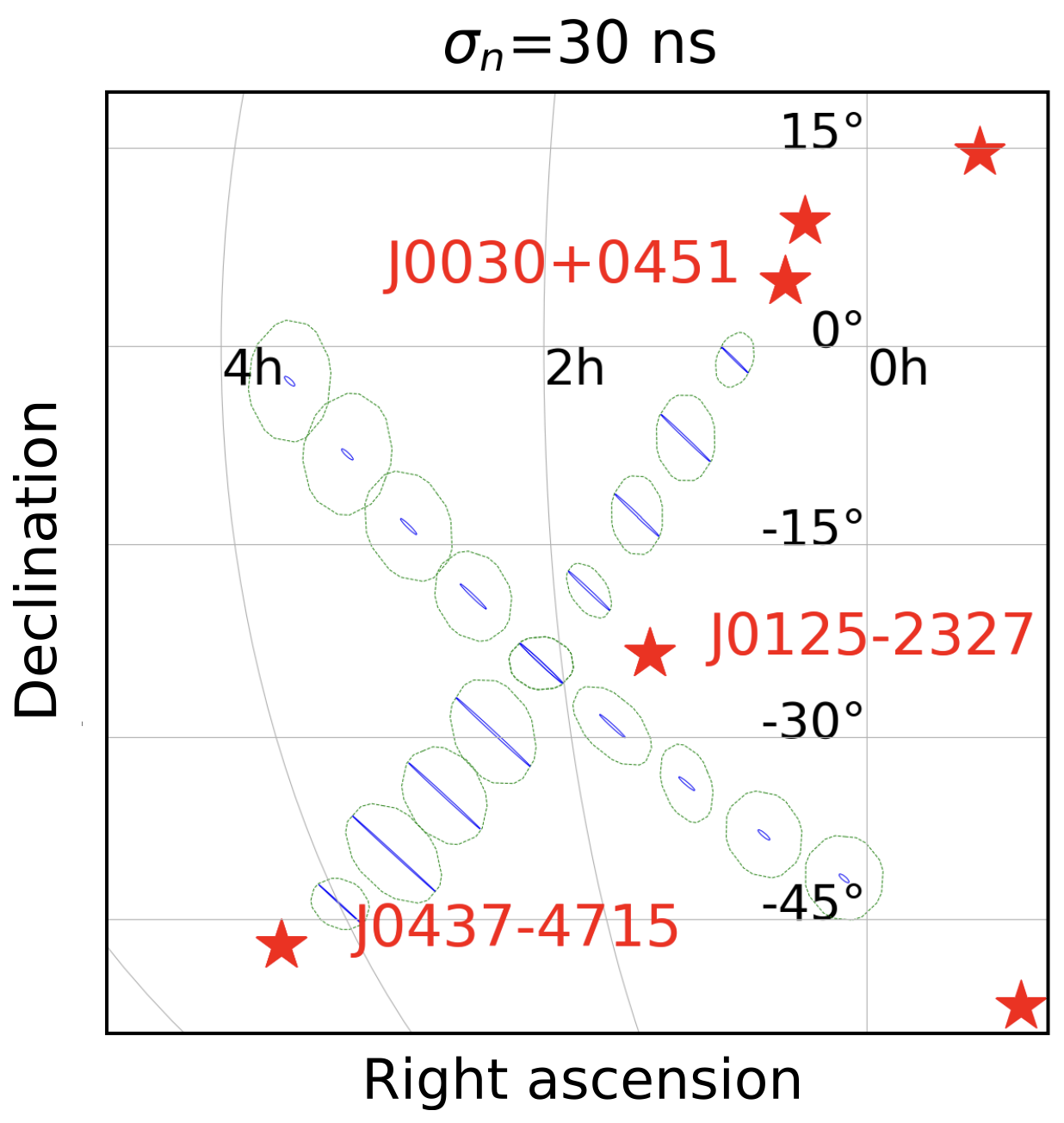}
\caption{ \label{fig:localization}
Expected localization precision (8 $\sigma$) of a single GW source around two pulsars, J0437-4715 and J0030+0451 \citep{2025arXiv250602819K}.
Source localizations are shown at different assumed positions. The outer contours (green dashed line) correspond to the case without distance information, while the inner contour (blue solid line) correspond to the case with distance errors expected in the SKA era.  The analysis assumes timing precision of $\sigma_n=30$\,ns can be achieved.  The red stars show the positions of pulsars.
}
\end{figure}

Building on this, \cite{2025arXiv250602819K} performed simulations with realistic sky distributions and actual distances for  $87$ IPTA pulsars.
They assumed distance precisions of $0.37$\,pc and $1.7$\,pc for the two nearby pulsars PSRs~J0437$-$4715 and J0030$+$0451, respectively, based on the precision levels anticipated with VLBI astrometry using the SKAO and assumed white noise levels for each pulsar of $\sigma_n=30$\,ns.
It was demonstrated that the GW source localization improves by two orders of magnitude near J0437$-$4715 and by more than one order near J0030$+$0451, with substantial improvement maintained over a broader region of the sky (Fig. \ref{fig:localization}).
These findings underline that even a limited set of high-precision pulsar distances can have a transformative impact on the capability of PTAs to pinpoint GW sources, facilitate host galaxy identification, and enable multi-messenger follow-up observations.
These results highlight the exceptional synergy between high-precision VLBI astrometry and SKAO PTA science, strongly motivating the prioritization of distance measurements for the most influential pulsars in the array.

\section{Connections to high-energy pulsar timing}
\label{sec:gptas}

Pulsar timing array efforts can also benefit from high energy observations of gamma-ray or X-ray bright MSPs.  
Millisecond pulsars emit prodigious high-energy  emission, with more than $140$ known to be gamma-ray bright \cite[][]{2023ApJ...958..191S}.
There are ongoing efforts to use observations of MSPs from the Fermi Large Area Telescope (Fermi-LAT) to search for gravitational waves \cite[][]{2022Sci...376..521F}, which have timing baselines extending to shortly after the launch of Fermi in 2008, as part of the Gamma-ray PTA (GPTA).
Using a 12.5 year data set, the 95\% limit on the amplitude of a GWB is $A_{\rm yr} < 10^{-14}$.  The GPTA is expected to be sensitive to CURN at an amplitude of $A_{\rm yr}=2\times10^{-15}$ with $\approx 26$\,yr observation, assuming the weak signal scaling  $A_{\rm yr} \propto T^{-13/6}$ \cite[][]{2022Sci...376..521F}.
While there are pulsars that are common between radio and gamma-ray PTAs, many of the brightest gamma-ray MSPs are faint radio sources. The inclusion of gamma-ray bright MSPs can increase the size and sensitivity of international efforts.

A handful of MSPs can also be timed with adequate precision for GW searches using X-ray timing, as demonstrated by the NICER mission \cite[][]{2019ApJ...874..160D}. Past searches have detected red noise in X-ray observations  of PSR~J1824$-$2452A \cite[][]{2022ApJ...928...67H}, and current analyses show evidence for red noise in X-ray data from the first MSP discovered, PSR~J1939$+$2134 (C. Wayt, et al. in prep).  

Even if a pulsar's timing precision makes it better suited to PTA searches at radio wavelengths, gamma-ray and X-ray observations can help with better noise characterization.
High-energy observations are not impacted by  propagation effects through intervening plasma (discussed in Section \ref{sec:noise}) that impact timing observations at radio frequencies. Thus, the joint study of pulsars at high energies and radio can be used to better characterize chromatic noise processes and distinguish them from achromatic processes, including gravitational waves and intrinsic spin noise \cite[][]{2015ApJ...814..128K}.
The detection of red noise in NICER MSP observations, confirms that the noise is intrinsic to these pulsars and not a misspecified chromatic noise.

Gamma-ray observations can also be used to understand the origin and impact of profile variability on pulsar timing. The gamma-ray emission mechanism, while likely related to radio emission, is different as it is more directly linked to the acceleration of charged particles \cite[][]{2005ApJ...622..531H}.
There has been no evidence for temporal variability in the gamma-ray emission of MSPs \cite[][]{2025ApJ...991..225K}; therefore, high-energy observations can be used to understand the nature of profile variability observed at radio wavelengths.

Observing pulsars in gamma-rays is fundamentally different than in radio. To produce pulse profiles with signal to noise ratios $\gg 1$ from gamma-ray photons requires days to months integration time.  However, instruments such as Fermi-LAT have wide fields of view of $\sim$ steradian, so monitoring of gamma-ray pulsars can be conducted commensal to survey observations. Additionally, ``photon-by-photon'' timing techniques \cite[][]{2011ApJ...732...38K,2015ApJ...809L...2C,2022Sci...376..521F} provide a more optimal approach to using gamma-ray data than traditional (radio-like) TOA-based modeling. Both the traditional and photon-by-photon methods were used in the first GPTA GW searches.

Gamma-ray PTA sensitivity will steadily improve with the Fermi-LAT's continued operation. The SKAO PTA will benefit greatly from contemporaneous gamma-ray, X-ray, and radio pulsar timing observations to better characterize both intrinsic and ISM-induced red noise processes, thereby improving the PTA's sensitivity to the nHz GWB. 

\section{Requirements for the SKAO PTA}
\label{sec:ska_requirements}

There are a number of key capabilities that are required to make the best use of the SKAO as a PTA instrument. 

\begin{itemize}
\item Observing bands: Because increased bandwidths minimize system temperature fluctuations and improve sensitivity, the precision to which a pulsar can be timed is determined in part by the available simultaneous bandwidth. As steep-spectrum emitters, most MSPs perform best at relatively low radio frequencies; current PTA experiments conduct most observations between $\sim$ 400 MHz and 4 GHz, though some instruments (such as LOFAR) significantly improve low-frequency coverage. Ultra-wideband receivers improve timing precision not only through increased bandwidth, but also by providing a simultaneous lever arm for characterizing and mitigating dispersion and scattering by the IISM. Such an ultra-wideband receiver covering SKAO-Mid bands 1-4 (or 5) would be ideal. Barring that, monitoring in bands 1 and 2 (350-1760\,MHz) is a baseline requirement for PTA science. 
\item Timing Precision: Pulsars timed to better than 1$\mu s$ precision are generally considered to improve sensitivity to the GWB \cite[][]{2013CQGra..30v4015S}. Although the SKAO will facilitate significantly better precision ($\sim$tens of ns) for bright sources, PTA sensitivity has a strong (linear) dependence on the number of pulsars timed, suggesting that the SKAO PTA will include plenty of 1$\mu s$ MSPs. Increased bandwidth, increased integration times (up to the jitter limit), high-quality polarization calibration, and improved system temperature are important contributions to the achievable timing precision. 
\item Polarization fidelity: High-quality polarimetry  enables the use of full Stokes information in pulsar timing, which could result in higher timing precision \cite[][]{2006ApJ...642.1004V}. Because polarization systematics introduced at the instrument level are correlated between all observed pulsars, controlling these effects can yield significant improvements not only in sensitivity, but also PTA performance. Regular observations of the pulsars themselves can be further used to improve polarization calibration, and may somewhat reduce the polarization purity requirements \cite[][]{2013ApJS..204...13V}.
\item Spectral purity: Leakage between spectral channels can introduce frequency-dependent artefacts and cause bias in integrated pulse profiles that significantly hamper high-precision timing. MeerKAT has critically sampled coarse channels. For pulsar timing modes it was necessary to develop sharply tapered filters for  MeerKAT, which reduced throughput close to the channel edges. This removed the artefacts while only reducing sensitivity by 5\% \cite[][]{2020PASA...37...28B}.  With its oversampled filters should not exhibit such artefacts.
\item RFI removal: RFI contamination decreases usable bandwidth and introduces other systematics that reduce the sensitivity of PTA observations. In addition to observatory-side improvements such as coordinated spectrum management, modernizing RFI excision algorithms could have a significant impact on PTA timing precision. 
\item Flexible sub-arraying and scheduling capability: Pulsar flux densities vary due to interstellar scintillation. \cite{2023MNRAS.526.3370G} found that an array could be made more sensitive if a dynamic scheduling strategy in which integration times are increased when pulsars were in bright scintillation states is implemented. For example, low-DM pulsars (which show pronounced scintillation at decimetre wavelengths) could be made a factor of four times more sensitive to the GWB, and potentially other GW sources. Additionally, sub-arraying capabilities can improve the efficiency of PTA programs and reduce the total on-sky time requirements.  The ability to quickly reconfigure sub-arrays would further improve efficiency by tailoring collecting area on individual pulsars to their intrinsic jitter limits and changing observing bands to maximize signal to noise ratio or collect broad band observations to mitigate propagation effects.  In any case the SKAO PTA would observe most efficiently if there are low overheads (slew times and system reconfigurations) between observations.   
\item Overlap with MeerKAT:  It is essential to have time overlap of PTA observations between the SKAO-Mid telescope and with MeerKAT, which will enable offsets between the two observatories to be measured. Offsets will likely be present in digital and analogue portions of the system.  Different responses of MeerKAT and SKAO receiving systems can potentially result in different observed pulse profiles.
It may be possible to measure absolute offsets  by correlating sky noise \cite[][]{2015ApJ...813...65N} or observations of Giant pulses in pulsars, such as the Crab pulsar \citep{2021PASA...38...17S}. 
Overlapping (and simultaneous) observations will verify the timing performance of SKAO. 

\item Ability to record data with multiple backends simultaneously. Ancillary science would benefit from data being recorded in multiple modes.  Single pulse analysis can be undertaken in baseband (voltage) format or in search mode (filterbank) format.  Such studies are useful for both understanding the nature of pulsar emission, but also for developing techniques to mitigate the effects of jitter noise.  Scintillation studies often benefit from higher frequency resolution than possible or desired in pulsar timing observations, or the use of cyclic spectroscopy \cite[][]{2011MNRAS.416.2821D} to descatter pulse profiles.   
\end{itemize}

Post processing:
\begin{itemize}
\item Pulsar timing observations are relatively computationally inexpensive.  Only one or a few tied array beams need to be formed.   Phase resolved pulse profiles can be integrated together for many seconds without impacting science goals of the SKAO-PTA, with science data products produced without requiring significant resources from the science data processor.     
\item Wide-band timing methods: Because pulse profiles evolve as a function of frequency, extracting TOAs from averaged profiles can result in measurement biases. Wide-band timing techniques, which use a frequency-dependent template to measure one or more TOAs per epoch, improve timing precision and reduce the required number of timing parameters when wide-bandwidth systems are in use \cite[][]{2014ApJ...790...93P,2014MNRAS.443.3752L,2023ApJ...944..128C,2024MNRAS.paladi.527..213P}.
\item Noise analysis and GW searches will use Bayesian codes \cite[e.g.,][]{2014MNRAS.437.3004L,2020zndo...4059815E} to search for GW signals and characterize the noise in the data set. These codes will be accelerated by hardware algorithm advances, as has been evidenced by increases in code speed in the 2020s.  These models are developed in concert with pulsar timing models developed in codes such as {\sc tempo2} \cite[][]{2006MNRAS.369..655H} or {\sc pint} \cite[][]{2021ApJ...911...45L}.
Given the sensitivity of the observations, it will be essential to continue to refine noise models.  This could be through for example, looking at the  DMCalc method  \citep{2021A&A...651A...5K,2025PASA...42..108R} for estimating dispersion measures.
\end{itemize}

\section{Connections to other SKAO science and other 2030 facilities}
\label{sec:other_science}

The large scale pulsar survey with SKAO \citep{Keane2025_SKA_SKAPTA} will help enlarge the pulsar sample for the SKAO PTA thereby making it the most sensitive PTA experiment. The regular cadence monitoring over long time baseline needed for an SKAO PTA will have benefits for other pulsar science. While the measurements of variations in DM, scatter broadening, and solar wind over such time baseline will provide a better understanding of the IISM \citep{Tiburzi2025_SKA_SKAPTA}, the estimate of timing noise will provide constraints for Equation of State \citep{Basu2025_SKA_EOS}. PTA observations will similarly contribute to NS measurements through new or improved Shapiro delay measurements. The jitter noise measurements over a wide frequency range will have implications for understanding the physics of the pulsar emission mechanism and magnetosphere \citep{Oswald2025_SKA_SKAPTA}. Additional constraints 
on magnetospheric physics will come from wideband observations of profile 
change events in some PTA MSPs. Overall, precision timing of MSPs may 
also be helpful in tests for gravitational theory \citep{Krishnan2025_SKA_SKAPTA} 
apart from constraints on the existence dipolar radiation and alternative 
polarizations in alternative gravity theories. These connections to other pulsar 
science with the SKA observatory are discussed in detail in the companion articles 
in this book.

Other next-generation radio instruments, such as the Deep Synoptic Array (DSA; formerly DSA-2000) and the next-generation Very Large Array (ngVLA), will be complementary to the SKA effort and further improve the GW sensitivity of IPTA data sets. Currently, the DSA project \cite[][]{2019BAAS...51g.255H} plans to dedicate $\sim$25\% of on-sky time towards the observation of 150--200 MSPs for PTA science, with observations beginning in 2028. Spanning a bandwidth of 750 MHz--2 GHz, the instrument will operate in an ideal range for radio pulsar observations and provide a substantial sensitivity and timing precision boost, especially for Northern hemisphere pulsars (with a declination limit of -39 degrees). On a longer timescale, the ngVLA \cite[][]{2018ASPC..517..751C} is slated to provide higher-frequency, high-sensitivity capabilities to PTAs --- which lessen the impact of IISM effects --- and support synergistic long-baseline science as described in Section~\ref{sec:ska_long_baseline}. These instruments, taken together with the SKAO, will provide the full-sky coverage necessary to maximize sensitivity to the GWB and other signal sources.

Several other 2030 facilities will be complementary to the SKAO PTA. 
The Laser Interferometer Space Antenna (LISA) is an ESA-NASA mission to launch a space-based GW observatory. 
LISA will open a new window on the millihertz GW frequency band and is slated to launch in 2035~\citep{LISARedBook:2024}.
Other missions, including TianQin~\citep{TianQin:2025} and Taiji~\citep{Taiji:2020} are planned on a similar timescale. 
This millihertz GW band will provide a complementary view on the evolution and growth of SMBHBs across cosmic time by capturing GWs from the mergers themselves~\citep{RhookWhithe:2005,SesanaEtAl:2011, KleinEtAl:2016,KatzEtAl:2020,BarausseEtAl:2020,ToubianaEtAl:2025}.
The potential of multiband SMBHB observations is already being explored with forecasting of LISA detection rates based on current PTA constraints on the SMBHB population~\citep[e.g.][]{SteinleEtAl:2023}. 
Detections of GWBs in multiple bands can potentially to understand the physical processes that cause inflation \cite[][]{2016PhRvX...6a1035L}.
Distinguishing between multiple sources of a cosmological GWB will  require coordinated activities, combining PTA measurements with observations of the cosmic microwave background, and future ground and space-based gravitational wave detectors such as Cosmic Explorer \cite[][]{2021arXiv210909882E}, the Einstein Telescope \cite[][]{2010CQGra..27s4002P} and LISA \cite[][]{LISARedBook:2024}.

\section{Conclusions}
\label{sec:conclusions}

The SKAO promises to be an excellent vehicle for an efficient and high impact PTA experiment.  The sensitivity, wide frequency coverage and availability of sub-arrays will ensure it can be optimized to maximize nanohertz-frequency GW science in the 2030s and beyond. Through the SKAO Pulsar Timing Array it will be able to make detailed maps of the nanohertz-frequency gravitational wave sky, and use them to chart the nature of gravitational wave sources, and the fundamental physics that powers the emission.

\section*{Acknowledgements}

Part of this work was undertaken as part of the Australian Research Council Centre of Excellence for Gravitational Wave Discovery (OzGrav) CE230100016.
HTC acknowledges support from the U.S. Naval Research Laboratory. Basic research in pulsar astronomy at NRL is supported by NASA, in particular via Fermi Guest Investigator award NNG22OB35A.
AG acknowledges support of the Department of Atomic Energy, Government of India, under Project Identification No. RTI 4002
KG acknowledges support from the International Max Planck Research School (IMPRS) for Astronomy and Astrophysics at the Universities of Bonn and Cologne and the Bonn-Cologne Graduate School of Physics and Astronomy.
All authors affiliated with the Max-Planck-Gesellschaft (MPG) acknowledge its constant support.
JSH acknowledges support from NSF CAREER award 2339728 and NSF Physics Frontiers Center award 2020265.
JSH, MTM and CMFM acknowledge support from the NANOGrav Collaboration's National Science Foundation (NSF) Physics Frontiers Center award  2020265.
BCJ acknowledges the support from Raja Ramanna Chair fellowship of the Department of Atomic Energy, Government of India (RRC – Track I Grant 3/3401 Atomic Energy Research 00 004 Research and Development 27 02 31 1002//2/2023/RRC/R\&D-II/13886 and 1002/2/2023/RRC/R\&D-II/14369).
RK is supported by JSPS KAKENHI Grant Number 24K17051.
KJL is supported by the National SKA Program of China (Grant No. 2020SKA012010).
HM is supported by the UK Space Agency, Grant No. ST/V002813/1.
 MTM and CMFM acknowledge support from the NANOGrav Collaboration's National Science Foundation (NSF) Physics Frontiers Center award  1430284.
CMFM was also supported in part by the NSF under Grant NSF PHY-1748958  and NASA LPS 80NSSC24K0440
AP acknowledges financial support from the European Research Council (ERC) starting grant ’GIGA’ (grant agreement number: 101116134) and through the NWO-I Veni fellowship.
GMS acknowledges the financial support provided under the European Union’s H2020 ERC Consolidator Grant B Massive (Grant Agreement: 818691) and Advanced Grant PINGU (Grant Agreement: 101142097).
KT is partially supported by JSPS KAKENHI Grant Numbers 20H00180, 21H01130, 21H04467, and 24H01813, and Bilateral Joint Research Projects of JSPS.
XX is funded by the grant CNS2023-143767. Grant CNS2023-143767  is funded by MICIU/AEI/10.13039/501100011033 and by European Union NextGenerationEU/PRTR. IFAE is partially funded by the CERCA program of the Generalitat de Catalunya.

\bibliographystyle{aasjournal}

\bibliography{ska_pta}

\begin{thebibliography}{}
\expandafter\ifx\csname natexlab\endcsname\relax\def\natexlab#1{#1}\fi
\providecommand{\url}[1]{\href{#1}{#1}}
\providecommand{\dodoi}[1]{doi:~\href{http://doi.org/#1}{\nolinkurl{#1}}}
\providecommand{\doeprint}[1]{\href{http://ascl.net/#1}{\nolinkurl{http://ascl.net/#1}}}
\providecommand{\doarXiv}[1]{\href{https://arxiv.org/abs/#1}{\nolinkurl{https://arxiv.org/abs/#1}}}

\bibitem[{{Abac} {et~al.}(2025){Abac}, {Abouelfettouh}, {Acernese}, {Ackley}, {Adamcewicz}, {Adhicary}, {Adhikari}, {Adhikari}, {Adhikari}, {Adkins}, {Afroz}, {Agapito}, {Agarwal}, {Agathos}, {Aggarwal}, {Aggarwal}, {Aguiar}, {Ahrend}, {Aiello}, {Ain}, {Ajith}, {Akutsu}, {Al-Kershi}, {Al-Shammari}, {Albanesi}, {Ali}, {All{\'e}n{\'e}}, {Allocca}, {Altin}, {Alvarez-Lopez}, {Amar}, {Amarasinghe}, {Amato}, {Amicucci}, {Amra}, {Ananyeva}, {Anderson}, {Anderson}, {Andia}, {Ando}, {Andr{\'e}s-Carcasona}, {Andri{\'c}}, {Anglin}, {Ansoldi}, {Antelis}, {Antier}, {Aoumi}, {Appavuravther}, {Appert}, {Apple}, {Arai}, {Araya}, {Araya}, {Arca Sedda}, {Areeda}, {Aritomi}, {Armato}, {Armstrong}, {Arnaud}, {Arogeti}, {Aronson}, {Ashton}, {Aso}, {Asprea}, {Assiduo}, {Assis de Souza Melo}, {Aston}, {Astone}, {Attadio}, {Aubin}, {Aultoneal}, {Avallone}, {Avila}, {Babak}, {Badger}, {Bae}, {Bagnasco}, {Baiotti}, {Bajpai}, {Baka}, {Baker}, {Baker}, {Baker}, {Baldi}, {Baldicchi}, {Ball}, {Ballardin}, {Ballmer}, {Banagiri},
  {Banerjee}, {Bankar}, {Baptiste}, {Baral}, {Baratti}, {Barayoga}, {Barish}, {Barker}, {Barman}, {Barneo}, {Barone}, {Barr}, {Barsotti}, {Barsuglia}, {Barta}, {Bartoletti}, {Barton}, {Bartos}, {Basalaev}, {Bassiri}, {Basti}, {Bawaj}, {Baxi}, {Bayley}, {Baylor}, {Baynard}, {Bazzan}, {Bedakihale}, {Beirnaert}, {Bejger}, {Belardinelli}, {Bell}, {Bellie}, {Bellizzi}, {Benoit}, {Bentara}, {Bentley}, {Ben Yaala}, {Bera}, {Bergamin}, {Berger}, {Bernuzzi}, {Beroiz}, {Berry}, {Bersanetti}, {Bertheas}, {Bertolini}, {Betzwieser}, {Beveridge}, {Bevilacqua}, {Bevins}, {Bhagwat}, {Bhandare}, {Bhatt}, {Bhattacharjee}, {Bhattacharyya}, {Bhaumik}, {Biancalana}, {Bianchi}, {Bilenko}, {Billingsley}, {Binetti}, {Bini}, {Binu}, {Biot}, {Birnholtz}, {Biscoveanu}, {Bisht}, {Bitossi}, {Bizouard}, {Blaber}, {Blackburn}, {Blagg}, {Blair}, {Blair}, {Bode}, {Boettner}, {Boileau}, {Boldrini}, {Bolingbroke}, {Bolliand}, {Bonavena}, {Bondarescu}, {Bondu}, {Bonilla}, {Bonilla}, {Bonino}, {Bonnand}, {Borchers}, {Borhanian}, {Boschi},
  {Bose}, {Bossilkov}, {Bothra}, {Boudon}, {Bourg}, {Boyle}, {Bozzi}, {Bradaschia}, {Brady}, {Branch}, {Branchesi}, {Braun}, {Briant}, {Brillet}, {Brinkmann}, {Brockill}, {Brockmueller}, {Brooks}, {Brown}, \& {Brown}}]{2025PhRvL.135k1403A}
{Abac}, A.~G., {Abouelfettouh}, I., {Acernese}, F., {et~al.} 2025, \prl, 135, 111403, \dodoi{10.1103/kw5g-d732}

\bibitem[{{Abbott} {et~al.}(2016){Abbott}, {Abbott}, {Abbott}, {Abernathy}, {Acernese}, {Ackley}, {Adams}, {Adams}, {Addesso}, {Adhikari}, {Adya}, {Affeldt}, {Agathos}, {Agatsuma}, {Aggarwal}, {Aguiar}, {Aiello}, {Ain}, {Ajith}, {Allen}, {Allocca}, {Altin}, {Anderson}, {Anderson}, {Arai}, {Arain}, {Araya}, {Arceneaux}, {Areeda}, {Arnaud}, {Arun}, {Ascenzi}, {Ashton}, {Ast}, {Aston}, {Astone}, {Aufmuth}, {Aulbert}, {Babak}, {Bacon}, {Bader}, {Baker}, {Baldaccini}, {Ballardin}, {Ballmer}, {Barayoga}, {Barclay}, {Barish}, {Barker}, {Barone}, {Barr}, {Barsotti}, {Barsuglia}, {Barta}, {Bartlett}, {Barton}, {Bartos}, {Bassiri}, {Basti}, {Batch}, {Baune}, {Bavigadda}, {Bazzan}, {Behnke}, {Bejger}, {Belczynski}, {Bell}, {Bell}, {Berger}, {Bergman}, {Bergmann}, {Berry}, {Bersanetti}, {Bertolini}, {Betzwieser}, {Bhagwat}, {Bhandare}, {Bilenko}, {Billingsley}, {Birch}, {Birney}, {Birnholtz}, {Biscans}, {Bisht}, {Bitossi}, {Biwer}, {Bizouard}, {Blackburn}, {Blair}, {Blair}, {Blair}, {Bloemen}, {Bock}, {Bodiya}, {Boer},
  {Bogaert}, {Bogan}, {Bohe}, {Bojtos}, {Bond}, {Bondu}, {Bonnand}, {Boom}, {Bork}, {Boschi}, {Bose}, {Bouffanais}, {Bozzi}, {Bradaschia}, {Brady}, {Braginsky}, {Branchesi}, {Brau}, {Briant}, {Brillet}, {Brinkmann}, {Brisson}, {Brockill}, {Brooks}, {Brown}, {Brown}, {Brown}, {Buchanan}, {Buikema}, {Bulik}, {Bulten}, {Buonanno}, {Buskulic}, {Buy}, {Byer}, {Cabero}, {Cadonati}, {Cagnoli}, {Cahillane}, {Bustillo}, {Callister}, {Calloni}, {Camp}, {Cannon}, {Cao}, {Capano}, {Capocasa}, {Carbognani}, {Caride}, {Diaz}, {Casentini}, {Caudill}, {Cavagli{\`a}}, {Cavalier}, {Cavalieri}, {Cella}, {Cepeda}, {Baiardi}, {Cerretani}, {Cesarini}, {Chakraborty}, {Chalermsongsak}, {Chamberlin}, {Chan}, {Chao}, {Charlton}, {Chassande-Mottin}, {Chen}, {Chen}, {Cheng}, {Chincarini}, {Chiummo}, {Cho}, {Cho}, {Chow}, {Christensen}, {Chu}, {Chua}, {Chung}, {Ciani}, {Clara}, {Clark}, {Cleva}, {Coccia}, {Cohadon}, {Colla}, {Collette}, {Cominsky}, {Constancio}, {Conte}, {Conti}, {Cook}, {Corbitt}, {Cornish}, {Corsi}, {Cortese}, {Costa},
  {Coughlin}, {Coughlin}, {Coulon}, {Countryman}, {Couvares}, {Cowan}, {Coward}, \& {Cowart}}]{2016PhRvL.116f1102A}
{Abbott}, B.~P., {Abbott}, R., {Abbott}, T.~D., {et~al.} 2016, \prl, 116, 061102, \dodoi{10.1103/PhysRevLett.116.061102}

\bibitem[{{Abbott} {et~al.}(2017{\natexlab{a}}){Abbott}, {Abbott}, {Abbott}, {Acernese}, {Ackley}, {Adams}, {Adams}, {Addesso}, {Adhikari}, {Adya}, {Affeldt}, {Afrough}, {Agarwal}, {Agathos}, {Agatsuma}, {Aggarwal}, {Aguiar}, {Aiello}, {Ain}, {Ajith}, {Allen}, {Allen}, {Allocca}, {Altin}, {Amato}, {Ananyeva}, {Anderson}, {Anderson}, {Angelova}, {Antier}, {Appert}, {Arai}, {Araya}, {Areeda}, {Arnaud}, {Arun}, {Ascenzi}, {Ashton}, {Ast}, {Aston}, {Astone}, {Atallah}, {Aufmuth}, {Aulbert}, {AultONeal}, {Austin}, {Avila-Alvarez}, {Babak}, {Bacon}, {Bader}, {Bae}, {Bailes}, {Baker}, {Baldaccini}, {Ballardin}, {Ballmer}, {Banagiri}, {Barayoga}, {Barclay}, {Barish}, {Barker}, {Barkett}, {Barone}, {Barr}, {Barsotti}, {Barsuglia}, {Barta}, {Barthelmy}, {Bartlett}, {Bartos}, {Bassiri}, {Basti}, {Batch}, {Bawaj}, {Bayley}, {Bazzan}, {B{\'e}csy}, {Beer}, {Bejger}, {Belahcene}, {Bell}, {Berger}, {Bergmann}, {Bernuzzi}, {Bero}, {Berry}, {Bersanetti}, {Bertolini}, {Betzwieser}, {Bhagwat}, {Bhandare}, {Bilenko},
  {Billingsley}, {Billman}, {Birch}, {Birney}, {Birnholtz}, {Biscans}, {Biscoveanu}, {Bisht}, {Bitossi}, {Biwer}, {Bizouard}, {Blackburn}, {Blackman}, {Blair}, {Blair}, {Blair}, {Bloemen}, {Bock}, {Bode}, {Boer}, {Bogaert}, {Bohe}, {Bondu}, {Bonilla}, {Bonnand}, {Boom}, {Bork}, {Boschi}, {Bose}, {Bossie}, {Bouffanais}, {Bozzi}, {Bradaschia}, {Brady}, {Branchesi}, {Brau}, {Briant}, {Brillet}, {Brinkmann}, {Brisson}, {Brockill}, {Broida}, {Brooks}, {Brown}, {Brown}, {Brunett}, {Buchanan}, {Buikema}, {Bulik}, {Bulten}, {Buonanno}, {Buskulic}, {Buy}, {Byer}, {Cabero}, {Cadonati}, {Cagnoli}, {Cahillane}, {Calder{\'o}n Bustillo}, {Callister}, {Calloni}, {Camp}, {Canepa}, {Canizares}, {Cannon}, {Cao}, {Cao}, {Capano}, {Capocasa}, {Carbognani}, {Caride}, {Carney}, {Carullo}, {Casanueva Diaz}, {Casentini}, {Caudill}, {Cavagli{\`a}}, {Cavalier}, {Cavalieri}, {Cella}, {Cepeda}, {Cerd{\'a}-Dur{\'a}n}, {Cerretani}, {Cesarini}, {Chamberlin}, {Chan}, {Chao}, {Charlton}, {Chase}, {Chassande-Mottin}, {Chatterjee},
  {Chatziioannou}, {Cheeseboro}, {Chen}, {Chen}, {Chen}, {Cheng}, {Chia}, {Chincarini}, {Chiummo}, {Chmiel}, {Cho}, {Cho}, {Chow}, {Christensen}, {Chu}, {Chua}, \& {Chua}}]{2017PhRvL.119p1101A}
---. 2017{\natexlab{a}}, \prl, 119, 161101, \dodoi{10.1103/PhysRevLett.119.161101}

\bibitem[{{Abbott} {et~al.}(2017{\natexlab{b}}){Abbott}, {Abbott}, {Abbott}, {Acernese}, {Ackley}, {Adams}, {Adams}, {Addesso}, {Adhikari}, {Adya}, {Affeldt}, {Afrough}, {Agarwal}, {Agathos}, {Agatsuma}, {Aggarwal}, {Aguiar}, {Aiello}, {Ain}, {Ajith}, {Allen}, {Allen}, {Allocca}, {Altin}, {Amato}, {Ananyeva}, {Anderson}, {Anderson}, {Angelova}, {Antier}, {Appert}, {Arai}, {Araya}, {Areeda}, {Arnaud}, {Arun}, {Ascenzi}, {Ashton}, {Ast}, {Aston}, {Astone}, {Atallah}, {Aufmuth}, {Aulbert}, {AultONeal}, {Austin}, {Avila-Alvarez}, {Babak}, {Bacon}, {Bader}, {Bae}, {Baker}, {Baldaccini}, {Ballardin}, {Ballmer}, {Banagiri}, {Barayoga}, {Barclay}, {Barish}, {Barker}, {Barkett}, {Barone}, {Barr}, {Barsotti}, {Barsuglia}, {Barta}, {Barthelmy}, {Bartlett}, {Bartos}, {Bassiri}, {Basti}, {Batch}, {Bawaj}, {Bayley}, {Bazzan}, {B{\'e}csy}, {Beer}, {Bejger}, {Belahcene}, {Bell}, {Berger}, {Bergmann}, {Bero}, {Berry}, {Bersanetti}, {Bertolini}, {Betzwieser}, {Bhagwat}, {Bhandare}, {Bilenko}, {Billingsley}, {Billman}, {Birch},
  {Birney}, {Birnholtz}, {Biscans}, {Biscoveanu}, {Bisht}, {Bitossi}, {Biwer}, {Bizouard}, {Blackburn}, {Blackman}, {Blair}, {Blair}, {Blair}, {Bloemen}, {Bock}, {Bode}, {Boer}, {Bogaert}, {Bohe}, {Bondu}, {Bonilla}, {Bonnand}, {Boom}, {Bork}, {Boschi}, {Bose}, {Bossie}, {Bouffanais}, {Bozzi}, {Bradaschia}, {Brady}, {Branchesi}, {Brau}, {Briant}, {Brillet}, {Brinkmann}, {Brisson}, {Brockill}, {Broida}, {Brooks}, {Brown}, {Brown}, {Brunett}, {Buchanan}, {Buikema}, {Bulik}, {Bulten}, {Buonanno}, {Buskulic}, {Buy}, {Byer}, {Cabero}, {Cadonati}, {Cagnoli}, {Cahillane}, {Calder{\'o}n Bustillo}, {Callister}, {Calloni}, {Camp}, {Canepa}, {Canizares}, {Cannon}, {Cao}, {Cao}, {Capano}, {Capocasa}, {Carbognani}, {Caride}, {Carney}, {Casanueva Diaz}, {Casentini}, {Caudill}, {Cavagli{\`a}}, {Cavalier}, {Cavalieri}, {Cella}, {Cepeda}, {Cerd{\'a}-Dur{\'a}n}, {Cerretani}, {Cesarini}, {Chamberlin}, {Chan}, {Chao}, {Charlton}, {Chase}, {Chassande-Mottin}, {Chatterjee}, {Chatziioannou}, {Cheeseboro}, {Chen}, {Chen}, {Chen},
  {Cheng}, {Chia}, {Chincarini}, {Chiummo}, {Chmiel}, {Cho}, {Cho}, {Chow}, {Christensen}, {Chu}, {Chua}, {Chua}, {Chung}, {Chung}, \& {Ciani}}]{2017ApJ...848L..12A}
---. 2017{\natexlab{b}}, \apjl, 848, L12, \dodoi{10.3847/2041-8213/aa91c9}

\bibitem[{{Abbott} {et~al.}(2017{\natexlab{c}}){Abbott}, {Abbott}, {Abbott}, {Abernathy}, {Acernese}, {Ackley}, {Adams}, {Adams}, {Addesso}, {Adhikari}, {Adya}, {Affeldt}, {Agathos}, {Agatsuma}, {Aggarwal}, {Aguiar}, {Aiello}, {Ain}, {Ajith}, {Allen}, {Allocca}, {Altin}, {Anderson}, {Anderson}, {Arai}, {Araya}, {Arceneaux}, {Areeda}, {Arnaud}, {Arun}, {Ascenzi}, {Ashton}, {Ast}, {Aston}, {Astone}, {Aufmuth}, {Aulbert}, {Babak}, {Bacon}, {Bader}, {Baldaccini}, {Ballardin}, {Ballmer}, {Barayoga}, {Barclay}, {Barish}, {Barker}, {Barone}, {Barr}, {Barsotti}, {Barsuglia}, {Barta}, {Bartlett}, {Bartos}, {Bassiri}, {Basti}, {Batch}, {Baune}, {Bavigadda}, {Bazzan}, {Bejger}, {Bell}, {Bergmann}, {Berry}, {Bersanetti}, {Bertolini}, {Betzwieser}, {Bhagwat}, {Bhandare}, {Bilenko}, {Billingsley}, {Birch}, {Birney}, {Birnholtz}, {Biscans}, {Bisht}, {Bitossi}, {Biwer}, {Bizouard}, {Blackburn}, {Blair}, {Blair}, {Blair}, {Bloemen}, {Bock}, {Boer}, {Bogaert}, {Bogan}, {Bohe}, {Bond}, {Bondu}, {Bonnand}, {Boom}, {Bork},
  {Boschi}, {Bose}, {Bouffanais}, {Bozzi}, {Bradaschia}, {Braginsky}, {Branchesi}, {Brau}, {Briant}, {Brillet}, {Brinkmann}, {Brisson}, {Brockill}, {Broida}, {Brooks}, {Brown}, {Brown}, {Brown}, {Brunett}, {Buchanan}, {Buikema}, {Bulik}, {Bulten}, {Buonanno}, {Buskulic}, {Buy}, {Byer}, {Cabero}, {Cadonati}, {Cagnoli}, {Cahillane}, {Bustillo}, {Callister}, {Calloni}, {Camp}, {Cannon}, {Cao}, {Capano}, {Capocasa}, {Carbognani}, {Caride}, {Casanueva Diaz}, {Casentini}, {Caudill}, {Cavagli{\`a}}, {Cavalier}, {Cavalieri}, {Cella}, {Cepeda}, {Cerboni Baiardi}, {Cerretani}, {Cesarini}, {Chamberlin}, {Chan}, {Chao}, {Charlton}, {Chassande-Mottin}, {Chen}, {Chen}, {Cheng}, {Chincarini}, {Chiummo}, {Cho}, {Cho}, {Chow}, {Christensen}, {Chu}, {Chua}, {Chung}, {Ciani}, {Clara}, {Clark}, {Cleva}, {Coccia}, {Cohadon}, {Colla}, {Collette}, {Cominsky}, {Constancio}, {Conte}, {Conti}, {Cook}, {Corbitt}, {Corsi}, {Cortese}, {Costa}, {Coughlin}, {Coughlin}, {Coulon}, {Countryman}, {Couvares}, {Cowan}, {Coward}, {Cowart},
  {Coyne}, {Coyne}, {Craig}, {Creighton}, {Cripe}, {Crowder}, {Cumming}, {Cunningham}, {Cuoco}, {Dal Canton}, {Danilishin}, \& {D'Antonio}}]{2017AnP...52900209A}
---. 2017{\natexlab{c}}, Annalen der Physik, 529, 1600209, \dodoi{10.1002/andp.201600209}

\bibitem[{{Acernese} {et~al.}(2015){Acernese}, {Agathos}, {Agatsuma}, {Aisa}, {Allemandou}, {Allocca}, {Amarni}, {Astone}, {Balestri}, {Ballardin}, {Barone}, {Baronick}, {Barsuglia}, {Basti}, {Basti}, {Bauer}, {Bavigadda}, {Bejger}, {Beker}, {Belczynski}, {Bersanetti}, {Bertolini}, {Bitossi}, {Bizouard}, {Bloemen}, {Blom}, {Boer}, {Bogaert}, {Bondi}, {Bondu}, {Bonelli}, {Bonnand}, {Boschi}, {Bosi}, {Bouedo}, {Bradaschia}, {Branchesi}, {Briant}, {Brillet}, {Brisson}, {Bulik}, {Bulten}, {Buskulic}, {Buy}, {Cagnoli}, {Calloni}, {Campeggi}, {Canuel}, {Carbognani}, {Cavalier}, {Cavalieri}, {Cella}, {Cesarini}, {Mottin}, {Chincarini}, {Chiummo}, {Chua}, {Cleva}, {Coccia}, {Cohadon}, {Colla}, {Colombini}, {Conte}, {Coulon}, {Cuoco}, {Dalmaz}, {D'Antonio}, {Dattilo}, {Davier}, {Day}, {Debreczeni}, {Degallaix}, {Del{\'e}glise}, {Pozzo}, {Dereli}, {Rosa}, {Fiore}, {Lieto}, {Virgilio}, {Doets}, {Dolique}, {Drago}, {Ducrot}, {Endr{\H{o}}czi}, {Fafone}, {Farinon}, {Ferrante}, {Ferrini}, {Fidecaro}, {Fiori}, {Flaminio},
  {Fournier}, {Franco}, {Frasca}, {Frasconi}, {Gammaitoni}, {Garufi}, {Gaspard}, {Gatto}, {Gemme}, {Gendre}, {Genin}, {Gennai}, {Ghosh}, {Giacobone}, {Giazotto}, {Gouaty}, {Granata}, {Greco}, {Groot}, {Guidi}, {Harms}, {Heidmann}, {Heitmann}, {Hello}, {Hemming}, {Hennes}, {Hofman}, {Jaranowski}, {Jonker}, {Kasprzack}, {K{\'e}f{\'e}lian}, {Kowalska}, {Kraan}, {Kr{\'o}lak}, {Kutynia}, {Lazzaro}, {Leonardi}, {Leroy}, {Letendre}, {Li}, {Lieunard}, {Lorenzini}, {Loriette}, {Losurdo}, {Magazz{\`u}}, {Majorana}, {Maksimovic}, {Malvezzi}, {Man}, {Mangano}, {Mantovani}, {Marchesoni}, {Marion}, {Marque}, {Martelli}, {Martellini}, {Masserot}, {Meacher}, {Meidam}, {Mezzani}, {Michel}, {Milano}, {Minenkov}, {Moggi}, {Mohan}, {Montani}, {Morgado}, {Mours}, {Mul}, {Nagy}, {Nardecchia}, {Naticchioni}, {Nelemans}, {Neri}, {Neri}, {Nocera}, {Pacaud}, {Palomba}, {Paoletti}, {Paoli}, {Pasqualetti}, {Passaquieti}, {Passuello}, {Perciballi}, {Petit}, {Pichot}, {Piergiovanni}, {Pillant}, {Piluso}, {Pinard}, {Poggiani}, {Prijatelj},
  {Prodi}, {Punturo}, {Puppo}, {Rabeling}, {R{\'a}cz}, {Rapagnani}, {Razzano}, {Re}, {Regimbau}, {Ricci}, {Robinet}, {Rocchi}, {Rolland}, {Romano}, {Rosi{\'n}ska}, {Ruggi}, \& {Saracco}}]{AdVirgo:2015}
{Acernese}, F., {Agathos}, M., {Agatsuma}, K., {et~al.} 2015, Classical and Quantum Gravity, 32, 024001, \dodoi{10.1088/0264-9381/32/2/024001}

\bibitem[{{Afzal} {et~al.}(2023){Afzal}, {Agazie}, {Anumarlapudi}, {Archibald}, {Arzoumanian}, {Baker}, {B{\'e}csy}, {Blanco-Pillado}, {Blecha}, {Boddy}, {Brazier}, {Brook}, {Burke-Spolaor}, {Burnette}, {Case}, {Charisi}, {Chatterjee}, {Chatziioannou}, {Cheeseboro}, {Chen}, {Cohen}, {Cordes}, {Cornish}, {Crawford}, {Cromartie}, {Crowter}, {Cutler}, {Decesar}, {Degan}, {Demorest}, {Deng}, {Dolch}, {Drachler}, {von Eckardstein}, {Ferrara}, {Fiore}, {Fonseca}, {Freedman}, {Garver-Daniels}, {Gentile}, {Gersbach}, {Glaser}, {Good}, {Guertin}, {G{\"u}ltekin}, {Hazboun}, {Hourihane}, {Islo}, {Jennings}, {Johnson}, {Jones}, {Kaiser}, {Kaplan}, {Kelley}, {Kerr}, {Key}, {Laal}, {Lam}, {Lamb}, {Lazio}, {Lee}, {Lewandowska}, {Lino Dos Santos}, {Littenberg}, {Liu}, {Lorimer}, {Luo}, {Lynch}, {Ma}, {Madison}, {McEwen}, {McKee}, {McLaughlin}, {McMann}, {Meyers}, {Meyers}, {Mingarelli}, {Mitridate}, {Nay}, {Natarajan}, {Ng}, {Nice}, {Ocker}, {Olum}, {Pennucci}, {Perera}, {Petrov}, {Pol}, {Radovan}, {Ransom}, {Ray}, {Romano},
  {Sardesai}, {Schmiedekamp}, {Schmiedekamp}, {Schmitz}, {Schr{\"o}der}, {Schult}, {Shapiro-Albert}, {Siemens}, {Simon}, {Siwek}, {Stairs}, {Stinebring}, {Stovall}, {Stratmann}, {Sun}, {Susobhanan}, {Swiggum}, {Taylor}, {Taylor}, {Trickle}, {Turner}, {Unal}, {Vallisneri}, {Verma}, {Vigeland}, {Wahl}, {Wang}, {Witt}, {Wright}, {Young}, {Zurek}, \& {Nanograv Collaboration}}]{2023ApJ...951L..11A}
{Afzal}, A., {Agazie}, G., {Anumarlapudi}, A., {et~al.} 2023, \apjl, 951, L11, \dodoi{10.3847/2041-8213/acdc91}

\bibitem[{Afzal {et~al.}(2023)Afzal, Agazie, Anumarlapudi, Archibald, Arzoumanian, Baker, B{\'e}csy, Blanco-Pillado, Blecha, Boddy, Brazier, Brook, Burke-Spolaor, Burnette, Case, Charisi, Chatterjee, Chatziioannou, Cheeseboro, Chen, Cohen, Cordes, Cornish, Crawford, Cromartie, Crowter, Cutler, Decesar, Degan, Demorest, Deng, Dolch, Drachler, von Eckardstein, Ferrara, Fiore, Fonseca, Freedman, Garver-Daniels, Gentile, Gersbach, Glaser, Good, Guertin, G{\"u}ltekin, Hazboun, Hourihane, Islo, Jennings, Johnson, Jones, Kaiser, Kaplan, Kelley, Kerr, Key, Laal, Lam, Lamb, Lazio, Lee, Lewandowska, Lino Dos~Santos, Littenberg, Liu, Lorimer, Luo, Lynch, Ma, Madison, McEwen, McKee, McLaughlin, McMann, Meyers, Meyers, Mingarelli, Mitridate, Nay, Natarajan, Ng, Nice, Ocker, Olum, Pennucci, Perera, Petrov, Pol, Radovan, Ransom, Ray, Romano, Sardesai, Schmiedekamp, Schmiedekamp, Schmitz, Schr{\"o}der, Schult, Shapiro-Albert, Siemens, Simon, Siwek, Stairs, Stinebring, Stovall, Stratmann, Sun, Susobhanan, Swiggum, Taylor,
  Taylor, Trickle, Turner, Unal, Vallisneri, Verma, Vigeland, Wahl, Wang, Witt, Wright, Young, Zurek, \& Collaboration}]{NANOGrav:2023NewPhysics}
Afzal, A., Agazie, G., Anumarlapudi, A., {et~al.} 2023, Astrophys. J. Lett., 951, L11, \dodoi{10.3847/2041-8213/acdc91}

\bibitem[{{Agarwal} {et~al.}(2025){Agarwal}, {Agazie}, {Anumarlapudi}, {Archibald}, {Arzoumanian}, {Baier}, {Baker}, {Becsy}, {Blecha}, {Brazier}, {Brook}, {Burke-Spolaor}, {Burnette}, {Case}, {Casey-Clyde}, {Chang}, {Charisi}, {Chatterjee}, {Cohen}, {Coppi}, {Cordes}, {Cornish}, {Crawford}, {Cromartie}, {Crowter}, {DeCesar}, {Demorest}, {Deng}, {Dey}, {Dolch}, {D'Orazio}, {Eisenberg}, {Ferrara}, {Fiore}, {Fonseca}, {Freedman}, {Gardiner}, {Garver-Daniels}, {Gentile}, {Gersbach}, {Glaser}, {Graham}, {Good}, {Gultekin}, {Harris}, {Hazboun}, {Hutchison}, {Jennings}, {Johnson}, {Jones}, {Kaplan}, {Kelley}, {Kerr}, {Key}, {Laal}, {Lam}, {Lamb}, {Larsen}, {Lazio}, {Lewandowska}, {Liu}, {Lorimer}, {Luo}, {Lynch}, {Ma}, {Madison}, {Matt}, {McEwen}, {McKee}, {McLaughlin}, {McMann}, {Meyers}, {Meyers}, {Mingarelli}, {Mitridate}, {Natarajan}, {Ng}, {Nice}, {Ocker}, {Olum}, {Pennucci}, {Perera}, {Petrov}, {Pol}, {Radovan}, {Ransom}, {Ray}, {Romano}, {Runnoe}, {Saffer}, {Sardesai}, {Schmiedekamp}, {Schmiedekamp},
  {Schmitz}, {Semenzato}, {Shapiro-Albert}, {Shivakumar}, {Siemens}, {Simon}, {Sosa Fiscella}, {Stairs}, {Stinebring}, {Stovall}, {Susobhanan}, {Swiggum}, {Taylor}, {Taylor}, {Thompson}, {Turner}, {Vallisneri}, {van Haasteren}, {Vigeland}, {Wahl}, {Willson}, {Wilson}, {Witt}, {Wright}, {Young}, \& {Zheng}}]{Agarwaletal2025}
{Agarwal}, N., {Agazie}, G., {Anumarlapudi}, A., {et~al.} 2025, arXiv e-prints, arXiv:2508.16534, \dodoi{10.48550/arXiv.2508.16534}

\bibitem[{{Agazie} {et~al.}(2023{\natexlab{a}}){Agazie}, {Alam}, {Anumarlapudi}, {Archibald}, {Arzoumanian}, {Baker}, {Blecha}, {Bonidie}, {Brazier}, {Brook}, {Burke-Spolaor}, {B{\'e}csy}, {Chapman}, {Charisi}, {Chatterjee}, {Cohen}, {Cordes}, {Cornish}, {Crawford}, {Cromartie}, {Crowter}, {Decesar}, {Demorest}, {Dolch}, {Drachler}, {Ferrara}, {Fiore}, {Fonseca}, {Freedman}, {Garver-Daniels}, {Gentile}, {Glaser}, {Good}, {G{\"u}ltekin}, {Hazboun}, {Jennings}, {Jessup}, {Johnson}, {Jones}, {Kaiser}, {Kaplan}, {Kelley}, {Kerr}, {Key}, {Kuske}, {Laal}, {Lam}, {Lamb}, {Lazio}, {Lewandowska}, {Lin}, {Liu}, {Lorimer}, {Luo}, {Lynch}, {Ma}, {Madison}, {Maraccini}, {McEwen}, {McKee}, {McLaughlin}, {McMann}, {Meyers}, {Mingarelli}, {Mitridate}, {Ng}, {Nice}, {Ocker}, {Olum}, {Panciu}, {Pennucci}, {Perera}, {Pol}, {Radovan}, {Ransom}, {Ray}, {Romano}, {Salo}, {Sardesai}, {Schmiedekamp}, {Schmiedekamp}, {Schmitz}, {Shapiro-Albert}, {Siemens}, {Simon}, {Siwek}, {Stairs}, {Stinebring}, {Stovall}, {Susobhanan}, {Swiggum},
  {Taylor}, {Turner}, {Unal}, {Vallisneri}, {Vigeland}, {Wahl}, {Wang}, {Witt}, {Young}, \& {Nanograv Collaboration}}]{2023ApJ...951L...9A}
{Agazie}, G., {Alam}, M.~F., {Anumarlapudi}, A., {et~al.} 2023{\natexlab{a}}, \apjl, 951, L9, \dodoi{10.3847/2041-8213/acda9a}

\bibitem[{{Agazie} {et~al.}(2023{\natexlab{b}}){Agazie}, {Anumarlapudi}, {Archibald}, {Arzoumanian}, {Baker}, {B{\'e}csy}, {Blecha}, {Brazier}, {Brook}, {Burke-Spolaor}, {Burnette}, {Case}, {Charisi}, {Chatterjee}, {Chatziioannou}, {Cheeseboro}, {Chen}, {Cohen}, {Cordes}, {Cornish}, {Crawford}, {Cromartie}, {Crowter}, {Cutler}, {Decesar}, {Degan}, {Demorest}, {Deng}, {Dolch}, {Drachler}, {Ellis}, {Ferrara}, {Fiore}, {Fonseca}, {Freedman}, {Garver-Daniels}, {Gentile}, {Gersbach}, {Glaser}, {Good}, {G{\"u}ltekin}, {Hazboun}, {Hourihane}, {Islo}, {Jennings}, {Johnson}, {Jones}, {Kaiser}, {Kaplan}, {Kelley}, {Kerr}, {Key}, {Klein}, {Laal}, {Lam}, {Lamb}, {Lazio}, {Lewandowska}, {Littenberg}, {Liu}, {Lommen}, {Lorimer}, {Luo}, {Lynch}, {Ma}, {Madison}, {Mattson}, {McEwen}, {McKee}, {McLaughlin}, {McMann}, {Meyers}, {Meyers}, {Mingarelli}, {Mitridate}, {Natarajan}, {Ng}, {Nice}, {Ocker}, {Olum}, {Pennucci}, {Perera}, {Petrov}, {Pol}, {Radovan}, {Ransom}, {Ray}, {Romano}, {Sardesai}, {Schmiedekamp}, {Schmiedekamp},
  {Schmitz}, {Schult}, {Shapiro-Albert}, {Siemens}, {Simon}, {Siwek}, {Stairs}, {Stinebring}, {Stovall}, {Sun}, {Susobhanan}, {Swiggum}, {Taylor}, {Taylor}, {Turner}, {Unal}, {Vallisneri}, {van Haasteren}, {Vigeland}, {Wahl}, {Wang}, {Witt}, {Young}, \& {Nanograv Collaboration}}]{2023ApJ...951L...8A}
{Agazie}, G., {Anumarlapudi}, A., {Archibald}, A.~M., {et~al.} 2023{\natexlab{b}}, \apjl, 951, L8, \dodoi{10.3847/2041-8213/acdac6}

\bibitem[{{Agazie} {et~al.}(2023{\natexlab{c}}){Agazie}, {Anumarlapudi}, {Archibald}, {Arzoumanian}, {Baker}, {B{\'e}csy}, {Blecha}, {Brazier}, {Brook}, {Burke-Spolaor}, {Case}, {Casey-Clyde}, {Charisi}, {Chatterjee}, {Cohen}, {Cordes}, {Cornish}, {Crawford}, {Cromartie}, {Crowter}, {Decesar}, {Demorest}, {Digman}, {Dolch}, {Drachler}, {Ferrara}, {Fiore}, {Fonseca}, {Freedman}, {Garver-Daniels}, {Gentile}, {Glaser}, {Good}, {G{\"u}ltekin}, {Hazboun}, {Hourihane}, {Jennings}, {Johnson}, {Jones}, {Kaiser}, {Kaplan}, {Kelley}, {Kerr}, {Key}, {Laal}, {Lam}, {Lamb}, {Lazio}, {Lewandowska}, {Liu}, {Lorimer}, {Luo}, {Lynch}, {Ma}, {Madison}, {McEwen}, {McKee}, {McLaughlin}, {McMann}, {Meyers}, {Meyers}, {Mingarelli}, {Mitridate}, {Ng}, {Nice}, {Ocker}, {Olum}, {Pennucci}, {Perera}, {Petrov}, {Pol}, {Radovan}, {Ransom}, {Ray}, {Romano}, {Sardesai}, {Schmiedekamp}, {Schmiedekamp}, {Schmitz}, {Shapiro-Albert}, {Siemens}, {Simon}, {Siwek}, {Stairs}, {Stinebring}, {Stovall}, {Susobhanan}, {Swiggum}, {Taylor}, {Taylor},
  {Turner}, {Unal}, {Vallisneri}, {van Haasteren}, {Vigeland}, {Wahl}, {Witt}, {Young}, \& {Nanograv Collaboration}}]{2023ApJ...951L..50A}
---. 2023{\natexlab{c}}, \apjl, 951, L50, \dodoi{10.3847/2041-8213/ace18a}

\bibitem[{{Agazie} {et~al.}(2023{\natexlab{d}}){Agazie}, {Anumarlapudi}, {Archibald}, {Arzoumanian}, {Baker}, {B{\'e}csy}, {Blecha}, {Brazier}, {Brook}, {Burke-Spolaor}, {Charisi}, {Chatterjee}, {Cohen}, {Cordes}, {Cornish}, {Crawford}, {Cromartie}, {Crowter}, {Decesar}, {Demorest}, {Dolch}, {Drachler}, {Ferrara}, {Fiore}, {Fonseca}, {Freedman}, {Garver-Daniels}, {Gentile}, {Glaser}, {Good}, {Guertin}, {G{\"u}ltekin}, {Hazboun}, {Jennings}, {Johnson}, {Jones}, {Kaiser}, {Kaplan}, {Kelley}, {Kerr}, {Key}, {Laal}, {Lam}, {Lamb}, {Lazio}, {Lewandowska}, {Liu}, {Lorimer}, {Luo}, {Lynch}, {Ma}, {Madison}, {McEwen}, {McKee}, {McLaughlin}, {McMann}, {Meyers}, {Mingarelli}, {Mitridate}, {Ng}, {Nice}, {Ocker}, {Olum}, {Pennucci}, {Perera}, {Pol}, {Radovan}, {Ransom}, {Ray}, {Romano}, {Sardesai}, {Schmiedekamp}, {Schmiedekamp}, {Schmitz}, {Shapiro-Albert}, {Siemens}, {Simon}, {Siwek}, {Stairs}, {Stinebring}, {Stovall}, {Susobhanan}, {Swiggum}, {Taylor}, {Turner}, {Unal}, {Vallisneri}, {Vigeland}, {Wahl}, {Witt},
  {Young}, \& {Nanograv Collaboration}}]{ng15detchar}
---. 2023{\natexlab{d}}, \apjl, 951, L10, \dodoi{10.3847/2041-8213/acda88}

\bibitem[{{Agazie} {et~al.}(2024{\natexlab{a}}){Agazie}, {Antoniadis}, {Anumarlapudi}, {Archibald}, {Arumugam}, {Arumugam}, {Arzoumanian}, {Askew}, {Babak}, {Bagchi}, {Bailes}, {Bak Nielsen}, {Baker}, {Bassa}, {Bathula}, {B{\'e}csy}, {Berthereau}, {Bhat}, {Blecha}, {Bonetti}, {Bortolas}, {Brazier}, {Brook}, {Burgay}, {Burke-Spolaor}, {Burnette}, {Caballero}, {Cameron}, {Case}, {Chalumeau}, {Champion}, {Chanlaridis}, {Charisi}, {Chatterjee}, {Chatziioannou}, {Cheeseboro}, {Chen}, {Chen}, {Cognard}, {Cohen}, {Coles}, {Cordes}, {Cornish}, {Crawford}, {Cromartie}, {Crowter}, {Cury{\l}o}, {Cutler}, {Dai}, {Dandapat}, {Deb}, {DeCesar}, {DeGan}, {Demorest}, {Deng}, {Desai}, {Desvignes}, {Dey}, {Dhanda-Batra}, {Di Marco}, {Dolch}, {Drachler}, {Dwivedi}, {Ellis}, {Falxa}, {Feng}, {Ferdman}, {Ferrara}, {Fiore}, {Fonseca}, {Franchini}, {Freedman}, {Gair}, {Garver-Daniels}, {Gentile}, {Gersbach}, {Glaser}, {Good}, {Goncharov}, {Gopakumar}, {Graikou}, {Griessmeier}, {Guillemot}, {G{\"u}ltekin}, {Guo}, {Gupta}, {Grunthal},
  {Hazboun}, {Hisano}, {Hobbs}, {Hourihane}, {Hu}, {Iraci}, {Islo}, {Izquierdo-Villalba}, {Jang}, {Jawor}, {Janssen}, {Jennings}, {Jessner}, {Johnson}, {Jones}, {Joshi}, {Kaiser}, {Kaplan}, {Kapur}, {Kareem}, {Karuppusamy}, {Keane}, {Keith}, {Kelley}, {Kerr}, {Key}, {Kharbanda}, {Kikunaga}, {Klein}, {Kolhe}, {Kramer}, {Krishnakumar}, {Kulkarni}, {Laal}, {Lackeos}, {Lam}, {Lamb}, {Larsen}, {Lazio}, {Lee}, {Levin}, {Lewandowska}, {Littenberg}, {Liu}, {Liu}, {Liu}, {Lommen}, {Lorimer}, {Lower}, {Luo}, {Luo}, {Lynch}, {Lyne}, {Ma}, {Maan}, {Madison}, {Main}, {Manchester}, {Mandow}, {Mattson}, {McEwen}, {McKee}, {McLaughlin}, {McMann}, {Meyers}, {Meyers}, {Mickaliger}, {Miles}, {Mingarelli}, {Mitridate}, {Natarajan}, {Nathan}, {Ng}, {Nice}, {Ni{\c{t}}u}, {Nobleson}, {Ocker}, {Olum}, {Os{\l}owski}, {Paladi}, {Parthasarathy}, {Pennucci}, {Perera}, {Perrodin}, {Petiteau}, {Petrov}, {Pol}, {Porayko}, {Possenti}, {Prabu}, {Quelquejay Leclere}, {Radovan}, {Rana}, {Ransom}, {Ray}, {Reardon}, {Rogers}, {Romano},
  {Russell}, {Samajdar}, {Sanidas}, {Sardesai}, {Schmiedekamp}, {Schmiedekamp}, {Schmitz}, {Schult}, {Sesana}, {Shaifullah}, {Shannon}, {Shapiro-Albert}, {Siemens}, {Simon}, \& {Singha}}]{2024ApJ...966..105A}
{Agazie}, G., {Antoniadis}, J., {Anumarlapudi}, A., {et~al.} 2024{\natexlab{a}}, \apj, 966, 105, \dodoi{10.3847/1538-4357/ad36be}

\bibitem[{{Agazie} {et~al.}(2024{\natexlab{b}}){Agazie}, {Arzoumanian}, {Baker}, {B{\'e}csy}, {Blecha}, {Blumer}, {Brazier}, {Brook}, {Burke-Spolaor}, {Casey-Clyde}, {Charisi}, {Chatterjee}, {Cheeseboro}, {Cohen}, {Cordes}, {Cornish}, {Crawford}, {Cromartie}, {Decesar}, {Demorest}, {Dey}, {Dolch}, {Ellis}, {Ferdman}, {Ferrara}, {Fiore}, {Fonseca}, {Freedman}, {Garver-Daniels}, {Gentile}, {Glaser}, {Good}, {Gopakumar}, {G{\"u}ltekin}, {Hazboun}, {Jennings}, {Johnson}, {Jones}, {Kaiser}, {Kaplan}, {Kelley}, {Key}, {Laal}, {Lam}, {Lamb}, {W. Lazio}, {Lewandowska}, {Liu}, {Lorimer}, {Luo}, {Lynch}, {Ma}, {Madison}, {McEwen}, {McKee}, {McLaughlin}, {Meyers}, {Mingarelli}, {Mitridate}, {Ng}, {Nice}, {Ocker}, {Olum}, {Pennucci}, {Pol}, {Radovan}, {Ransom}, {Ray}, {Romano}, {Sardesai}, {Schmitz}, {Siemens}, {Simon}, {Siwek}, {Sosa Fiscella}, {Spiewak}, {Stairs}, {Stinebring}, {Stovall}, {Susobhanan}, {Swiggum}, {Taylor}, {Turner}, {Unal}, {Vallisneri}, {Vigeland}, {Witt}, {Young}, \& {NANOGrav
  Collaboration}}]{N15ecc}
{Agazie}, G., {Arzoumanian}, Z., {Baker}, P.~T., {et~al.} 2024{\natexlab{b}}, \apj, 963, 144, \dodoi{10.3847/1538-4357/ad1f61}

\bibitem[{{Agazie} {et~al.}(2025{\natexlab{a}}){Agazie}, {Anumarlapudi}, {Archibald}, {Arzoumanian}, {Baier}, {Baker}, {B{\'e}csy}, {Blecha}, {Brazier}, {Brook}, {Brown}, {Burke-Spolaor}, {Casey-Clyde}, {Charisi}, {Chatterjee}, {Cohen}, {Cordes}, {Cornish}, {Crawford}, {Cromartie}, {Crowter}, {DeCesar}, {Demorest}, {Deng}, {Dolch}, {Ferrara}, {Fiore}, {Fonseca}, {Freedman}, {Garver-Daniels}, {Gentile}, {Glaser}, {Good}, {G{\"u}ltekin}, {Hazboun}, {Jennings}, {Johnson}, {Jones}, {Kaiser}, {Kaplan}, {Kelley}, {Kerr}, {Key}, {Laal}, {Lam}, {Lamb}, {Larsen}, {Lazio}, {Lewandowska}, {Liu}, {Lorimer}, {Luo}, {Lynch}, {Ma}, {Madison}, {McEwen}, {McKee}, {McLaughlin}, {McMann}, {Meyers}, {Meyers}, {Mingarelli}, {Mitridate}, {Natarajan}, {Ng}, {Nice}, {Ocker}, {Olum}, {Pennucci}, {Perera}, {Pol}, {Radovan}, {Ransom}, {Ray}, {Romano}, {Runnoe}, {Sardesai}, {Schmiedekamp}, {Schmiedekamp}, {Schmitz}, {Shapiro-Albert}, {Siemens}, {Simon}, {Siwek}, {Sosa Fiscella}, {Stairs}, {Stinebring}, {Stovall}, {Susobhanan},
  {Swiggum}, {Taylor}, {Turner}, {Unal}, {Vallisneri}, {Vigeland}, {Wahl}, {Willson}, {Witt}, {Wright}, \& {Young}}]{2025ApJ...978...31A}
{Agazie}, G., {Anumarlapudi}, A., {Archibald}, A.~M., {et~al.} 2025{\natexlab{a}}, \apj, 978, 31, \dodoi{10.3847/1538-4357/ad93d5}

\bibitem[{{Agazie} {et~al.}(2025{\natexlab{b}}){Agazie}, {Anumarlapudi}, {Archibald}, {Arzoumanian}, {Baier}, {Baker}, {B{\'e}csy}, {Blecha}, {Brazier}, {Brook}, {Brown}, {Burke-Spolaor}, {Casey-Clyde}, {Charisi}, {Chatterjee}, {Cohen}, {Cordes}, {Cornish}, {Crawford}, {Cromartie}, {Crowter}, {DeCesar}, {Demorest}, {Deng}, {Dolch}, {Ferrara}, {Fiore}, {Fonseca}, {Freedman}, {Garver-Daniels}, {Gentile}, {Glaser}, {Good}, {G{\"u}ltekin}, {Hazboun}, {Jennings}, {Johnson}, {Jones}, {Kaiser}, {Kaplan}, {Kelley}, {Kerr}, {Key}, {Laal}, {Lam}, {Lamb}, {Larsen}, {Lazio}, {Lewandowska}, {Liu}, {Lorimer}, {Luo}, {Lynch}, {Ma}, {Madison}, {McEwen}, {McKee}, {McLaughlin}, {McMann}, {Meyers}, {Meyers}, {Mingarelli}, {Mitridate}, {Natarajan}, {Ng}, {Nice}, {Ocker}, {Olum}, {Pennucci}, {Perera}, {Pol}, {Radovan}, {Ransom}, {Ray}, {Romano}, {Runnoe}, {Sardesai}, {Schmiedekamp}, {Schmiedekamp}, {Schmitz}, {Shapiro-Albert}, {Siemens}, {Simon}, {Siwek}, {Sosa Fiscella}, {Stairs}, {Stinebring}, {Stovall}, {Susobhanan},
  {Swiggum}, {Taylor}, {Turner}, {Unal}, {Vallisneri}, {Vigeland}, {Wahl}, {Willson}, {Witt}, {Wright}, \& {Young}}]{ng15-discreteness}
---. 2025{\natexlab{b}}, \apj, 978, 31, \dodoi{10.3847/1538-4357/ad93d5}

\bibitem[{{Agazie} {et~al.}(2025{\natexlab{c}}){Agazie}, {Kaplan}, {Susobhanan}, {Stairs}, {Good}, {Meyers}, {Fonseca}, {Pennucci}, {Anumarlapudi}, {Archibald}, {Arzoumanian}, {Baker}, {Brook}, {Cassity}, {Cromartie}, {Crowter}, {DeCesar}, {Demorest}, {Dolch}, {Dong}, {Ferrara}, {Fiore}, {Freedman}, {Garver-Daniels}, {Gentile}, {Glaser}, {Hazboun}, {Jennings}, {Jones}, {Kerr}, {Lam}, {Lorimer}, {Luo}, {Lynch}, {McEwen}, {McKee}, {McLaughlin}, {McMann}, {Ng}, {Nice}, {Perera}, {Pol}, {Radovan}, {Ransom}, {Ray}, {Saffer}, {Schmiedekamp}, {Schmiedekamp}, {Shapiro-Albert}, {Stovall}, {Swiggum}, {Thompson}, \& {Wahl}}]{2025arXiv251016668A}
{Agazie}, G., {Kaplan}, D.~L., {Susobhanan}, A., {et~al.} 2025{\natexlab{c}}, arXiv e-prints, arXiv:2510.16668, \dodoi{10.48550/arXiv.2510.16668}

\bibitem[{{Akutsu} {et~al.}(2021){Akutsu}, {Ando}, {Arai}, {Arai}, {Araki}, {Araya}, {Aritomi}, {Asada}, {Aso}, {Bae}, {Bae}, {Baiotti}, {Bajpai}, {Barton}, {Cannon}, {Cao}, {Capocasa}, {Chan}, {Chen}, {Chen}, {Chen}, {Chiang}, {Chu}, {Chu}, {Eguchi}, {Enomoto}, {Flaminio}, {Fujii}, {Fujikawa}, {Fukunaga}, {Fukushima}, {Gao}, {Ge}, {Ha}, {Hagiwara}, {Haino}, {Han}, {Hasegawa}, {Hattori}, {Hayakawa}, {Hayama}, {Himemoto}, {Hiranuma}, {Hirata}, {Hirose}, {Hong}, {Hsieh}, {Huang}, {Huang}, {Huang}, {Huang}, {Huang}, {Hui}, {Ide}, {Ikenoue}, {Imam}, {Inayoshi}, {Inoue}, {Ioka}, {Ito}, {Itoh}, {Izumi}, {Jeon}, {Jin}, {Jung}, {Jung}, {Kaihotsu}, {Kajita}, {Kakizaki}, {Kamiizumi}, {Kanda}, {Kang}, {Kashiyama}, {Kawaguchi}, {Kawai}, {Kawasaki}, {Kim}, {Kim}, {Kim}, {Kim}, {Kim}, {Kimura}, {Kita}, {Kitazawa}, {Kojima}, {Kokeyama}, {Komori}, {Kong}, {Kotake}, {Kozakai}, {Kozu}, {Kumar}, {Kume}, {Kuo}, {Kuo}, {Kuromiya}, {Kuroyanagi}, {Kusayanagi}, {Kwak}, {Lee}, {Lee}, {Lee}, {Leonardi}, {Li}, {Li}, {Lin}, {Lin},
  {Lin}, {Lin}, {Lin}, {Liu}, {Luo}, {Majorana}, {Marchio}, {Michimura}, {Mio}, {Miyakawa}, {Miyamoto}, {Miyazaki}, {Miyo}, {Miyoki}, {Mori}, {Morisaki}, {Moriwaki}, {Nagano}, {Nagano}, {Nakamura}, {Nakano}, {Nakano}, {Nakashima}, {Nakayama}, {Narikawa}, {Naticchioni}, {Negishi}, {Nguyen Quynh}, {Ni}, {Nishizawa}, {Nozaki}, {Obuchi}, {Ogaki}, {Oh}, {Oh}, {Oh}, {Ohashi}, {Ohishi}, {Ohkawa}, {Ohta}, {Okutani}, {Okutomi}, {Oohara}, {Ooi}, {Oshino}, {Otabe}, {Pan}, {Pang}, {Parisi}, {Park}, {Pe na Arellano}, {Pinto}, {Sago}, {Saito}, {Saito}, {Sakai}, {Sakai}, {Sakuno}, {Sato}, {Sato}, {Sawada}, {Sekiguchi}, {Sekiguchi}, {Shao}, {Shibagaki}, {Shimizu}, {Shimoda}, {Shimode}, {Shinkai}, {Shishido}, {Shoda}, {Somiya}, {Son}, {Sotani}, {Sugimoto}, {Suresh}, {Suzuki}, {Suzuki}, {Tagoshi}, {Takahashi}, {Takahashi}, {Takamori}, {Takano}, {Takeda}, {Takeda}, {Tanaka}, {Tanaka}, {Tanaka}, {Tanaka}, {Tanaka}, {Tanioka}, {Tapia San Martin}, \& {Telada}}]{KAGRA:2021}
{Akutsu}, T., {Ando}, M., {Arai}, K., {et~al.} 2021, Progress of Theoretical and Experimental Physics, 2021, 05A103, \dodoi{10.1093/ptep/ptaa120}

\bibitem[{{Ali-Ha{\"\i}moud} {et~al.}(2020){Ali-Ha{\"\i}moud}, {Smith}, \& {Mingarelli}}]{alihaimoud2020}
{Ali-Ha{\"\i}moud}, Y., {Smith}, T.~L., \& {Mingarelli}, C. M.~F. 2020, \prd, 102, 122005, \dodoi{10.1103/PhysRevD.102.122005}

\bibitem[{{Ali-Ha{\"\i}moud} {et~al.}(2021){Ali-Ha{\"\i}moud}, {Smith}, \& {Mingarelli}}]{alihaimoud2021}
---. 2021, \prd, 103, 042009, \dodoi{10.1103/PhysRevD.103.042009}

\bibitem[{Allen(1996)}]{Allen:1996vm}
Allen, B. 1996, in {Les Houches School of Physics: Astrophysical Sources of Gravitational Radiation}, 373--417.
\newblock \doarXiv{gr-qc/9604033}

\bibitem[{{Allen} {et~al.}(2023){Allen}, {Dhurandhar}, {Gupta}, {McLaughlin}, {Natarajan}, {Shannon}, {Thrane}, \& {Vecchio}}]{2023arXiv230404767A}
{Allen}, B., {Dhurandhar}, S., {Gupta}, Y., {et~al.} 2023, arXiv e-prints, arXiv:2304.04767, \dodoi{10.48550/arXiv.2304.04767}

\bibitem[{Allen \& Shellard(1992)}]{Allen:1991bk}
Allen, B., \& Shellard, E. P.~S. 1992, Phys. Rev. D, 45, 1898, \dodoi{10.1103/PhysRevD.45.1898}

\bibitem[{{Alpar} {et~al.}(1982){Alpar}, {Cheng}, {Ruderman}, \& {Shaham}}]{1982Natur.300..728A}
{Alpar}, M.~A., {Cheng}, A.~F., {Ruderman}, M.~A., \& {Shaham}, J. 1982, \nat, 300, 728, \dodoi{10.1038/300728a0}

\bibitem[{{Amaro-Seoane} {et~al.}(2010){Amaro-Seoane}, {Sesana}, {Hoffman}, {Benacquista}, {Eichhorn}, {Makino}, \& {Spurzem}}]{2010MNRAS.402.2308A}
{Amaro-Seoane}, P., {Sesana}, A., {Hoffman}, L., {et~al.} 2010, \mnras, 402, 2308, \dodoi{10.1111/j.1365-2966.2009.16104.x}

\bibitem[{Ananda {et~al.}(2007)Ananda, Clarkson, \& Wands}]{Ananda:2006af}
Ananda, K.~N., Clarkson, C., \& Wands, D. 2007, Phys. Rev. D, 75, 123518, \dodoi{10.1103/PhysRevD.75.123518}

\bibitem[{{Armstrong} {et~al.}(1995){Armstrong}, {Rickett}, \& {Spangler}}]{1995ApJ...443..209A}
{Armstrong}, J.~W., {Rickett}, B.~J., \& {Spangler}, S.~R. 1995, \apj, 443, 209, \dodoi{10.1086/175515}

\bibitem[{Arvanitaki {et~al.}(2015)Arvanitaki, Huang, \& Van~Tilburg}]{Arvanitaki:2015iga}
Arvanitaki, A., Huang, J., \& Van~Tilburg, K. 2015, \prd, 91, 015015, \dodoi{10.1103/PhysRevD.91.015015}

\bibitem[{{Arzoumanian} {et~al.}(2018){Arzoumanian}, {Baker}, {Brazier}, {Burke-Spolaor}, {Chamberlin}, {Chatterjee}, {Christy}, {Cordes}, {Cornish}, {Crawford}, \& {NANOGrav Collaboration}}]{NG11}
{Arzoumanian}, Z., {Baker}, P.~T., {Brazier}, A., {et~al.} 2018, \apj, 859, 47, \dodoi{10.3847/1538-4357/aabd3b}

\bibitem[{{Arzoumanian} {et~al.}(2020{\natexlab{a}}){Arzoumanian}, {Baker}, {Blumer}, {B{\'e}csy}, {Brazier}, {Brook}, {Burke-Spolaor}, {Chatterjee}, {Chen}, {Cordes}, {Cornish}, {Crawford}, {Cromartie}, {Decesar}, {Demorest}, {Dolch}, {Ellis}, {Ferrara}, {Fiore}, {Fonseca}, {Garver-Daniels}, {Gentile}, {Good}, {Hazboun}, {Holgado}, {Islo}, {Jennings}, {Jones}, {Kaiser}, {Kaplan}, {Kelley}, {Key}, {Laal}, {Lam}, {Lazio}, {Lorimer}, {Luo}, {Lynch}, {Madison}, {McLaughlin}, {Mingarelli}, {Ng}, {Nice}, {Pennucci}, {Pol}, {Ransom}, {Ray}, {Shapiro-Albert}, {Siemens}, {Simon}, {Spiewak}, {Stairs}, {Stinebring}, {Stovall}, {Sun}, {Swiggum}, {Taylor}, {Turner}, {Vallisneri}, {Vigeland}, {Witt}, \& {Nanograv Collaboration}}]{2020ApJ...905L..34A}
{Arzoumanian}, Z., {Baker}, P.~T., {Blumer}, H., {et~al.} 2020{\natexlab{a}}, \apjl, 905, L34, \dodoi{10.3847/2041-8213/abd401}

\bibitem[{{Arzoumanian} {et~al.}(2020{\natexlab{b}}){Arzoumanian}, {Baker}, {Brazier}, {Brook}, {Burke-Spolaor}, {B{\'e}csy}, {Charisi}, {Chatterjee}, {Cordes}, {Cornish}, {Crawford}, {Cromartie}, {Crowter}, {Decesar}, {Demorest}, {Dolch}, {Elliott}, {Ellis}, {Ferdman}, {Ferrara}, {Fonseca}, {Garver-Daniels}, {Gentile}, {Good}, {Hazboun}, {Islo}, {Jennings}, {Jones}, {Kaiser}, {Kaplan}, {Kelley}, {Key}, {Lam}, {Lazio}, {Levin}, {Luo}, {Lynch}, {Madison}, {McLaughlin}, {Mingarelli}, {Ng}, {Nice}, {Pennucci}, {Pol}, {Ransom}, {Ray}, {Shapiro-Albert}, {Siemens}, {Simon}, {Spiewak}, {Stairs}, {Stinebring}, {Stovall}, {Swiggum}, {Taylor}, {Vallisneri}, {Vigeland}, {Witt}, {Zhu}, \& {NANOGrav Collaboration}}]{Arzoumanianetal2020}
{Arzoumanian}, Z., {Baker}, P.~T., {Brazier}, A., {et~al.} 2020{\natexlab{b}}, \apj, 900, 102, \dodoi{10.3847/1538-4357/ababa1}

\bibitem[{Athron {et~al.}(2024)Athron, Balazs, Jacobson, \& Wu}]{Athron:2023hgm}
Athron, P., Balazs, C., Jacobson, D., \& Wu, Y.-L.~S. 2024, Physical Review Letters, 132, 221001, \dodoi{10.1103/PhysRevLett.132.221001}

\bibitem[{{Backer}(1970)}]{1970Natur.228.1297B}
{Backer}, D.~C. 1970, \nat, 228, 1297, \dodoi{10.1038/2281297a0}

\bibitem[{{Backer} {et~al.}(1982){Backer}, {Kulkarni}, {Heiles}, {Davis}, \& {Goss}}]{1982Natur.300..615B}
{Backer}, D.~C., {Kulkarni}, S.~R., {Heiles}, C., {Davis}, M.~M., \& {Goss}, W.~M. 1982, \nat, 300, 615, \dodoi{10.1038/300615a0}

\bibitem[{{Bailes} {et~al.}(2020){Bailes}, {Jameson}, {Abbate}, {Barr}, {Bhat}, {Bondonneau}, {Burgay}, {Buchner}, {Camilo}, {Champion}, {Cognard}, {Demorest}, {Freire}, {Gautam}, {Geyer}, {Griessmeier}, {Guillemot}, {Hu}, {Jankowski}, {Johnston}, {Karastergiou}, {Karuppusamy}, {Kaur}, {Keith}, {Kramer}, {van Leeuwen}, {Lower}, {Maan}, {McLaughlin}, {Meyers}, {Os{\l}owski}, {Oswald}, {Parthasarathy}, {Pennucci}, {Posselt}, {Possenti}, {Ransom}, {Reardon}, {Ridolfi}, {Schollar}, {Serylak}, {Shaifullah}, {Shamohammadi}, {Shannon}, {Sobey}, {Song}, {Spiewak}, {Stairs}, {Stappers}, {van Straten}, {Szary}, {Theureau}, {Venkatraman Krishnan}, {Weltevrede}, {Wex}, {Abbott}, {Adams}, {Burger}, {Gamatham}, {Gouws}, {Horn}, {Hugo}, {Joubert}, {Manley}, {McAlpine}, {Passmoor}, {Peens-Hough}, {Ramudzuli}, {Rust}, {Salie}, {Schwardt}, {Siebrits}, {Van Tonder}, {Van Tonder}, \& {Welz}}]{2020PASA...37...28B}
{Bailes}, M., {Jameson}, A., {Abbate}, F., {et~al.} 2020, \pasa, 37, e028, \dodoi{10.1017/pasa.2020.19}

\bibitem[{{Barausse} {et~al.}(2020){Barausse}, {Dvorkin}, {Tremmel}, {Volonteri}, \& {Bonetti}}]{BarausseEtAl:2020}
{Barausse}, E., {Dvorkin}, I., {Tremmel}, M., {Volonteri}, M., \& {Bonetti}, M. 2020, \apj, 904, 16, \dodoi{10.3847/1538-4357/abba7f}

\bibitem[{{Basu} {et~al.}(2025){Basu}, {Graber}, {Lower}, {Antonelli}, {Antonopoulou}, {Bagchi}, {Char}, {Freire}, {Haskell}, {Hu}, {Jones}, {Mukhopadhyay}, {Oertel}, {Rea}, {Sagun}, {Shaw}, {Singha}, {Stappers}, {Thongmeearkom}, {Watts}, {Weltevrede}, \& {{The SKA Pulsar Science Working Group}}}]{Basu2025_SKA_EOS}
{Basu}, A., {Graber}, V., {Lower}, M.~E., {et~al.} 2025

\bibitem[{{Baumann} {et~al.}(2007){Baumann}, {Steinhardt}, {Takahashi}, \& {Ichiki}}]{2007PhRvD..76h4019B}
{Baumann}, D., {Steinhardt}, P., {Takahashi}, K., \& {Ichiki}, K. 2007, \prd, 76, 084019, \dodoi{10.1103/PhysRevD.76.084019}

\bibitem[{{Begelman} {et~al.}(1980){Begelman}, {Blandford}, \& {Rees}}]{1980Natur.287..307B}
{Begelman}, M.~C., {Blandford}, R.~D., \& {Rees}, M.~J. 1980, \nat, 287, 307, \dodoi{10.1038/287307a0}

\bibitem[{{Bhat} {et~al.}(2016){Bhat}, {Ord}, {Tremblay}, {McSweeney}, \& {Tingay}}]{2016ApJ...818...86B}
{Bhat}, N.~D.~R., {Ord}, S.~M., {Tremblay}, S.~E., {McSweeney}, S.~J., \& {Tingay}, S.~J. 2016, \apj, 818, 86, \dodoi{10.3847/0004-637X/818/1/86}

\bibitem[{{Bhat} {et~al.}(2018){Bhat}, {Tremblay}, {Kirsten}, {Meyers}, {Sokolowski}, {van Straten}, {McSweeney}, {Ord}, {Shannon}, {Beardsley}, {Crosse}, {Emrich}, {Franzen}, {Horsley}, {Johnston-Hollitt}, {Kaplan}, {Kenney}, {Morales}, {Pallot}, {Steele}, {Tingay}, {Trott}, {Walker}, {Wayth}, {Williams}, \& {Wu}}]{2018ApJS..238....1B}
{Bhat}, N.~D.~R., {Tremblay}, S.~E., {Kirsten}, F., {et~al.} 2018, \apjs, 238, 1, \dodoi{10.3847/1538-4365/aad37c}

\bibitem[{{Bird} {et~al.}(1980){Bird}, {Schruefer}, {Volland}, \& {Sieber}}]{Bird1980}
{Bird}, M.~K., {Schruefer}, E., {Volland}, H., \& {Sieber}, W. 1980, \nat, 283, 459, \dodoi{10.1038/283459a0}

\bibitem[{Blas {et~al.}(2017)Blas, Nacir, \& Sibiryakov}]{Blas:2016ddr}
Blas, D., Nacir, D.~L., \& Sibiryakov, S. 2017, \prl, 118, 261102, \dodoi{10.1103/PhysRevLett.118.261102}

\bibitem[{{Bondonneau} {et~al.}(2021){Bondonneau}, {Grie{\ss}meier}, {Theureau}, {Cognard}, {Brionne}, {Kondratiev}, {Bilous}, {McKee}, {Zarka}, {Viou}, {Guillemot}, {Chen}, {Main}, {Pilia}, {Possenti}, {Serylak}, {Shaifullah}, {Tiburzi}, {Verbiest}, {Wu}, {Wucknitz}, {Yerin}, {Briand}, {Cecconi}, {Corbel}, {Dallier}, {Girard}, {Loh}, {Martin}, {Tagger}, \& {Tasse}}]{2021A&A...652A..34B}
{Bondonneau}, L., {Grie{\ss}meier}, J.-M., {Theureau}, G., {et~al.} 2021, \aap, 652, A34, \dodoi{10.1051/0004-6361/202039339}

\bibitem[{{Boyle} \& {Pen}(2012)}]{Boyle_Pen_2012}
{Boyle}, L., \& {Pen}, U.-L. 2012, \prd, 86, 124028, \dodoi{10.1103/PhysRevD.86.124028}

\bibitem[{Boyle \& Steinhardt(2008)}]{Boyle:2005se}
Boyle, L.~A., \& Steinhardt, P.~J. 2008, \prd, 77, 063504, \dodoi{10.1103/PhysRevD.77.063504}

\bibitem[{{Caballero} {et~al.}(2018){Caballero}, {Guo}, {Lee}, {Lazarus}, {Champion}, {Desvignes}, {Kramer}, {Plant}, {Arzoumanian}, {Bailes}, {Bassa}, {Bhat}, {Brazier}, {Burgay}, {Burke-Spolaor}, {Chamberlin}, {Chatterjee}, {Cognard}, {Cordes}, {Dai}, {Demorest}, {Dolch}, {Ferdman}, {Fonseca}, {Gair}, {Garver-Daniels}, {Gentile}, {Gonzalez}, {Graikou}, {Guillemot}, {Hobbs}, {Janssen}, {Karuppusamy}, {Keith}, {Kerr}, {Lam}, {Lasky}, {Lazio}, {Levin}, {Liu}, {Lommen}, {Lorimer}, {Lynch}, {Madison}, {Manchester}, {McKee}, {McLaughlin}, {McWilliams}, {Mingarelli}, {Nice}, {Os{\l}owski}, {Palliyaguru}, {Pennucci}, {Perera}, {Perrodin}, {Possenti}, {Ransom}, {Reardon}, {Sanidas}, {Sesana}, {Shaifullah}, {Shannon}, {Siemens}, {Simon}, {Spiewak}, {Stairs}, {Stappers}, {Stinebring}, {Stovall}, {Swiggum}, {Taylor}, {Theureau}, {Tiburzi}, {Toomey}, {van Haasteren}, {van Straten}, {Verbiest}, {Wang}, {Zhu}, \& {Zhu}}]{Caballero2018}
{Caballero}, R.~N., {Guo}, Y.~J., {Lee}, K.~J., {et~al.} 2018, \mnras, 481, 5501, \dodoi{10.1093/mnras/sty2632}

\bibitem[{Cai {et~al.}(2021)Cai, Pi, \& Wang}]{Cai:2021uup}
Cai, Y.-F., Pi, S., \& Wang, M.-Z. 2021, \jcap, 2021, 008, \dodoi{10.1088/1475-7516/2021/09/008}

\bibitem[{{Campeti} \& {Komatsu}(2022)}]{Campeti2022}
{Campeti}, P., \& {Komatsu}, E. 2022, \apj, 941, 110, \dodoi{10.3847/1538-4357/ac9ea3}

\bibitem[{Caprini {et~al.}(2009)Caprini, Durrer, \& Servant}]{Caprini:2009yp}
Caprini, C., Durrer, R., \& Servant, G. 2009, JCAP, 12, 024, \dodoi{10.1088/1475-7516/2009/12/024}

\bibitem[{Caprini \& Figueroa(2018)}]{Caprini:2018mtu}
Caprini, C., \& Figueroa, D.~G. 2018, \cqg, 35, 163001, \dodoi{10.1088/1361-6382/aac608}

\bibitem[{Caprini {et~al.}(2020)}]{Caprini:2019egz}
Caprini, C., {et~al.} 2020, JCAP, 03, 024, \dodoi{10.1088/1475-7516/2020/03/024}

\bibitem[{Caputo {et~al.}(2019)Caputo, Sberna, Frias, Blas, Pani, Shao, \& Yan}]{Caputo:2019tms}
Caputo, A., Sberna, L., Frias, M., {et~al.} 2019, Phys. Rev. D, 100, 063515, \dodoi{10.1103/PhysRevD.100.063515}

\bibitem[{Carroll {et~al.}(1990)Carroll, Field, \& Jackiw}]{Carroll:1989vb}
Carroll, S.~M., Field, G.~B., \& Jackiw, R. 1990, Phys. Rev. D, 41, 1231, \dodoi{10.1103/PhysRevD.41.1231}

\bibitem[{{Casey-Clyde} {et~al.}(2025){Casey-Clyde}, {Mingarelli}, {Greene}, {Goulding}, {Chen}, \& {Trump}}]{caseyclyde2025}
{Casey-Clyde}, J.~A., {Mingarelli}, C. M.~F., {Greene}, J.~E., {et~al.} 2025, \apj, 987, 106, \dodoi{10.3847/1538-4357/adce05}

\bibitem[{Castillo {et~al.}(2022)Castillo, Martin-Camalich, Terol-Calvo, Blas, Caputo, Santos, Sberna, Peel, \& Rubi{\~n}o-Mart{\'\i}n}]{Castillo:2022zfl}
Castillo, A., Martin-Camalich, J., Terol-Calvo, J., {et~al.} 2022, JCAP, 06, 014, \dodoi{10.1088/1475-7516/2022/06/014}

\bibitem[{{Champion} {et~al.}(2010){Champion}, {Hobbs}, {Manchester}, {et~al.}}]{Champion2010}
{Champion}, D.~J., {Hobbs}, G.~B., {Manchester}, R.~N., {et~al.} 2010, ApJL, 720, L201, \dodoi{10.1088/2041-8205/720/2/L201}

\bibitem[{{Chatterjee}(2018)}]{2018ASPC..517..751C}
{Chatterjee}, S. 2018, in Astronomical Society of the Pacific Conference Series, Vol. 517, Science with a Next Generation Very Large Array, ed. E.~{Murphy}, 751

\bibitem[{{Chen} {et~al.}(2017){Chen}, {Sesana}, \& {Del Pozzo}}]{2017MNRAS.470.1738C}
{Chen}, S., {Sesana}, A., \& {Del Pozzo}, W. 2017, \mnras, 470, 1738, \dodoi{10.1093/mnras/stx1093}

\bibitem[{{Chen} {et~al.}(2021){Chen}, {Caballero}, {Guo}, {Chalumeau}, {Liu}, {Shaifullah}, {Lee}, {Babak}, {Desvignes}, {Parthasarathy}, {Hu}, {van der Wateren}, {Antoniadis}, {Bak Nielsen}, {Bassa}, {Berthereau}, {Burgay}, {Champion}, {Cognard}, {Falxa}, {Ferdman}, {Freire}, {Gair}, {Graikou}, {Guillemot}, {Jang}, {Janssen}, {Karuppusamy}, {Keith}, {Kramer}, {Liu}, {Lyne}, {Main}, {McKee}, {Mickaliger}, {Perera}, {Perrodin}, {Petiteau}, {Porayko}, {Possenti}, {Samajdar}, {Sanidas}, {Sesana}, {Speri}, {Stappers}, {Theureau}, {Tiburzi}, {Vecchio}, {Verbiest}, {Wang}, {Wang}, \& {Xu}}]{2021MNRAS.508.4970C}
{Chen}, S., {Caballero}, R.~N., {Guo}, Y.~J., {et~al.} 2021, \mnras, 508, 4970, \dodoi{10.1093/mnras/stab2833}

\bibitem[{Chern \& Simons(1974)}]{d648afea-bdd6-33d6-bac5-32fb94e6cccf}
Chern, S.-S., \& Simons, J. 1974, Annals of Mathematics, 99, 48

\bibitem[{{Clark} {et~al.}(2015){Clark}, {Pletsch}, {Wu}, {Guillemot}, {Ackermann}, {Allen}, {de Angelis}, {Aulbert}, {Baldini}, {Ballet}, {Barbiellini}, {Bastieri}, {Bellazzini}, {Bissaldi}, {Bock}, {Bonino}, {Bottacini}, {Brandt}, {Bregeon}, {Bruel}, {Buson}, {Caliandro}, {Cameron}, {Caragiulo}, {Caraveo}, {Cecchi}, {Champion}, {Charles}, {Chekhtman}, {Chiang}, {Chiaro}, {Ciprini}, {Claus}, {Cohen-Tanugi}, {Cu{\'e}llar}, {Cutini}, {D'Ammando}, {Desiante}, {Drell}, {Eggenstein}, {Favuzzi}, {Fehrmann}, {Ferrara}, {Focke}, {Franckowiak}, {Fusco}, {Gargano}, {Gasparrini}, {Giglietto}, {Giordano}, {Glanzman}, {Godfrey}, {Grenier}, {Grove}, {Guiriec}, {Harding}, {Hays}, {Hewitt}, {Hill}, {Horan}, {Hou}, {Jogler}, {Johnson}, {J{\'o}hannesson}, {Kramer}, {Krauss}, {Kuss}, {Laffon}, {Larsson}, {Latronico}, {Li}, {Li}, {Longo}, {Loparco}, {Lovellette}, {Lubrano}, {Machenschalk}, {Manfreda}, {Marelli}, {Mayer}, {Mazziotta}, {Michelson}, {Mizuno}, {Monzani}, {Morselli}, {Moskalenko}, {Murgia}, {Nuss}, {Ohsugi},
  {Orienti}, {Orlando}, {de Palma}, {Paneque}, {Pesce-Rollins}, {Piron}, {Pivato}, {Rain{\`o}}, {Rando}, {Razzano}, {Reimer}, {Saz Parkinson}, {Schaal}, {Schulz}, {Sgr{\`o}}, {Siskind}, {Spada}, {Spandre}, {Spinelli}, {Suson}, {Takahashi}, {Thayer}, {Tibaldo}, {Torne}, {Torres}, {Tosti}, {Troja}, {Vianello}, {Wood}, {Wood}, \& {Yassine}}]{2015ApJ...809L...2C}
{Clark}, C.~J., {Pletsch}, H.~J., {Wu}, J., {et~al.} 2015, \apjl, 809, L2, \dodoi{10.1088/2041-8205/809/1/L2}

\bibitem[{Co {et~al.}(2022)Co, Garcia-Cely, Harigaya, Lu, \& Murayama}]{Co:2022bqq}
Co, R.~T., Garcia-Cely, C., Harigaya, K., Lu, P., \& Murayama, H. 2022, \jcap, 2022, 058, \dodoi{10.1088/1475-7516/2022/11/058}

\bibitem[{{Colpi} {et~al.}(2024){Colpi}, {Danzmann}, {Hewitson}, {Holley-Bockelmann}, {Jetzer}, {Nelemans}, {Petiteau}, {Shoemaker}, {Sopuerta}, {Stebbins}, {Tanvir}, {Ward}, {Weber}, {Thorpe}, {Daurskikh}, {Deep}, {Fern{\'a}ndez N{\'u}{\~n}ez}, {Garc{\'\i}a Marirrodriga}, {Gehler}, {Halain}, {Jennrich}, {Lammers}, {Larra{\~n}aga}, {Lieser}, {L{\"u}tzgendorf}, {Martens}, {Mondin}, {Piris Ni{\~n}o}, {Amaro-Seoane}, {Arca Sedda}, {Auclair}, {Babak}, {Baghi}, {Baibhav}, {Baker}, {Bayle}, {Berry}, {Berti}, {Boileau}, {Bonetti}, {Brito}, {Buscicchio}, {Calcagni}, {Capelo}, {Caprini}, {Caputo}, {Castelli}, {Chen}, {Chen}, {Chua}, {Davies}, {Derdzinski}, {Domcke}, {Doneva}, {Dvorkin}, {Mar{\'\i}a Ezquiaga}, {Gair}, {Haiman}, {Harry}, {Hartwig}, {Hees}, {Heffernan}, {Husa}, {Izquierdo-Villalba}, {Karnesis}, {Klein}, {Korol}, {Korsakova}, {Kupfer}, {Laghi}, {Lamberts}, {Larson}, {Le Jeune}, {Lewicki}, {Littenberg}, {Madge}, {Mangiagli}, {Marsat}, {Vilchez}, {Maselli}, {Mathews}, {van de Meent}, {Muratore}, {Nardini},
  {Pani}, {Peloso}, {Pieroni}, {Pound}, {Quelquejay-Leclere}, {Ricciardone}, {Rossi}, {Sartirana}, {Savalle}, {Sberna}, {Sesana}, {Shoemaker}, {Slutsky}, {Sotiriou}, {Speri}, {Staab}, {Steer}, {Tamanini}, {Tasinato}, {Torrado}, {Torres-Orjuela}, {Toubiana}, {Vallisneri}, {Vecchio}, {Volonteri}, {Yagi}, \& {Zwick}}]{LISARedBook:2024}
{Colpi}, M., {Danzmann}, K., {Hewitson}, M., {et~al.} 2024, arXiv e-prints, arXiv:2402.07571, \dodoi{10.48550/arXiv.2402.07571}

\bibitem[{{Cordes} \& {Jenet}(2012)}]{2012ApJ...752...54C}
{Cordes}, J.~M., \& {Jenet}, F.~A. 2012, \apj, 752, 54, \dodoi{10.1088/0004-637X/752/1/54}

\bibitem[{{Cordes} {et~al.}(2004){Cordes}, {Kramer}, {Lazio}, {Stappers}, {Backer}, \& {Johnston}}]{2004NewAR..48.1413C}
{Cordes}, J.~M., {Kramer}, M., {Lazio}, T.~J.~W., {et~al.} 2004, \nar, 48, 1413, \dodoi{10.1016/j.newar.2004.09.040}

\bibitem[{{Cordes} \& {Shannon}(2010)}]{2010arXiv1010.3785C}
{Cordes}, J.~M., \& {Shannon}, R.~M. 2010, arXiv e-prints, arXiv:1010.3785, \dodoi{10.48550/arXiv.1010.3785}

\bibitem[{{Counselman} \& {Rankin}(1972)}]{Counselman1972}
{Counselman}, III, C.~C., \& {Rankin}, J.~M. 1972, \apj, 175, 843, \dodoi{10.1086/151604}

\bibitem[{{Cury{\l}o} {et~al.}(2023){Cury{\l}o}, {Pennucci}, {Bailes}, {Bhat}, {Cameron}, {Dai}, {Hobbs}, {Kapur}, {Manchester}, {Mandow}, {Miles}, {Russell}, {Reardon}, {Shannon}, {Spiewak}, {van Straten}, {Zhu}, \& {Zic}}]{2023ApJ...944..128C}
{Cury{\l}o}, M., {Pennucci}, T.~T., {Bailes}, M., {et~al.} 2023, \apj, 944, 128, \dodoi{10.3847/1538-4357/aca535}

\bibitem[{{Dandapat} {et~al.}(2024){Dandapat}, {Susobhanan}, {Dey}, {Gopakumar}, {Baker}, \& {Jetzer}}]{SDN125hyp}
{Dandapat}, S., {Susobhanan}, A., {Dey}, L., {et~al.} 2024, \prd, 109, 103018, \dodoi{10.1103/PhysRevD.109.103018}

\bibitem[{{Dandapat} {et~al.}(2023){Dandapat}, {Ebersold}, {Susobhanan}, {Rana}, {Gopakumar}, {Tiwari}, {Haney}, {Lee}, \& {Kolhe}}]{SD23hyp}
{Dandapat}, S., {Ebersold}, M., {Susobhanan}, A., {et~al.} 2023, \prd, 108, 024013, \dodoi{10.1103/PhysRevD.108.024013}

\bibitem[{{Demorest}(2011)}]{2011MNRAS.416.2821D}
{Demorest}, P.~B. 2011, \mnras, 416, 2821, \dodoi{10.1111/j.1365-2966.2011.19230.x}

\bibitem[{{Demorest} {et~al.}(2013){Demorest}, {Ferdman}, {Gonzalez}, {Nice}, {Ransom}, {Stairs}, {Arzoumanian}, {Brazier}, {Burke-Spolaor}, {Chamberlin}, {Cordes}, {Ellis}, {Finn}, {Freire}, {Giampanis}, {Jenet}, {Kaspi}, {Lazio}, {Lommen}, {McLaughlin}, {Palliyaguru}, {Perrodin}, {Shannon}, {Siemens}, {Stinebring}, {Swiggum}, \& {Zhu}}]{2013ApJ...762...94D}
{Demorest}, P.~B., {Ferdman}, R.~D., {Gonzalez}, M.~E., {et~al.} 2013, \apj, 762, 94, \dodoi{10.1088/0004-637X/762/2/94}

\bibitem[{{Deneva} {et~al.}(2019){Deneva}, {Ray}, {Lommen}, {Ransom}, {Bogdanov}, {Kerr}, {Wood}, {Arzoumanian}, {Black}, {Doty}, {Gendreau}, {Guillot}, {Harding}, {Lewandowska}, {Malacaria}, {Markwardt}, {Price}, {Winternitz}, {Wolff}, {Guillemot}, {Cognard}, {Baker}, {Blumer}, {Brook}, {Cromartie}, {Demorest}, {DeCesar}, {Dolch}, {Ellis}, {Ferdman}, {Ferrara}, {Fonseca}, {Garver-Daniels}, {Gentile}, {Jones}, {Lam}, {Lorimer}, {Lynch}, {McLaughlin}, {Ng}, {Nice}, {Pennucci}, {Spiewak}, {Stairs}, {Stovall}, {Swiggum}, {Vigeland}, \& {Zhu}}]{2019ApJ...874..160D}
{Deneva}, J.~S., {Ray}, P.~S., {Lommen}, A., {et~al.} 2019, \apj, 874, 160, \dodoi{10.3847/1538-4357/ab0966}

\bibitem[{{Detweiler}(1979)}]{1979ApJ...234.1100D}
{Detweiler}, S. 1979, \apj, 234, 1100, \dodoi{10.1086/157593}

\bibitem[{{Dey} {et~al.}(2018){Dey}, {Valtonen}, {Gopakumar}, {Zola}, {Hudec}, {Pihajoki}, {Ciprini}, {Matsumoto}, {Sadakane}, {Kidger}, {Nilsson}, {Mikkola}, {Sillanp{\"a}{\"a}}, {Takalo}, {Lehto}, {Berdyugin}, {Piirola}, {Jermak}, {Baliyan}, {Pursimo}, {Caton}, {Alicavus}, {Baransky}, {Blay}, {Boumis}, {Boyd}, {Campas Torrent}, {Campos}, {Carrillo G{\'o}mez}, {Chandra}, {Chavushyan}, {Dalessio}, {Debski}, {Drozdz}, {Er}, {Erdem}, {Escartin P{\'e}rez}, {Fallah Ramazani}, {Filippenko}, {Gafton}, {Ganesh}, {Garcia}, {Gazeas}, {Godunova}, {G{\'o}mez Pinilla}, {Gopinathan}, {Haislip}, {Harmanen}, {Hurst}, {Jan{\'\i}k}, {Jelinek}, {Joshi}, {Kagitani}, {Karjalainen}, {Kaur}, {Keel}, {Kouprianov}, {Kundera}, {Kurowski}, {Kvammen}, {LaCluyze}, {Lee}, {Liakos}, {Lindfors}, {Lozano de Haro}, {Mugrauer}, {Naves Nogues}, {Neely}, {Nelson}, {Ogloza}, {Okano}, {Pajdosz-{\'S}mierciak}, {Pandey}, {Perri}, {Poyner}, {Provencal}, {Raj}, {Reichart}, {Reinthal}, {Reynolds}, {Saario}, {Sadegi}, {Sakanoi}, {Salto Gonz{\'a}lez},
  {Sameer}, {Schweyer}, {Simon}, {Siwak}, {Sold{\'a}n Alfaro}, {Sonbas}, {Steele}, {Stocke}, {Strobl}, {Tomov}, {Tremosa Espasa}, {Valdes}, {Valero P{\'e}rez}, {Verrecchia}, {Vasylenko}, {Webb}, {Yoneda}, {Zejmo}, {Zheng}, \& {Zielinski}}]{LD18}
{Dey}, L., {Valtonen}, M.~J., {Gopakumar}, A., {et~al.} 2018, \apj, 866, 11, \dodoi{10.3847/1538-4357/aadd95}

\bibitem[{Domènech(2021)}]{Domenech:2021ztg}
Domènech, G. 2021, Universe, 7, 398, \dodoi{10.3390/universe7110398}

\bibitem[{{Donner} {et~al.}(2019){Donner}, {Verbiest}, {Tiburzi}, {Os{\l}owski}, {Michilli}, {Serylak}, {Anderson}, {Horneffer}, {Kramer}, {Grie{\ss}meier}, {K{\"u}nsem{\"o}ller}, {Hessels}, {Hoeft}, \& {Miskolczi}}]{Donner:2019AandA624A22}
{Donner}, J.~Y., {Verbiest}, J.~P.~W., {Tiburzi}, C., {et~al.} 2019, \aap, 624, A22, \dodoi{10.1051/0004-6361/201834059}

\bibitem[{{Donner} {et~al.}(2020{\natexlab{a}}){Donner}, {Verbiest}, {Tiburzi}, {Os{\l}owski}, {K{\"u}nsem{\"o}ller}, {Bak Nielsen}, {Grie{\ss}meier}, {Serylak}, {Kramer}, {Anderson}, {Wucknitz}, {Keane}, {Kondratiev}, {Sobey}, {McKee}, {Bilous}, {Breton}, {Br{\"u}ggen}, {Ciardi}, {Hoeft}, {van Leeuwen}, \& {Vocks}}]{Donner:2020AandA644A153}
---. 2020{\natexlab{a}}, \aap, 644, A153, \dodoi{10.1051/0004-6361/202039517}

\bibitem[{{Donner} {et~al.}(2020{\natexlab{b}}){Donner}, {Verbiest}, {Tiburzi}, {Os{\l}owski}, {K{\"u}nsem{\"o}ller}, {Bak Nielsen}, {Grie{\ss}meier}, {Serylak}, {Kramer}, {Anderson}, {Wucknitz}, {Keane}, {Kondratiev}, {Sobey}, {McKee}, {Bilous}, {Breton}, {Br{\"u}ggen}, {Ciardi}, {Hoeft}, {van Leeuwen}, \& {Vocks}}]{2020A&A...644A.153D}
---. 2020{\natexlab{b}}, \aap, 644, A153, \dodoi{10.1051/0004-6361/202039517}

\bibitem[{Donner {et~al.}(2020)Donner, Verbiest, Tiburzi, Osłowski, Künsemöller, Bak~Nielsen, Grießmeier, Serylak, Kramer, Anderson, Wucknitz, Keane, Kondratiev, Sobey, McKee, Bilous, Breton, Brüggen, Ciardi, Hoeft, van Leeuwen, \& Vocks}]{Donner:2020AandA639A77}
Donner, J.~Y., Verbiest, J. P.~W., Tiburzi, C., {et~al.} 2020, \aap, 639, A77, \dodoi{10.1051/0004-6361/202039517}

\bibitem[{{Edwards} {et~al.}(2006){Edwards}, {Hobbs}, \& {Manchester}}]{2006MNRAS.372.1549E}
{Edwards}, R.~T., {Hobbs}, G.~B., \& {Manchester}, R.~N. 2006, \mnras, 372, 1549, \dodoi{10.1111/j.1365-2966.2006.10870.x}

\bibitem[{Ellis {et~al.}(2024)Ellis, Lewicki, Vaskonen, \& Veermäe}]{Ellis:2024AandA}
Ellis, J., Lewicki, M., Vaskonen, V., \& Veermäe, H. 2024, \aap, 685, A94, \dodoi{10.1051/0004-6361/202348224}

\bibitem[{{Ellis} {et~al.}(2020){Ellis}, {Vallisneri}, {Taylor}, \& {Baker}}]{2020zndo...4059815E}
{Ellis}, J.~A., {Vallisneri}, M., {Taylor}, S.~R., \& {Baker}, P.~T. 2020, {ENTERPRISE: Enhanced Numerical Toolbox Enabling a Robust PulsaR Inference SuitE}, v3.0.0,  Zenodo, \dodoi{10.5281/zenodo.4059815}

\bibitem[{{EPTA Collaboration} {et~al.}(2023{\natexlab{a}}){EPTA Collaboration}, {Antoniadis}, {Babak}, {Bak Nielsen}, {Bassa}, {Berthereau}, {Bonetti}, {Bortolas}, {Brook}, {Burgay}, {Caballero}, {Chalumeau}, {Champion}, {Chanlaridis}, {Chen}, {Cognard}, {Desvignes}, {Falxa}, {Ferdman}, {Franchini}, {Gair}, {Goncharov}, {Graikou}, {Grie{\ss}meier}, {Guillemot}, {Guo}, {Hu}, {Iraci}, {Izquierdo-Villalba}, {Jang}, {Jawor}, {Janssen}, {Jessner}, {Karuppusamy}, {Keane}, {Keith}, {Kramer}, {Krishnakumar}, {Lackeos}, {Lee}, {Liu}, {Liu}, {Lyne}, {McKee}, {Main}, {Mickaliger}, {Ni{\c{t}}u}, {Parthasarathy}, {Perera}, {Perrodin}, {Petiteau}, {Porayko}, {Possenti}, {Quelquejay Leclere}, {Samajdar}, {Sanidas}, {Sesana}, {Shaifullah}, {Speri}, {Spiewak}, {Stappers}, {Susarla}, {Theureau}, {Tiburzi}, {van der Wateren}, {Vecchio}, {Venkatraman Krishnan}, {Verbiest}, {Wang}, {Wang}, \& {Wu}}]{EPTA+2023a}
{EPTA Collaboration}, {Antoniadis}, J., {Babak}, S., {et~al.} 2023{\natexlab{a}}, \aap, 678, A48, \dodoi{10.1051/0004-6361/202346841}

\bibitem[{{EPTA Collaboration} {et~al.}(2023{\natexlab{b}}){EPTA Collaboration}, {InPTA Collaboration}, {Antoniadis}, {Arumugam}, {Arumugam}, {Babak}, {Bagchi}, {Nielsen}, {Bassa}, {Bathula}, {Berthereau}, {Bonetti}, {Bortolas}, {Brook}, {Burgay}, {Caballero}, {Chalumeau}, {Champion}, {Chanlaridis}, {Chen}, {Cognard}, {Dandapat}, {Deb}, {Desai}, {Desvignes}, {Dhanda-Batra}, {Dwivedi}, {Falxa}, {Ferdman}, {Franchini}, {Gair}, {Goncharov}, {Gopakumar}, {Graikou}, {Grie{\ss}meier}, {Guillemot}, {Guo}, {Gupta}, {Hisano}, {Hu}, {Iraci}, {Izquierdo-Villalba}, {Jang}, {Jawor}, {Janssen}, {Jessner}, {Joshi}, {Kareem}, {Karuppusamy}, {Keane}, {Keith}, {Kharbanda}, {Kikunaga}, {Kolhe}, {Kramer}, {Krishnakumar}, {Lackeos}, {Lee}, {Liu}, {Liu}, {Lyne}, {McKee}, {Maan}, {Main}, {Mickaliger}, {Ni{\c{t}}u}, {Nobleson}, {Paladi}, {Parthasarathy}, {Perera}, {Perrodin}, {Petiteau}, {Porayko}, {Possenti}, {Prabu}, {Leclere}, {Rana}, {Samajdar}, {Sanidas}, {Sesana}, {Shaifullah}, {Singha}, {Speri}, {Spiewak}, {Srivastava},
  {Stappers}, {Surnis}, {Susarla}, {Susobhanan}, {Takahashi}, {Tarafdar}, {Theureau}, {Tiburzi}, {van der Wateren}, {Vecchio}, {Krishnan}, {Verbiest}, {Wang}, {Wang}, \& {Wu}}]{2023A&A...678A..49E}
{EPTA Collaboration}, {InPTA Collaboration}, {Antoniadis}, J., {et~al.} 2023{\natexlab{b}}, \aap, 678, A49, \dodoi{10.1051/0004-6361/202346842}

\bibitem[{{EPTA Collaboration} {et~al.}(2024{\natexlab{a}}){EPTA Collaboration}, {InPTA Collaboration}, {Antoniadis}, {Arumugam}, {Arumugam}, {Babak}, {Bagchi}, {Bak Nielsen}, {Bassa}, {Bathula}, {Berthereau}, {Bonetti}, {Bortolas}, {Brook}, {Burgay}, {Caballero}, {Chalumeau}, {Champion}, {Chanlaridis}, {Chen}, {Cognard}, {Dandapat}, {Deb}, {Desai}, {Desvignes}, {Dhanda-Batra}, {Dwivedi}, {Falxa}, {Ferranti}, {Ferdman}, {Franchini}, {Gair}, {Goncharov}, {Gopakumar}, {Graikou}, {Grie{\ss}meier}, {Guillemot}, {Guo}, {Gupta}, {Hisano}, {Hu}, {Iraci}, {Izquierdo-Villalba}, {Jang}, {Jawor}, {Janssen}, {Jessner}, {Joshi}, {Kareem}, {Karuppusamy}, {Keane}, {Keith}, {Kharbanda}, {Kikunaga}, {Kolhe}, {Kramer}, {Krishnakumar}, {Lackeos}, {Lee}, {Liu}, {Liu}, {Lyne}, {McKee}, {Maan}, {Main}, {Manzini}, {Mickaliger}, {Ni{\c{t}}u}, {Nobleson}, {Paladi}, {Parthasarathy}, {Perera}, {Perrodin}, {Petiteau}, {Porayko}, {Possenti}, {Prabu}, {Quelquejay Leclere}, {Rana}, {Samajdar}, {Sanidas}, {Sesana}, {Shaifullah}, {Singha},
  {Speri}, {Spiewak}, {Srivastava}, {Stappers}, {Surnis}, {Susarla}, {Susobhanan}, {Takahashi}, {Tarafdar}, {Theureau}, {Tiburzi}, {van der Wateren}, {Vecchio}, {Venkatraman Krishnan}, {Verbiest}, {Wang}, {Wang}, \& {Wu}}]{2024A&A...690A.118E}
---. 2024{\natexlab{a}}, \aap, 690, A118, \dodoi{10.1051/0004-6361/202348568}

\bibitem[{{EPTA Collaboration} {et~al.}(2024{\natexlab{b}}){EPTA Collaboration}, {InPTA Collaboration}, Antoniadis, Arumugam, Arumugam, Babak, Bagchi, Bak~Nielsen, Bassa, Bathula, Berthereau, Bonetti, Bortolas, Brook, Burgay, Caballero, Chalumeau, Champion, Chanlaridis, Chen, Cognard, Dandapat, Deb, Desai, Desvignes, Dhanda-Batra, Dwivedi, Falxa, Ferdman, Franchini, Gair, Goncharov, Gopakumar, Graikou, Grie{\ss}meier, Gualandris, Guillemot, Guo, Gupta, Hisano, Hu, Iraci, Izquierdo-Villalba, Jang, Jawor, Janssen, Jessner, Joshi, Kareem, Karuppusamy, Keane, Keith, Kharbanda, Kikunaga, Kolhe, Kramer, Krishnakumar, Lackeos, Lee, Liu, Liu, Lyne, McKee, Maan, Main, Mickaliger, Ni{\c{t}}u, Nobleson, Paladi, Parthasarathy, Perera, Perrodin, Petiteau, Porayko, Possenti, Prabu, Quelquejay~Leclere, Rana, Samajdar, Sanidas, Sesana, Shaifullah, Singha, Speri, Spiewak, Srivastava, Stappers, Surnis, Susarla, Susobhanan, Takahashi, Tarafdar, Theureau, Tiburzi, van~der Wateren, Vecchio, Venkatraman~Krishnan, Verbiest, Wang,
  Wang, Wu, Auclair, Barausse, Caprini, Crisostomi, Fastidio, Khizriev, Middleton, Neronov, Postnov, Roper~Pol, Semikoz, Smarra, Steer, Truant, \& Valtolina}]{EPTA:2024DR2-IV}
{EPTA Collaboration}, {InPTA Collaboration}, Antoniadis, J., {et~al.} 2024{\natexlab{b}}, \aap, 685, A94, \dodoi{10.1051/0004-6361/202347433}

\bibitem[{{EPTA+InPTA Collaboration} {et~al.}(2023){EPTA+InPTA Collaboration}, {InPTA Collaboration}, {Antoniadis}, {Arumugam}, {Arumugam}, {Babak}, {Bagchi}, {Bak Nielsen}, {Bassa}, {Bathula}, {Berthereau}, {Bonetti}, {Bortolas}, {Brook}, {Burgay}, {Caballero}, {Chalumeau}, {Champion}, {Chanlaridis}, {Chen}, {Cognard}, {Dandapat}, {Deb}, {Desai}, {Desvignes}, {Dhanda-Batra}, {Dwivedi}, {Falxa}, {Ferdman}, {Franchini}, {Gair}, {Goncharov}, {Gopakumar}, {Graikou}, {Grie{\ss}meier}, {Guillemot}, {Guo}, {Gupta}, {Hisano}, {Hu}, {Iraci}, {Izquierdo-Villalba}, {Jang}, {Jawor}, {Janssen}, {Jessner}, {Joshi}, {Kareem}, {Karuppusamy}, {Keane}, {Keith}, {Kharbanda}, {Kikunaga}, {Kolhe}, {Kramer}, {Krishnakumar}, {Lackeos}, {Lee}, {Liu}, {Liu}, {Lyne}, {McKee}, {Maan}, {Main}, {Mickaliger}, {Ni{\c{t}}u}, {Nobleson}, {Paladi}, {Parthasarathy}, {Perera}, {Perrodin}, {Petiteau}, {Porayko}, {Possenti}, {Prabu}, {Quelquejay Leclere}, {Rana}, {Samajdar}, {Sanidas}, {Sesana}, {Shaifullah}, {Singha}, {Speri}, {Spiewak},
  {Srivastava}, {Stappers}, {Surnis}, {Susarla}, {Susobhanan}, {Takahashi}, {Tarafdar}, {Theureau}, {Tiburzi}, {van der Wateren}, {Vecchio}, {Venkatraman Krishnan}, {Verbiest}, {Wang}, {Wang}, \& {Wu}}]{2023A&A...678A..50E}
{EPTA+InPTA Collaboration}, {InPTA Collaboration}, {Antoniadis}, J., {et~al.} 2023, \aap, 678, A50, \dodoi{10.1051/0004-6361/202346844}

\bibitem[{Espinosa {et~al.}(2018)Espinosa, Racco, \& Riotto}]{Espinosa:2018eve}
Espinosa, J.~R., Racco, D., \& Riotto, A. 2018, JCAP, 09, 012, \dodoi{10.1088/1475-7516/2018/09/012}

\bibitem[{{Evans} {et~al.}(2021){Evans}, {Adhikari}, {Afle}, {Ballmer}, {Biscoveanu}, {Borhanian}, {Brown}, {Chen}, {Eisenstein}, {Gruson}, {Gupta}, {Hall}, {Huxford}, {Kamai}, {Kashyap}, {Kissel}, {Kuns}, {Landry}, {Lenon}, {Lovelace}, {McCuller}, {Ng}, {Nitz}, {Read}, {Sathyaprakash}, {Shoemaker}, {Slagmolen}, {Smith}, {Srivastava}, {Sun}, {Vitale}, \& {Weiss}}]{2021arXiv210909882E}
{Evans}, M., {Adhikari}, R.~X., {Afle}, C., {et~al.} 2021, arXiv e-prints, arXiv:2109.09882, \dodoi{10.48550/arXiv.2109.09882}

\bibitem[{{Favata}(2010)}]{2010CQGra..27h4036F}
{Favata}, M. 2010, Classical and Quantum Gravity, 27, 084036, \dodoi{10.1088/0264-9381/27/8/084036}

\bibitem[{{FERMI-LAT Collaboration} {et~al.}(2022){FERMI-LAT Collaboration}, {Ajello}, {Atwood}, {Baldini}, {Ballet}, {Barbiellini}, {Bastieri}, {Bellazzini}, {Berretta}, {Bhattacharyya}, {Bissaldi}, {Blandford}, {Bloom}, {Bonino}, {Bruel}, {Buehler}, {Burns}, {Buson}, {Cameron}, {Caraveo}, {Cavazzuti}, {Cibrario}, {Ciprini}, {Clark}, {Cognard}, {Coronado-Bl{\'a}zquez}, {Crnogorcevic}, {Cromartie}, {Crowter}, {Cutini}, {D'Ammando}, {De Gaetano}, {de Palma}, {Digel}, {Di Lalla}, {Fana Dirirsa}, {Di Venere}, {Dom{\'\i}nguez}, {Ferrara}, {Fiori}, {Franckowiak}, {Fukazawa}, {Funk}, {Fusco}, {Gammaldi}, {Gargano}, {Gasparrini}, {Giglietto}, {Giordano}, {Giroletti}, {Green}, {Grenier}, {Guillemot}, {Guiriec}, {Gustafsson}, {Harding}, {Hays}, {Hewitt}, {Horan}, {Hou}, {J{\'o}hannesson}, {Keith}, {Kerr}, {Kramer}, {Kuss}, {Larsson}, {Latronico}, {Li}, {Longo}, {Loparco}, {Lovellette}, {Lubrano}, {Maldera}, {Manfreda}, {Mart{\'\i}-Devesa}, {Mazziotta}, {Mereu}, {Michelson}, {Mirabal}, {Mitthumsiri}, {Mizuno},
  {Monzani}, {Morselli}, {Negro}, {Nieder}, {Ojha}, {Omodei}, {Orienti}, {Orlando}, {Ormes}, {Paneque}, {Parthasarathy}, {Pei}, {Persic}, {Pesce-Rollins}, {Pillera}, {Poon}, {Porter}, {Principe}, {Racusin}, {Rain{\`o}}, {Rando}, {Rani}, {Ransom}, {Ray}, {Razzano}, {Razzaque}, {Reimer}, {Reimer}, {Roy}, {S{\'a}nchez-Conde}, {Saz Parkinson}, {Scargle}, {Scotton}, {Serini}, {Sgr{\`o}}, {Siskind}, {Smith}, {Spandre}, {Spiewak}, {Spinelli}, {Stairs}, {Suson}, {Swihart}, {Tabassum}, {Thayer}, {Theureau}, {Torres}, {Troja}, {Valverde}, {Wadiasingh}, {Wood}, \& {Zaharijas}}]{2022Sci...376..521F}
{FERMI-LAT Collaboration}, {Ajello}, M., {Atwood}, W.~B., {et~al.} 2022, Science, 376, 521, \dodoi{10.1126/science.abm3231}

\bibitem[{{Ferranti} {et~al.}(2025){Ferranti}, {Falxa}, {Sesana}, {Chalumeau}, {Porayko}, {Shaifullah}, {Cognard}, {Guillemot}, {Kramer}, {Liu}, \& {Theureau}}]{Ferranti:2025AandA}
{Ferranti}, I., {Falxa}, M., {Sesana}, A., {et~al.} 2025, \aap, 694, A38, \dodoi{10.1051/0004-6361/202452805}

\bibitem[{Ferreira {et~al.}(2023)Ferreira, Notari, Pujolas, \& Rompineve}]{Ferreira:2022zzo}
Ferreira, R.~Z., Notari, A., Pujolas, O., \& Rompineve, F. 2023, JCAP, 02, 001, \dodoi{10.1088/1475-7516/2023/02/001}

\bibitem[{Ferreira {et~al.}(2024)Ferreira, Notari, \& Sloth}]{Ferreira:2024prl}
Ferreira, R.~Z., Notari, A., \& Sloth, M.~S. 2024, \prl, 132, 171002, \dodoi{10.1103/PhysRevLett.132.171002}

\bibitem[{{Foster} \& {Backer}(1990)}]{1990ApJ...361..300F}
{Foster}, R.~S., \& {Backer}, D.~C. 1990, \apj, 361, 300, \dodoi{10.1086/169195}

\bibitem[{{Gersbach} {et~al.}(2025{\natexlab{a}}){Gersbach}, {Taylor}, {B{\'e}csy}, {Lemke}, {Mitridate}, \& {Pol}}]{gersbach2025fopts}
{Gersbach}, K.~A., {Taylor}, S.~R., {B{\'e}csy}, B., {et~al.} 2025{\natexlab{a}}, arXiv e-prints, arXiv:2509.07090, \dodoi{10.48550/arXiv.2509.07090}

\bibitem[{{Gersbach} {et~al.}(2025{\natexlab{b}}){Gersbach}, {Taylor}, {Meyers}, \& {Romano}}]{Gersbach2025fopt}
{Gersbach}, K.~A., {Taylor}, S.~R., {Meyers}, P.~M., \& {Romano}, J.~D. 2025{\natexlab{b}}, \prd, 111, 023027, \dodoi{10.1103/PhysRevD.111.023027}

\bibitem[{{Gitika} {et~al.}(2025){Gitika}, {Shannon}, {Bailes}, {Reardon}, {Miles}, {Champion}, \& {Grunthal}}]{2025arXiv251003139G}
{Gitika}, P., {Shannon}, R.~M., {Bailes}, M., {et~al.} 2025, \pasa, 42, e146, \dodoi{10.1017/pasa.2025.10107}

\bibitem[{{Gitika} {et~al.}(2023){Gitika}, {Bailes}, {Shannon}, {Reardon}, {Cameron}, {Shamohammadi}, {Miles}, {Flynn}, {Corongiu}, \& {Kramer}}]{2023MNRAS.526.3370G}
{Gitika}, P., {Bailes}, M., {Shannon}, R.~M., {et~al.} 2023, \mnras, 526, 3370, \dodoi{10.1093/mnras/stad2841}

\bibitem[{{Goncharov} {et~al.}(2021{\natexlab{a}}){Goncharov}, {Shannon}, {Reardon}, {Hobbs}, {Zic}, {Bailes}, {Cury{\l}o}, {Dai}, {Kerr}, {Lower}, {Manchester}, {Mandow}, {Middleton}, {Miles}, {Parthasarathy}, {Thrane}, {Thyagarajan}, {Xue}, {Zhu}, {Cameron}, {Feng}, {Luo}, {Russell}, {Sarkissian}, {Spiewak}, {Wang}, {Wang}, {Zhang}, \& {Zhang}}]{2021ApJ...917L..19G}
{Goncharov}, B., {Shannon}, R.~M., {Reardon}, D.~J., {et~al.} 2021{\natexlab{a}}, \apjl, 917, L19, \dodoi{10.3847/2041-8213/ac17f4}

\bibitem[{{Goncharov} {et~al.}(2021{\natexlab{b}}){Goncharov}, {Reardon}, {Shannon}, {Zhu}, {Thrane}, {Bailes}, {Bhat}, {Dai}, {Hobbs}, {Kerr}, {Manchester}, {Os{\l}owski}, {Parthasarathy}, {Russell}, {Spiewak}, {Thyagarajan}, \& {Wang}}]{2021MNRAS.502..478G}
{Goncharov}, B., {Reardon}, D.~J., {Shannon}, R.~M., {et~al.} 2021{\natexlab{b}}, \mnras, 502, 478, \dodoi{10.1093/mnras/staa3411}

\bibitem[{{Goncharov} {et~al.}(2022){Goncharov}, {Thrane}, {Shannon}, {Harms}, {Bhat}, {Hobbs}, {Kerr}, {Manchester}, {Reardon}, {Russell}, {Zhu}, \& {Zic}}]{2022ApJ...932L..22G}
{Goncharov}, B., {Thrane}, E., {Shannon}, R.~M., {et~al.} 2022, \apjl, 932, L22, \dodoi{10.3847/2041-8213/ac76bb}

\bibitem[{Gouttenoire {et~al.}(2020)Gouttenoire, Servant, \& Simakachorn}]{Gouttenoire:2019kij}
Gouttenoire, Y., Servant, G., \& Simakachorn, P. 2020, JCAP, 07, 032, \dodoi{10.1088/1475-7516/2020/07/032}

\bibitem[{{Grishchuk}(1974)}]{1974ZhETF..67..825G}
{Grishchuk}, L.~P. 1974, Zhurnal Eksperimentalnoi i Teoreticheskoi Fiziki, 67, 825

\bibitem[{{Grishchuk}(1977)}]{Grishchuk:1977ky}
---. 1977, Soviet Physics Uspekhi, 20, 319, \dodoi{10.1070/PU1977v020n04ABEH005327}

\bibitem[{{Groth}(1975)}]{1975ApJS...29..453G}
{Groth}, E.~J. 1975, \apjs, 29, 453, \dodoi{10.1086/190354}

\bibitem[{{Grunthal} {et~al.}(in prep.){Grunthal}, {Champion}, {Thrane}, {Nathan}, {Kramer}, \& {Miles}}]{Grunthal_inprep}
{Grunthal}, K., {Champion}, D.~J., {Thrane}, E., {et~al.} in prep.

\bibitem[{{Grunthal} {et~al.}(2025){Grunthal}, {Nathan}, {Thrane}, {Champion}, {Miles}, {Shannon}, {Kulkarni}, {Abbate}, {Buchner}, {Cameron}, {Geyer}, {Gitika}, {Keith}, {Kramer}, {Lasky}, {Parthasarathy}, {Reardon}, {Singha}, \& {Venkatraman Krishnan}}]{2025MNRAS.536.1501G}
{Grunthal}, K., {Nathan}, R.~S., {Thrane}, E., {et~al.} 2025, \mnras, 536, 1501, \dodoi{10.1093/mnras/stae2573}

\bibitem[{Guo {et~al.}(2021)Guo, Sinha, Vagie, \& White}]{Guo:2020grp}
Guo, H.-K., Sinha, K., Vagie, D., \& White, G. 2021, JCAP, 01, 001, \dodoi{10.1088/1475-7516/2021/01/001}

\bibitem[{{Guo} {et~al.}(2024){Guo}, {Caballero}, {Champion}, \& {Lee}}]{Guo2024}
{Guo}, Y.~J., {Caballero}, R.~N., {Champion}, D.~J., \& {Lee}, K.~J. 2024, \mnras, 532, 2943, \dodoi{10.1093/mnras/stae1660}

\bibitem[{{Guo} {et~al.}(2018){Guo}, {Lee}, \& {Caballero}}]{Guo2018}
{Guo}, Y.~J., {Lee}, K.~J., \& {Caballero}, R.~N. 2018, \mnras, 475, 3644, \dodoi{10.1093/mnras/stx3326}

\bibitem[{{Guo} {et~al.}(2019){Guo}, {Li}, {Lee}, \& {Caballero}}]{Guo2019}
{Guo}, Y.~J., {Li}, G.~Y., {Lee}, K.~J., \& {Caballero}, R.~N. 2019, \mnras, 489, 5573, \dodoi{10.1093/mnras/stz2515}

\bibitem[{{Guth}(1981)}]{1981PhRvD..23..347G}
{Guth}, A.~H. 1981, \prd, 23, 347, \dodoi{10.1103/PhysRevD.23.347}

\bibitem[{{Hallinan} {et~al.}(2019){Hallinan}, {Ravi}, {Weinreb}, {Kocz}, {Huang}, {Woody}, {Lamb}, {D'Addario}, {Catha}, {Law}, {Kulkarni}, {Phinney}, {Eastwood}, {Bouman}, {McLaughlin}, {Ransom}, {Siemens}, {Cordes}, {Lynch}, {Kaplan}, {Brazier}, {Bhatnagar}, {Myers}, {Walter}, \& {Gaensler}}]{2019BAAS...51g.255H}
{Hallinan}, G., {Ravi}, V., {Weinreb}, S., {et~al.} 2019, in Bulletin of the American Astronomical Society, Vol.~51, 255, \dodoi{10.48550/arXiv.1907.07648}

\bibitem[{{Harding} {et~al.}(2005){Harding}, {Usov}, \& {Muslimov}}]{2005ApJ...622..531H}
{Harding}, A.~K., {Usov}, V.~V., \& {Muslimov}, A.~G. 2005, \apj, 622, 531, \dodoi{10.1086/427840}

\bibitem[{{Hazboun} {et~al.}(2019){Hazboun}, {Romano}, \& {Smith}}]{2019PhRvD.100j4028H}
{Hazboun}, J.~S., {Romano}, J.~D., \& {Smith}, T.~L. 2019, \prd, 100, 104028, \dodoi{10.1103/PhysRevD.100.104028}

\bibitem[{{Hazboun} {et~al.}(2022{\natexlab{a}}){Hazboun}, {Simon}, {Madison}, {Arzoumanian}, {Cromartie}, {Crowter}, {Decesar}, {Demorest}, {Dolch}, {Ellis}, {Ferdman}, {Ferrara}, {Fonseca}, {Gentile}, {Jones}, {Jones}, {Lam}, {Levin}, {Lorimer}, {Lynch}, {McLaughlin}, {Ng}, {Nice}, {Pennucci}, {Ransom}, {Ray}, {Spiewak}, {Stairs}, {Stovall}, {Swiggum}, {Zhu}, \& {The Nanograv Collaboration}}]{2022ApJ...929...39H}
{Hazboun}, J.~S., {Simon}, J., {Madison}, D.~R., {et~al.} 2022{\natexlab{a}}, \apj, 929, 39, \dodoi{10.3847/1538-4357/ac5829}

\bibitem[{{Hazboun} {et~al.}(2022{\natexlab{b}}){Hazboun}, {Crump}, {Lommen}, {Montano}, {Berry}, {Zeldes}, {Teng}, {Ray}, {Kerr}, {Arzoumanian}, {Bogdanov}, {Deneva}, {Lewandowska}, {Markwardt}, {Ransom}, {Enoto}, {Wood}, {Gendreau}, {Howe}, \& {Parthasarathy}}]{2022ApJ...928...67H}
{Hazboun}, J.~S., {Crump}, J., {Lommen}, A.~N., {et~al.} 2022{\natexlab{b}}, \apj, 928, 67, \dodoi{10.3847/1538-4357/ac54ae}

\bibitem[{{Hellings} \& {Downs}(1983)}]{1983ApJ...265L..39H}
{Hellings}, R.~W., \& {Downs}, G.~S. 1983, \apjl, 265, L39, \dodoi{10.1086/183954}

\bibitem[{Hindmarsh {et~al.}(2014)Hindmarsh, Huber, Rummukainen, \& Weir}]{Hindmarsh:2013xza}
Hindmarsh, M., Huber, S.~J., Rummukainen, K., \& Weir, D.~J. 2014, Phys. Rev. Lett., 112, 041301, \dodoi{10.1103/PhysRevLett.112.041301}

\bibitem[{Hindmarsh {et~al.}(2021)Hindmarsh, L{\"u}ben, Lumma, \& Pauly}]{Hindmarsh_2021}
Hindmarsh, M., L{\"u}ben, M., Lumma, J., \& Pauly, M. 2021, SciPost Physics Lecture Notes, 024, \dodoi{10.48550/arXiv.2008.09136}

\bibitem[{Hiramatsu {et~al.}(2014)Hiramatsu, Kawasaki, \& Saikawa}]{Hiramatsu:2013qaa}
Hiramatsu, T., Kawasaki, M., \& Saikawa, K. 2014, JCAP, 02, 031, \dodoi{10.1088/1475-7516/2014/02/031}

\bibitem[{{Hirata} {et~al.}(1987){Hirata}, {Kajita}, {Koshiba}, {Nakahata}, {Oyama}, {Sato}, {Suzuki}, {Takita}, {Totsuka}, {Kifune}, {Suda}, {Takahashi}, {Tanimori}, {Miyano}, {Yamada}, {Beier}, {Feldscher}, {Kim}, {Mann}, {Newcomer}, {van}, {Zhang}, \& {Cortez}}]{SN1987ASNnu}
{Hirata}, K., {Kajita}, T., {Koshiba}, M., {et~al.} 1987, \prl, 58, 1490, \dodoi{10.1103/PhysRevLett.58.1490}

\bibitem[{{Hobbs} {et~al.}(2010){Hobbs}, {Archibald}, {Arzoumanian}, {Backer}, {Bailes}, {Bhat}, {Burgay}, {Burke-Spolaor}, {Champion}, {Cognard}, {Coles}, {Cordes}, {Demorest}, {Desvignes}, {Ferdman}, {Finn}, {Freire}, {Gonzalez}, {Hessels}, {Hotan}, {Janssen}, {Jenet}, {Jessner}, {Jordan}, {Kaspi}, {Kramer}, {Kondratiev}, {Lazio}, {Lazaridis}, {Lee}, {Levin}, {Lommen}, {Lorimer}, {Lynch}, {Lyne}, {Manchester}, {McLaughlin}, {Nice}, {Oslowski}, {Pilia}, {Possenti}, {Purver}, {Ransom}, {Reynolds}, {Sanidas}, {Sarkissian}, {Sesana}, {Shannon}, {Siemens}, {Stairs}, {Stappers}, {Stinebring}, {Theureau}, {van Haasteren}, {van Straten}, {Verbiest}, {Yardley}, \& {You}}]{2010CQGra..27h4013H}
{Hobbs}, G., {Archibald}, A., {Arzoumanian}, Z., {et~al.} 2010, Classical and Quantum Gravity, 27, 084013, \dodoi{10.1088/0264-9381/27/8/084013}

\bibitem[{{Hobbs} {et~al.}(2012){Hobbs}, {Coles}, {Manchester}, {Keith}, {Shannon}, {Chen}, {Bailes}, {Bhat}, {Burke-Spolaor}, {Champion}, {Chaudhary}, {Hotan}, {Khoo}, {Kocz}, {Levin}, {Oslowski}, {Preisig}, {Ravi}, {Reynolds}, {Sarkissian}, {van Straten}, {Verbiest}, {Yardley}, \& {You}}]{2012MNRAS.427.2780H}
{Hobbs}, G., {Coles}, W., {Manchester}, R.~N., {et~al.} 2012, \mnras, 427, 2780, \dodoi{10.1111/j.1365-2966.2012.21946.x}

\bibitem[{{Hobbs} {et~al.}(2020){Hobbs}, {Guo}, {Caballero}, {Coles}, {Lee}, {Manchester}, {Reardon}, {Matsakis}, {Tong}, {Arzoumanian}, {Bailes}, {Bassa}, {Bhat}, {Brazier}, {Burke-Spolaor}, {Champion}, {Chatterjee}, {Cognard}, {Dai}, {Desvignes}, {Dolch}, {Ferdman}, {Graikou}, {Guillemot}, {Janssen}, {Keith}, {Kerr}, {Kramer}, {Lam}, {Liu}, {Lyne}, {Lazio}, {Lynch}, {McKee}, {McLaughlin}, {Mingarelli}, {Nice}, {Os{\l}owski}, {Pennucci}, {Perera}, {Perrodin}, {Possenti}, {Russell}, {Sanidas}, {Sesana}, {Shaifullah}, {Shannon}, {Simon}, {Spiewak}, {Stairs}, {Stappers}, {Swiggum}, {Taylor}, {Theureau}, {Toomey}, {van Haasteren}, {Wang}, {Wang}, \& {Zhu}}]{Hobbs2020}
{Hobbs}, G., {Guo}, L., {Caballero}, R.~N., {et~al.} 2020, \mnras, 491, 5951, \dodoi{10.1093/mnras/stz3071}

\bibitem[{{Hobbs} {et~al.}(2006){Hobbs}, {Edwards}, \& {Manchester}}]{2006MNRAS.369..655H}
{Hobbs}, G.~B., {Edwards}, R.~T., \& {Manchester}, R.~N. 2006, \mnras, 369, 655, \dodoi{10.1111/j.1365-2966.2006.10302.x}

\bibitem[{{Hogan}(1986)}]{Hogan:1986qda}
{Hogan}, C.~J. 1986, \mnras, 218, 629, \dodoi{10.1093/mnras/218.4.629}

\bibitem[{{Hulse} \& {Taylor}(1975)}]{1975ApJ...195L..51H}
{Hulse}, R.~A., \& {Taylor}, J.~H. 1975, \apjl, 195, L51, \dodoi{10.1086/181708}

\bibitem[{{IceCube Collaboration} {et~al.}(2018){IceCube Collaboration}, {Aartsen}, {Ackermann}, {Adams}, {Aguilar}, {Ahlers}, {Ahrens}, {Samarai}, {Altmann}, {Andeen}, {Anderson}, {Ansseau}, {Anton}, {Arg{\"u}elles}, {Arsioli}, {Auffenberg}, {Axani}, {Bagherpour}, {Bai}, {Barron}, {Barwick}, {Baum}, {Bay}, {Beatty}, {Becker Tjus}, {Becker}, {BenZvi}, {Berley}, {Bernardini}, {Besson}, {Binder}, {Bindig}, {Blaufuss}, {Blot}, {Bohm}, {B{\"o}rner}, {Bos}, {B{\"o}ser}, {Botner}, {Bourbeau}, {Bourbeau}, {Bradascio}, {Braun}, {Brenzke}, {Bretz}, {Bron}, {Brostean-Kaiser}, {Burgman}, {Busse}, {Carver}, {Cheung}, {Chirkin}, {Christov}, {Clark}, {Classen}, {Coenders}, {Collin}, {Conrad}, {Coppin}, {Correa}, {Cowen}, {Cross}, {Dave}, {Day}, {de Andr{\'e}}, {De Clercq}, {DeLaunay}, {Dembinski}, {DeRidder}, {Desiati}, {de Vries}, {de Wasseige}, {de With}, {DeYoung}, {D{\'\i}az-V{\'e}lez}, {di Lorenzo}, {Dujmovic}, {Dumm}, {Dunkman}, {Dvorak}, {Eberhardt}, {Ehrhardt}, {Eichmann}, {Eller}, {Evenson}, {Fahey}, {Fazely},
  {Felde}, {Filimonov}, {Finley}, {Flis}, {Franckowiak}, {Friedman}, {Fritz}, {Gaisser}, {Gallagher}, {Gerhardt}, {Ghorbani}, {Giommi}, {Glauch}, {Gl{\"u}senkamp}, {Goldschmidt}, {Gonzalez}, {Grant}, {Griffith}, {Haack}, {Hallgren}, {Halzen}, {Hanson}, {Hebecker}, {Heereman}, {Helbing}, {Hellauer}, {Hickford}, {Hignight}, {Hill}, {Hoffman}, {Hoffmann}, {Hoinka}, {Hokanson-Fasig}, {Hoshina}, {Huang}, {Huber}, {Hultqvist}, {H{\"u}nnefeld}, {Hussain}, {In}, {Iovine}, {Ishihara}, {Jacobi}, {Japaridze}, {Jeong}, {Jero}, {Jones}, {Kalaczynski}, {Kang}, {Kappes}, {Kappesser}, {Karg}, {Karle}, {Katz}, {Kauer}, {Keivani}, {Kelley}, {Kheirandish}, {Kim}, {Kim}, {Kintscher}, {Kiryluk}, {Kittler}, {Klein}, {Koirala}, {Kolanoski}, {K{\"o}pke}, {Kopper}, {Kopper}, {Koschinsky}, {Koskinen}, {Kowalski}, {Krammer}, {Krings}, {Kroll}, {Kr{\"u}ckl}, {Kunwar}, {Kurahashi}, {Kuwabara}, {Kyriacou}, {Labare}, {Lanfranchi}, {Larson}, {Lauber}, {Leonard}, {Lesiak-Bzdak}, {Leuermann}, {Liu}, {Lozano Mariscal}, {Lu}, {L{\"u}nemann},
  {Luszczak}, {Madsen}, {Maggi}, {Mahn}, {Mancina}, {Maruyama}, {Mase}, {Maunu}, {Meagher}, {Medici}, {Meier}, {Menne}, {Merino}, {Meures}, {Miarecki}, {Micallef}, {Moment{\'e}}, {Montaruli}, {Moore}, {Morse}, {Moulai}, \& {Nahnhauer}}]{TXS18}
{IceCube Collaboration}, {Aartsen}, M.~G., {Ackermann}, M., {et~al.} 2018, Science, 361, 147, \dodoi{10.1126/science.aat2890}

\bibitem[{Inomata {et~al.}(2021)Inomata, Kohri, \& Terada}]{Inomata:2021zel}
Inomata, K., Kohri, K., \& Terada, T. 2021, \prd, 104, 123525, \dodoi{10.1103/PhysRevD.104.123525}

\bibitem[{{Iraci} {et~al.}(2024){Iraci}, {Chalumeau}, {Tiburzi}, {Verbiest}, {Possenti}, {Shaifullah}, {Susarla}, {Krishnakumar}, {Lam}, {Cromartie}, {Kerr}, \& {Grie{\ss}meier}}]{Iraci:2024AandA}
{Iraci}, F., {Chalumeau}, A., {Tiburzi}, C., {et~al.} 2024, \aap, 692, A170, \dodoi{10.1051/0004-6361/202450740}

\bibitem[{{Iraci} {et~al.}(2025){Iraci}, {Chalumeau}, {Tiburzi}, {Verbiest}, {Possenti}, {Susarla}, {Krishnakumar}, {Shaifullah}, {Antoniadis}, {Bagchi}, {Bassa}, {Caballero}, {Cecconi}, {Chen}, {Chowdhury}, {Ciardi}, {Cognard}, {Corbel}, {Desai}, {Deb}, {Girard}, {Golden}, {Grie{\ss}meier}, {Guillemot}, {Hoeft}, {Hu}, {Jankowski}, {Janssen}, {Joshi}, {Kala}, {Keane}, {Nobelson}, {Konovalenko}, {Kravtsov}, {Kramer}, {Liu}, {Parthasarathy}, {Rana}, {Schwarz}, {Singha}, {Srivastava}, {Takahashi}, {Tarafdar}, {Theureau}, {Ulyanov}, {Vocks}, {Wang}, {Zakharenko}, \& {Zarka}}]{Iraci2025}
---. 2025, arXiv e-prints, arXiv:2510.04639, \dodoi{10.48550/arXiv.2510.04639}

\bibitem[{{Izquierdo-Villalba} {et~al.}(2022){Izquierdo-Villalba}, {Sesana}, {Bonoli}, \& {Colpi}}]{2022MNRAS.509.3488I}
{Izquierdo-Villalba}, D., {Sesana}, A., {Bonoli}, S., \& {Colpi}, M. 2022, \mnras, 509, 3488, \dodoi{10.1093/mnras/stab3239}

\bibitem[{{Jaffe} \& {Backer}(2003)}]{2003ApJ...583..616J}
{Jaffe}, A.~H., \& {Backer}, D.~C. 2003, \apj, 583, 616, \dodoi{10.1086/345443}

\bibitem[{{Janssen} {et~al.}(2015){Janssen}, {Hobbs}, {McLaughlin}, {Bassa}, {Deller}, {Kramer}, {Lee}, {Mingarelli}, {Rosado}, {Sanidas}, {Sesana}, {Shao}, {Stairs}, {Stappers}, \& {Verbiest}}]{2015aska.confE..37J}
{Janssen}, G., {Hobbs}, G., {McLaughlin}, M., {et~al.} 2015, in Advancing Astrophysics with the Square Kilometre Array (AASKA14), 37, \dodoi{10.22323/1.215.0037}

\bibitem[{{Jennings} {et~al.}(2024){Jennings}, {Cordes}, {Chatterjee}, {McLaughlin}, {Demorest}, {Arzoumanian}, {Baker}, {Blumer}, {Brook}, {Cohen}, {Crawford}, {Cromartie}, {DeCesar}, {Dolch}, {Ferrara}, {Fonseca}, {Good}, {Hazboun}, {Jones}, {Kaplan}, {Lam}, {Lazio}, {Lorimer}, {Luo}, {Lynch}, {McKee}, {Madison}, {Meyers}, {Mingarelli}, {Nice}, {Pennucci}, {Perera}, {Pol}, {Ransom}, {Ray}, {Shapiro-Albert}, {Siemens}, {Stairs}, {Stinebring}, {Swiggum}, {Tan}, {Taylor}, {Vigeland}, \& {Witt}}]{2024ApJ...964..179J}
{Jennings}, R.~J., {Cordes}, J.~M., {Chatterjee}, S., {et~al.} 2024, \apj, 964, 179, \dodoi{10.3847/1538-4357/ad2930}

\bibitem[{{Joshi} {et~al.}(2022){Joshi}, {Gopakumar}, {Pandian}, {Prabu}, {Dey}, {Bagchi}, {Desai}, {Tarafdar}, {Rana}, {Maan}, {Batra}, {Girgaonkar}, {Agarwal}, {Arumugam}, {Basu}, {Bathula}, {Dandapat}, {Gupta}, {Hisano}, {Kato}, {Kharbanda}, {Kikunaga}, {Kolhe}, {Krishnakumar}, {Manoharan}, {Marmat}, {Naidu}, {Banik}, {Nobleson}, {Paladi}, {Pathak}, {Singha}, {Srivastava}, {Surnis}, {Susarla}, {Susobhanan}, \& {Takahashi}}]{2022JApA...43...98J}
{Joshi}, B.~C., {Gopakumar}, A., {Pandian}, A., {et~al.} 2022, Journal of Astrophysics and Astronomy, 43, 98, \dodoi{10.1007/s12036-022-09869-w}

\bibitem[{Kamionkowski {et~al.}(1994)Kamionkowski, Kosowsky, \& Turner}]{Kamionkowski:1993fg}
Kamionkowski, M., Kosowsky, A., \& Turner, M.~S. 1994, Phys. Rev. D, 49, 2837, \dodoi{10.1103/PhysRevD.49.2837}

\bibitem[{{Kato} \& {Takahashi}(2023)}]{2023PhRvD.108l3535K}
{Kato}, R., \& {Takahashi}, K. 2023, \prd, 108, 123535, \dodoi{10.1103/PhysRevD.108.123535}

\bibitem[{{Kato} \& {Takahashi}(2025)}]{2025arXiv250602819K}
---. 2025, arXiv e-prints, arXiv:2506.02819, \dodoi{10.48550/arXiv.2506.02819}

\bibitem[{{Katz} {et~al.}(2020){Katz}, {Kelley}, {Dosopoulou}, {Berry}, {Blecha}, \& {Larson}}]{KatzEtAl:2020}
{Katz}, M.~L., {Kelley}, L.~Z., {Dosopoulou}, F., {et~al.} 2020, \mnras, 491, 2301, \dodoi{10.1093/mnras/stz3102}

\bibitem[{{Kaur} {et~al.}(2022){Kaur}, {Ramesh Bhat}, {Dai}, {McSweeney}, {Shannon}, {Kudale}, \& {van Straten}}]{2022ApJ...930L..27K}
{Kaur}, D., {Ramesh Bhat}, N.~D., {Dai}, S., {et~al.} 2022, \apjl, 930, L27, \dodoi{10.3847/2041-8213/ac64ff}

\bibitem[{{Kaur} {et~al.}(2019){Kaur}, {Bhat}, {Tremblay}, {Shannon}, {McSweeney}, {Ord}, {Beardsley}, {Crosse}, {Emrich}, {Franzen}, {Horsley}, {Johnston-Hollitt}, {Kaplan}, {Kenney}, {Morales}, {Pallot}, {Steele}, {Tingay}, {Trott}, {Walker}, {Wayth}, {Williams}, \& {Wu}}]{2019ApJ...882..133K}
{Kaur}, D., {Bhat}, N.~D.~R., {Tremblay}, S.~E., {et~al.} 2019, \apj, 882, 133, \dodoi{10.3847/1538-4357/ab338f}

\bibitem[{{Keane} {et~al.}(2025)}]{Keane2025_SKA_SKAPTA}
{Keane}, E., {et~al.} 2025

\bibitem[{{Kerr}(2011)}]{2011ApJ...732...38K}
{Kerr}, M. 2011, \apj, 732, 38, \dodoi{10.1088/0004-637X/732/1/38}

\bibitem[{{Kerr}(2015)}]{ker15}
---. 2015, \mnras, 452, 607, \dodoi{10.1093/mnras/stv1296}

\bibitem[{{Kerr}(2025)}]{2025ApJ...991..225K}
---. 2025, \apj, 991, 225, \dodoi{10.3847/1538-4357/adfa95}

\bibitem[{{Kerr} {et~al.}(2015){Kerr}, {Ray}, {Johnston}, {Shannon}, \& {Camilo}}]{2015ApJ...814..128K}
{Kerr}, M., {Ray}, P.~S., {Johnston}, S., {Shannon}, R.~M., \& {Camilo}, F. 2015, \apj, 814, 128, \dodoi{10.1088/0004-637X/814/2/128}

\bibitem[{Khmelnitsky \& Rubakov(2014)}]{Khmelnitsky:2013lxt}
Khmelnitsky, A., \& Rubakov, V. 2014, JCAP, 02, 019, \dodoi{10.1088/1475-7516/2014/02/019}

\bibitem[{{Khusid} {et~al.}(2023){Khusid}, {Mingarelli}, {Natarajan}, {Casey-Clyde}, \& {Barnacka}}]{KhusidEtAl2023}
{Khusid}, N.~M., {Mingarelli}, C. M.~F., {Natarajan}, P., {Casey-Clyde}, J.~A., \& {Barnacka}, A. 2023, \apj, 955, 25, \dodoi{10.3847/1538-4357/ace16f}

\bibitem[{Kibble(1976)}]{Kibble:1976sj}
Kibble, T. W.~B. 1976, J. Phys. A, 9, 1387, \dodoi{10.1088/0305-4470/9/8/029}

\bibitem[{{Kiehlmann} {et~al.}(2025){Kiehlmann}, {de la Parra}, {Sullivan}, {Synani}, {Liodakis}, {Mr{\'o}z}, {N{\ae}ss}, {Readhead}, {Begelman}, {Blandford}, {Chatziioannou}, {Ding}, {Graham}, {Harrison}, {Homan}, {Hovatta}, {Kulkarni}, {Lister}, {Maiolino}, {Max-Moerbeck}, {Molina}, {O'Dea}, {Pavlidou}, {Pearson}, {Aller}, {Lawrence}, {Lazio}, {O'Neill}, {Prince}, {Ravi}, {Reeves}, {Tassis}, {Vallisneri}, \& {Zensus}}]{PKS25}
{Kiehlmann}, S., {de la Parra}, P.~V., {Sullivan}, A.~G., {et~al.} 2025, \apj, 985, 59, \dodoi{10.3847/1538-4357/adc567}

\bibitem[{{Kikunaga} {et~al.}(2024){Kikunaga}, {Hisano}, {Batra}, {Desai}, {Joshi}, {Bagchi}, {Prabu}, {Takahashi}, {Arumugam}, {Bathula}, {Dandapat}, {Deb}, {Dwivedi}, {Gupta}, {Jacob}, {Kareem}, {Nobleson}, {Mamidipaka}, {Paladi}, {Pandian}, {Rana}, {Singha}, {Srivastava}, {Surnis}, \& {Tarafdar}}]{2024PASA...41...36K}
{Kikunaga}, T., {Hisano}, S., {Batra}, N.~D., {et~al.} 2024, \pasa, 41, e036, \dodoi{10.1017/pasa.2024.30}

\bibitem[{Kim \& Mitridate(2024)}]{Kim:2023kyy}
Kim, H., \& Mitridate, A. 2024, Phys. Rev. D, 109, 055017, \dodoi{10.1103/PhysRevD.109.055017}

\bibitem[{{Klein} {et~al.}(2016){Klein}, {Barausse}, {Sesana}, {Petiteau}, {Berti}, {Babak}, {Gair}, {Aoudia}, {Hinder}, {Ohme}, \& {Wardell}}]{KleinEtAl:2016}
{Klein}, A., {Barausse}, E., {Sesana}, A., {et~al.} 2016, \prd, 93, 024003, \dodoi{10.1103/PhysRevD.93.024003}

\bibitem[{{Kocsis} \& {Sesana}(2011)}]{2011MNRAS.411.1467K}
{Kocsis}, B., \& {Sesana}, A. 2011, \mnras, 411, 1467, \dodoi{10.1111/j.1365-2966.2010.17782.x}

\bibitem[{Kohri \& Terada(2018)}]{Kohri:2018awv}
Kohri, K., \& Terada, T. 2018, Phys. Rev. D, 97, 123532, \dodoi{10.1103/PhysRevD.97.123532}

\bibitem[{{Kormendy} \& {Ho}(2013)}]{2013ARA&A..51..511K}
{Kormendy}, J., \& {Ho}, L.~C. 2013, \araa, 51, 511, \dodoi{10.1146/annurev-astro-082708-101811}

\bibitem[{{Kramer} \& {Champion}(2013)}]{2013CQGra..30v4009K}
{Kramer}, M., \& {Champion}, D.~J. 2013, Classical and Quantum Gravity, 30, 224009, \dodoi{10.1088/0264-9381/30/22/224009}

\bibitem[{{Krishnakumar} {et~al.}(2021){Krishnakumar}, {Manoharan}, {Joshi}, {Girgaonkar}, {Desai}, {Bagchi}, {Nobleson}, {Dey}, {Susobhanan}, {Susarla}, {Surnis}, {Maan}, {Gopakumar}, {Basu}, {Batra}, {Choudhary}, {De}, {Gupta}, {Naidu}, {Pathak}, {Singha}, \& {Prabu}}]{2021A&A...651A...5K}
{Krishnakumar}, M.~A., {Manoharan}, P.~K., {Joshi}, B.~C., {et~al.} 2021, \aap, 651, A5, \dodoi{10.1051/0004-6361/202140340}

\bibitem[{{Kulkarni} {et~al.}(2025){Kulkarni}, {Shannon}, {Reardon}, \& {Miles}}]{2025arXiv251103185K}
{Kulkarni}, A.~D., {Shannon}, R.~M., {Reardon}, D.~J., \& {Miles}, M.~T. 2025, \mnras, 544, 2795, \dodoi{10.1093/mnras/staf1930}

\bibitem[{{Kulkarni} {et~al.}(2024){Kulkarni}, {Shannon}, {Reardon}, {Miles}, {Bailes}, \& {Shamohammadi}}]{2024MNRAS.528.3658K}
{Kulkarni}, A.~D., {Shannon}, R.~M., {Reardon}, D.~J., {et~al.} 2024, \mnras, 528, 3658, \dodoi{10.1093/mnras/stae041}

\bibitem[{Kuroyanagi {et~al.}(2015)Kuroyanagi, Takahashi, \& Yokoyama}]{Kuroyanagi:2014nba}
Kuroyanagi, S., Takahashi, T., \& Yokoyama, S. 2015, JCAP, 02, 003, \dodoi{10.1088/1475-7516/2015/02/003}

\bibitem[{{Laine} {et~al.}(2020){Laine}, {Dey}, {Valtonen}, {Gopakumar}, {Zola}, {Komossa}, {Kidger}, {Pihajoki}, {G{\'o}mez}, {Caton}, {Ciprini}, {Drozdz}, {Gazeas}, {Godunova}, {Haque}, {Hildebrandt}, {Hudec}, {Jermak}, {Kong}, {Lehto}, {Liakos}, {Matsumoto}, {Mugrauer}, {Pursimo}, {Reichart}, {Simon}, {Siwak}, \& {Sonbas}}]{Spitzer20}
{Laine}, S., {Dey}, L., {Valtonen}, M., {et~al.} 2020, \apjl, 894, L1, \dodoi{10.3847/2041-8213/ab79a4}

\bibitem[{{Lam} {et~al.}(2016{\natexlab{a}}){Lam}, {Cordes}, {Chatterjee}, {Jones}, {McLaughlin}, \& {Armstrong}}]{2016ApJ...821...66L}
{Lam}, M.~T., {Cordes}, J.~M., {Chatterjee}, S., {et~al.} 2016{\natexlab{a}}, \apj, 821, 66, \dodoi{10.3847/0004-637X/821/1/66}

\bibitem[{{Lam} {et~al.}(2018){Lam}, {McLaughlin}, {Cordes}, {Chatterjee}, \& {Lazio}}]{2018ApJ...861...12L}
{Lam}, M.~T., {McLaughlin}, M.~A., {Cordes}, J.~M., {Chatterjee}, S., \& {Lazio}, T.~J.~W. 2018, \apj, 861, 12, \dodoi{10.3847/1538-4357/aac48d}

\bibitem[{{Lam} {et~al.}(2016{\natexlab{b}}){Lam}, {Cordes}, {Chatterjee}, {Arzoumanian}, {Crowter}, {Demorest}, {Dolch}, {Ellis}, {Ferdman}, {Fonseca}, {Gonzalez}, {Jones}, {Jones}, {Levin}, {Madison}, {McLaughlin}, {Nice}, {Pennucci}, {Ransom}, {Siemens}, {Stairs}, {Stovall}, {Swiggum}, \& {Zhu}}]{Lam2016}
{Lam}, M.~T., {Cordes}, J.~M., {Chatterjee}, S., {et~al.} 2016{\natexlab{b}}, \apj, 819, 155, \dodoi{10.3847/0004-637X/819/2/155}

\bibitem[{{Larsen} {et~al.}(2024){Larsen}, {Mingarelli}, {Hazboun}, {Chalumeau}, {Good}, {Simon}, {Agazie}, {Anumarlapudi}, {Archibald}, {Arzoumanian}, {Baker}, {Brook}, {Cromartie}, {Crowter}, {DeCesar}, {Demorest}, {Dolch}, {Ferrara}, {Fiore}, {Fonseca}, {Freedman}, {Garver-Daniels}, {Gentile}, {Glaser}, {Jennings}, {Jones}, {Kaplan}, {Kerr}, {Lam}, {Lorimer}, {Luo}, {Lynch}, {McEwen}, {McLaughlin}, {McMann}, {Meyers}, {Ng}, {Nice}, {Pennucci}, {Perera}, {Pol}, {Radovan}, {Ransom}, {Ray}, {Schmiedekamp}, {Schmiedekamp}, {Shapiro-Albert}, {Stairs}, {Stovall}, {Susobhanan}, {Swiggum}, {Wahl}, {Champion}, {Cognard}, {Guillemot}, {Hu}, {Keith}, {Liu}, {McKee}, {Parthasarathy}, {Perrodin}, {Possenti}, {Shaifullah}, \& {Theureau}}]{2024ApJ...972...49L}
{Larsen}, B., {Mingarelli}, C. M.~F., {Hazboun}, J.~S., {et~al.} 2024, \apj, 972, 49, \dodoi{10.3847/1538-4357/ad5291}

\bibitem[{{Lasky} {et~al.}(2016{\natexlab{a}}){Lasky}, {Mingarelli}, {Smith}, {Giblin}, {Thrane}, {Reardon}, {Caldwell}, {Bailes}, {Bhat}, {Burke-Spolaor}, {Dai}, {Dempsey}, {Hobbs}, {Kerr}, {Levin}, {Manchester}, {Os{\l}owski}, {Ravi}, {Rosado}, {Shannon}, {Spiewak}, {van Straten}, {Toomey}, {Wang}, {Wen}, {You}, \& {Zhu}}]{2016PhRvX...6a1035L}
{Lasky}, P.~D., {Mingarelli}, C. M.~F., {Smith}, T.~L., {et~al.} 2016{\natexlab{a}}, Physical Review X, 6, 011035, \dodoi{10.1103/PhysRevX.6.011035}

\bibitem[{{Lasky} {et~al.}(2016{\natexlab{b}}){Lasky}, {Mingarelli}, {Smith}, {Giblin}, {Thrane}, {Reardon}, {Caldwell}, {Bailes}, {Bhat}, {Burke-Spolaor}, {Dai}, {Dempsey}, {Hobbs}, {Kerr}, {Levin}, {Manchester}, {Os{\l}owski}, {Ravi}, {Rosado}, {Shannon}, {Spiewak}, {van Straten}, {Toomey}, {Wang}, {Wen}, {You}, \& {Zhu}}]{Lasky_2016}
---. 2016{\natexlab{b}}, Physical Review X, 6, 011035, \dodoi{10.1103/PhysRevX.6.011035}

\bibitem[{{Lee} {et~al.}(2025){Lee}, {Bhat}, {Meyers}, {McSweeney}, {van Straten}, {Tan}, {Xue}, {Swainston}, {Ord}, {Sleap}, {Tremblay}, \& {Williams}}]{2025PASA...42..117L}
{Lee}, C.~P., {Bhat}, N.~D.~R., {Meyers}, B.~W., {et~al.} 2025, \pasa, 42, e117, \dodoi{10.1017/pasa.2025.10081}

\bibitem[{{Lee}(2016)}]{2016ASPC..502...19L}
{Lee}, K.~J. 2016, in Astronomical Society of the Pacific Conference Series, Vol. 502, Frontiers in Radio Astronomy and FAST Early Sciences Symposium 2015, ed. L.~{Qain} \& D.~{Li}, 19

\bibitem[{{Lee} {et~al.}(2012){Lee}, {Bassa}, {Janssen}, {Karuppusamy}, {Kramer}, {Smits}, \& {Stappers}}]{2012MNRAS.423.2642L}
{Lee}, K.~J., {Bassa}, C.~G., {Janssen}, G.~H., {et~al.} 2012, \mnras, 423, 2642, \dodoi{10.1111/j.1365-2966.2012.21070.x}

\bibitem[{{Lee} {et~al.}(2011){Lee}, {Wex}, {Kramer}, {Stappers}, {Bassa}, {Janssen}, {Karuppusamy}, \& {Smits}}]{2011MNRAS.414.3251L}
{Lee}, K.~J., {Wex}, N., {Kramer}, M., {et~al.} 2011, \mnras, 414, 3251, \dodoi{10.1111/j.1365-2966.2011.18622.x}

\bibitem[{{Lee} {et~al.}(2014){Lee}, {Bassa}, {Janssen}, {Karuppusamy}, {Kramer}, {Liu}, {Perrodin}, {Smits}, {Stappers}, {van Haasteren}, \& {Lentati}}]{2014MNRAS.441.2831L}
{Lee}, K.~J., {Bassa}, C.~G., {Janssen}, G.~H., {et~al.} 2014, \mnras, 441, 2831, \dodoi{10.1093/mnras/stu664}

\bibitem[{Lee {et~al.}(2021)Lee, Mitridate, Trickle, \& Zurek}]{Lee:2020wfn}
Lee, V. S.~H., Mitridate, A., Trickle, T., \& Zurek, K.~M. 2021, JHEP, 06, 028, \dodoi{10.1007/JHEP06(2021)028}

\bibitem[{{Lentati} {et~al.}(2014){Lentati}, {Alexander}, {Hobson}, {Feroz}, {van Haasteren}, {Lee}, \& {Shannon}}]{2014MNRAS.437.3004L}
{Lentati}, L., {Alexander}, P., {Hobson}, M.~P., {et~al.} 2014, \mnras, 437, 3004, \dodoi{10.1093/mnras/stt2122}

\bibitem[{{Lentati} {et~al.}(2017){Lentati}, {Kerr}, {Dai}, {Hobson}, {Shannon}, {Hobbs}, {Bailes}, {Bhat}, {Burke-Spolaor}, {Coles}, {Dempsey}, {Lasky}, {Levin}, {Manchester}, {Os{\l}owski}, {Ravi}, {Reardon}, {Rosado}, {Spiewak}, {van Straten}, {Toomey}, {Wang}, {Wen}, {You}, \& {Zhu}}]{2017MNRAS.466.3706L}
{Lentati}, L., {Kerr}, M., {Dai}, S., {et~al.} 2017, \mnras, 466, 3706, \dodoi{10.1093/mnras/stw3359}

\bibitem[{{Li} {et~al.}(2025){Li}, {Liu}, {Chen}, {Dai}, {Goncharov}, {Hu}, {Huang}, {Liu}, {Ren}, {Wu}, {Xue}, \& {Zhu}}]{Li:2025xlr}
{Li}, X., {Liu}, Y., {Chen}, Z.-C., {et~al.} 2025, arXiv e-prints, arXiv:2506.04871, \dodoi{10.48550/arXiv.2506.04871}

\bibitem[{{LIGO Scientific Collaboration} {et~al.}(2015){LIGO Scientific Collaboration}, {Aasi}, {Abbott}, {Abbott}, {Abbott}, {Abernathy}, {Ackley}, {Adams}, {Adams}, {Addesso}, {Adhikari}, {Adya}, {Affeldt}, {Aggarwal}, {Aguiar}, {Ain}, {Ajith}, {Alemic}, {Allen}, {Amariutei}, {Anderson}, {Anderson}, {Arai}, {Araya}, {Arceneaux}, {Areeda}, {Ashton}, {Ast}, {Aston}, {Aufmuth}, {Aulbert}, {Aylott}, {Babak}, {Baker}, {Ballmer}, {Barayoga}, {Barbet}, {Barclay}, {Barish}, {Barker}, {Barr}, {Barsotti}, {Bartlett}, {Barton}, {Bartos}, {Bassiri}, {Batch}, {Baune}, {Behnke}, {Bell}, {Bell}, {Benacquista}, {Bergman}, {Bergmann}, {Berry}, {Betzwieser}, {Bhagwat}, {Bhandare}, {Bilenko}, {Billingsley}, {Birch}, {Biscans}, {Biwer}, {Blackburn}, {Blackburn}, {Blair}, {Blair}, {Bock}, {Bodiya}, {Bojtos}, {Bond}, {Bork}, {Born}, {Bose}, {Brady}, {Braginsky}, {Brau}, {Bridges}, {Brinkmann}, {Brooks}, {Brown}, {Brown}, {Brown}, {Buchman}, {Buikema}, {Buonanno}, {Cadonati}, {Calder{\'o}n Bustillo}, {Camp}, {Cannon}, {Cao},
  {Capano}, {Caride}, {Caudill}, {Cavagli{\`a}}, {Cepeda}, {Chakraborty}, {Chalermsongsak}, {Chamberlin}, {Chao}, {Charlton}, {Chen}, {Cho}, {Cho}, {Chow}, {Christensen}, {Chu}, {Chung}, {Ciani}, {Clara}, {Clark}, {Collette}, {Cominsky}, {Constancio}, {Cook}, {Corbitt}, {Cornish}, {Corsi}, {Costa}, {Coughlin}, {Countryman}, {Couvares}, {Coward}, {Cowart}, {Coyne}, {Coyne}, {Craig}, {Creighton}, {Creighton}, {Cripe}, {Crowder}, {Cumming}, {Cunningham}, {Cutler}, {Dahl}, {Dal Canton}, {Damjanic}, {Danilishin}, {Danzmann}, {Dartez}, {Dave}, {Daveloza}, {Davies}, {Daw}, {DeBra}, {Del Pozzo}, {Denker}, {Dent}, {Dergachev}, {DeRosa}, {DeSalvo}, {Dhurandhar}, {D{\textasciiacute}{\i}az}, {Di Palma}, {Dojcinoski}, {Dominguez}, {Donovan}, {Dooley}, {Doravari}, {Douglas}, {Downes}, {Driggers}, {Du}, {Dwyer}, {Eberle}, {Edo}, {Edwards}, {Edwards}, {Effler}, {Eggenstein}, {Ehrens}, {Eichholz}, {Eikenberry}, {Essick}, {Etzel}, {Evans}, {Evans}, {Factourovich}, {Fairhurst}, {Fan}, {Fang}, {Farr}, {Farr}, {Favata}, {Fays},
  {Fehrmann}, {Fejer}, {Feldbaum}, {Ferreira}, {Fisher}, {Frei}, {Freise}, {Frey}, {Fricke}, {Fritschel}, {Frolov}, {Fuentes-Tapia}, {Fulda}, {Fyffe}, \& {Gair}}]{AdLIGO:2015}
{LIGO Scientific Collaboration}, {Aasi}, J., {Abbott}, B.~P., {et~al.} 2015, Classical and Quantum Gravity, 32, 074001, \dodoi{10.1088/0264-9381/32/7/074001}

\bibitem[{{LIGO Scientific Collaboration} {et~al.}(2025){LIGO Scientific Collaboration}, {the Virgo Collaboration}, {the KAGRA Collaboration}, {Abac}, {Abouelfettouh}, {Acernese}, {Ackley}, {Adhicary}, {Adhikari}, {Adhikari}, {Adhikari}, {Adkins}, {Afroz}, {Agarwal}, {Agathos}, {Aghaei Abchouyeh}, {Aguiar}, {Ahmadzadeh}, {Aiello}, {Ain}, {Ajith}, {Akcay}, {Akutsu}, {Albanesi}, {Alfaidi}, {Al-Jodah}, {All{\'e}n{\'e}}, {Allocca}, {Al-Shammari}, {Altin}, {Alvarez-Lopez}, {Amarasinghe}, {Amato}, {Amra}, {Ananyeva}, {Anderson}, {Anderson}, {Andia}, {Ando}, {Andrade}, {Andr{\'e}s-Carcasona}, {Andri{\'c}}, {Anglin}, {Ansoldi}, {Antelis}, {Antier}, {Aoumi}, {Appavuravther}, {Appert}, {Apple}, {Arai}, {Araya}, {Araya}, {Arca Sedda}, {Areeda}, {Argianas}, {Aritomi}, {Armato}, {Armstrong}, {Arnaud}, {Arogeti}, {Aronson}, {Ashton}, {Aso}, {Assiduo}, {Assis de Souza Melo}, {Aston}, {Astone}, {Attadio}, {Aubin}, {AultONeal}, {Avallone}, {Babak}, {Badaracco}, {Badger}, {Bae}, {Bagnasco}, {Bagui}, {Baiotti}, {Bajpai},
  {Baka}, {Baker}, {Ball}, {Ballardin}, {Ballmer}, {Banagiri}, {Banerjee}, {Bankar}, {Baptiste}, {Baral}, {Barayoga}, {Barish}, {Barker}, {Barman}, {Barneo}, {Barone}, {Barr}, {Barsotti}, {Barsuglia}, {Barta}, {Bartoletti}, {Barton}, {Bartos}, {Basak}, {Basalaev}, {Bassiri}, {Basti}, {Bates}, {Bawaj}, {Baxi}, {Bayley}, {Baylor}, {Baynard}, {Bazzan}, {Bedakihale}, {Beirnaert}, {Bejger}, {Belardinelli}, {Bell}, {Bellie}, {Bellizzi}, {Benoit}, {Bentara}, {Bentley}, {Ben Yaala}, {Bera}, {Bergamin}, {Berger}, {Bernuzzi}, {Beroiz}, {Berry}, {Bersanetti}, {Bertolini}, {Betzwieser}, {Beveridge}, {Bevilacqua}, {Bevins}, {Bhandare}, {Bhatt}, {Bhattacharjee}, {Bhaumik}, {Bhowmick}, {Biancalana}, {Bianchi}, {Bilenko}, {Billingsley}, {Binetti}, {Bini}, {Binu}, {Birnholtz}, {Biscoveanu}, {Bisht}, {Bitossi}, {Bizouard}, {Blaber}, {Blackburn}, {Blagg}, {Blair}, {Blair}, {Bobba}, {Bode}, {Boileau}, {Boldrini}, {Bolingbroke}, {Bolliand}, {Bonavena}, {Bondarescu}, {Bondu}, {Bonilla}, {Bonilla}, {Bonino}, {Bonnand}, {Booker},
  {Borchers}, {Borhanian}, {Boschi}, {Bose}, {Bossilkov}, {Boudon}, {Bozzi}, {Bradaschia}, {Brady}, {Branch}, {Branchesi}, {Braun}, {Briant}, {Brillet}, {Brinkmann}, {Brockill}, {Brockmueller}, {Brooks}, {Brown}, {Brown}, {Brozzetti}, {Brunett}, {Bruno}, {Bruntz}, {Bryant}, {Bu}, \& {Bucci}}]{GWTC4}
{LIGO Scientific Collaboration}, {the Virgo Collaboration}, {the KAGRA Collaboration}, {et~al.} 2025, arXiv e-prints, arXiv:2508.18080, \dodoi{10.48550/arXiv.2508.18080}

\bibitem[{{Linde}(1982)}]{1982PhLB..108..389L}
{Linde}, A.~D. 1982, Physics Letters B, 108, 389, \dodoi{10.1016/0370-2693(82)91219-9}

\bibitem[{{Liu} {et~al.}(2012){Liu}, {Keane}, {Lee}, {Kramer}, {Cordes}, \& {Purver}}]{lkl+12}
{Liu}, K., {Keane}, E.~F., {Lee}, K.~J., {et~al.} 2012, MNRAS, 420, 361, \dodoi{10.1111/j.1365-2966.2011.20041.x}

\bibitem[{{Liu} {et~al.}(2011){Liu}, {Verbiest}, {Kramer}, {Stappers}, {van Straten}, \& {Cordes}}]{lvk+11}
{Liu}, K., {Verbiest}, J.~P.~W., {Kramer}, M., {et~al.} 2011, MNRAS, 417, 2916, \dodoi{10.1111/j.1365-2966.2011.19452.x}

\bibitem[{{Liu} {et~al.}(2014){Liu}, {Desvignes}, {Cognard}, {Stappers}, {Verbiest}, {Lee}, {Champion}, {Kramer}, {Freire}, \& {Karuppusamy}}]{2014MNRAS.443.3752L}
{Liu}, K., {Desvignes}, G., {Cognard}, I., {et~al.} 2014, \mnras, 443, 3752, \dodoi{10.1093/mnras/stu1420}

\bibitem[{{Liu} {et~al.}(2015){Liu}, {Karuppusamy}, {Lee}, {Stappers}, {Kramer}, {Smits}, {Purver}, {Janssen}, \& {Perrodin}}]{lkl+15}
{Liu}, K., {Karuppusamy}, R., {Lee}, K.~J., {et~al.} 2015, MNRAS, 449, 1158, \dodoi{10.1093/mnras/stv397}

\bibitem[{{Liu} {et~al.}(2016){Liu}, {Bassa}, {Janssen}, {Karuppusamy}, {McKee}, {Kramer}, {Lee}, {Perrodin}, {Purver}, {Sanidas}, {Smits}, {Stappers}, {Weltevrede}, \& {Zhu}}]{lbj+16}
{Liu}, K., {Bassa}, C.~G., {Janssen}, G.~H., {et~al.} 2016, \mnras, 463, 3239, \dodoi{10.1093/mnras/stw2223}

\bibitem[{{Liu} {et~al.}(2022){Liu}, {Antoniadis}, {Bassa}, {Chen}, {Cognard}, {Gaikwad}, {Hu}, {Jang}, {Janssen}, {Karuppusamy}, {Kramer}, {Lee}, {Main}, {Mall}, {McKee}, {Mickaliger}, {Perrodin}, {Sanidas}, {Stappers}, {Wang}, {Zhu}, {Burgay}, {Concu}, {Corongiu}, {Melis}, {Pilia}, \& {Possenti}}]{lab+22}
{Liu}, K., {Antoniadis}, J., {Bassa}, C.~G., {et~al.} 2022, \mnras, 513, 4037, \dodoi{10.1093/mnras/stac1082}

\bibitem[{{Liu} {et~al.}(2025){Liu}, {Parthasarathy}, {Keith}, {Tiburzi}, {Susarla}, {Antoniadis}, {Chalumeau}, {Chen}, {Cognard}, {Golden}, {Grie{\ss}meier}, {Guillemot}, {Janssen}, {Keane}, {Kramer}, {McKee}, {Mickaliger}, {Theureau}, \& {Wang}}]{Liu2025}
{Liu}, K., {Parthasarathy}, A., {Keith}, M., {et~al.} 2025, \mnras, 536, 2603, \dodoi{10.1093/mnras/stae2727}

\bibitem[{Liu {et~al.}(2023)Liu, Cai, \& Wang}]{Liu:2023ymk}
Liu, L., Cai, Y.-F., \& Wang, M.-Z. 2023, Physical Review D, 108, 063530, \dodoi{10.1103/PhysRevD.108.063530}

\bibitem[{{Liu} {et~al.}(2023){Liu}, {Lou}, \& {Ren}}]{2023PhRvL.130l1401L}
{Liu}, T., {Lou}, X., \& {Ren}, J. 2023, \prl, 130, 121401, \dodoi{10.1103/PhysRevLett.130.121401}

\bibitem[{Liu {et~al.}(2020)Liu, Smoot, \& Zhao}]{Liu:2019brz}
Liu, T., Smoot, G., \& Zhao, Y. 2020, Phys. Rev. D, 101, 063012, \dodoi{10.1103/PhysRevD.101.063012}

\bibitem[{{Liu} \& {Vigeland}(2021)}]{2021ApJ...921..178L}
{Liu}, T., \& {Vigeland}, S.~J. 2021, \apj, 921, 178, \dodoi{10.3847/1538-4357/ac1da9}

\bibitem[{Lorenz {et~al.}(2010)Lorenz, Ringeval, \& Sakellariadou}]{Lorenz:2010sm}
Lorenz, L., Ringeval, C., \& Sakellariadou, M. 2010, JCAP, 10, 003, \dodoi{10.1088/1475-7516/2010/10/003}

\bibitem[{{Lorimer} \& {Kramer}(2004)}]{lk2004}
{Lorimer}, D.~R., \& {Kramer}, M. 2004, {Handbook of Pulsar Astronomy} (Cambridge University Press)

\bibitem[{{Lower} {et~al.}(2025){Lower}, {Karastergiou}, {Johnston}, {Brook}, {Dai}, {Kerr}, {Manchester}, {Oswald}, {Shannon}, {Sobey}, \& {Weltevrede}}]{2025arXiv250103500L}
{Lower}, M.~E., {Karastergiou}, A., {Johnston}, S., {et~al.} 2025, arXiv e-prints, arXiv:2501.03500, \dodoi{10.48550/arXiv.2501.03500}

\bibitem[{{Luo} {et~al.}(2021){Luo}, {Ransom}, {Demorest}, {Ray}, {Archibald}, {Kerr}, {Jennings}, {Bachetti}, {van Haasteren}, {Champagne}, {Colen}, {Phillips}, {Zimmerman}, {Stovall}, {Lam}, \& {Jenet}}]{2021ApJ...911...45L}
{Luo}, J., {Ransom}, S., {Demorest}, P., {et~al.} 2021, \apj, 911, 45, \dodoi{10.3847/1538-4357/abe62f}

\bibitem[{{Luo} {et~al.}(2025){Luo}, {Bai}, {Bai}, {Cai}, {Dang}, {Dong}, {Duan}, {Du}, {Fan}, {Fu}, {Gao}, {Gou}, {Gu}, {Guo}, {Hong}, {Hu}, {Hu}, {Hu}, {Hu}, {Huang}, {Ji}, {Jiang}, {Li}, {Li}, {Li}, {Li}, {Li}, {Li}, {Lian}, {Liang}, {Lin}, {Liu}, {Liu}, {Liu}, {Liu}, {Liu}, {Liu}, {Liu}, {Liu}, {Lu}, {Luo}, {Luo}, {Ma}, {Ma}, {Meng}, {Milyukov}, {Peng}, {Postnov}, {Qu}, {Shan}, {Shao}, {Shi}, {Song}, {Song}, {Su}, {Tan}, {Tan}, {Tan}, {Tan}, {Tong}, {Tu}, {Wang}, {Wang}, {Wang}, {Wang}, {Wang}, {Wang}, {Wang}, {Wang}, {Wei}, {Wu}, {Xu}, {Xu}, {Xue}, {Yan}, {Yan}, {Yang}, {Yang}, {Yeh}, {Yi}, {Yin}, {Yu}, {Yuan}, {Zhang}, {Zhang}, {Zhang}, {Zhang}, {Zhang}, {Zhang}, {Zhao}, {Zhao}, {Zhao}, {Zhou}, {Zhou}, {Zhou}, {Zhou}, {Zhou}, {Zhu}, {Zhu}, {Zhu}, {Zou}, \& {Mei}}]{TianQin:2025}
{Luo}, J., {Bai}, S., {Bai}, Y., {et~al.} 2025, Classical and Quantum Gravity, 42, 173001, \dodoi{10.1088/1361-6382/adda8a}

\bibitem[{Machado {et~al.}(2019)Machado, Ratzinger, Schwaller, \& Stefanek}]{Machado:2018nqk}
Machado, C.~S., Ratzinger, W., Schwaller, P., \& Stefanek, B.~A. 2019, JHEP, 01, 053, \dodoi{10.1007/JHEP01(2019)053}

\bibitem[{{Madison} {et~al.}(2019){Madison}, {Cordes}, {Arzoumanian}, {Chatterjee}, {Crowter}, {DeCesar}, {Demorest}, {Dolch}, {Ellis}, {Ferdman}, {Ferrara}, {Fonseca}, {Gentile}, {Jones}, {Jones}, {Lam}, {Levin}, {Lorimer}, {Lynch}, {McLaughlin}, {Mingarelli}, {Ng}, {Nice}, {Pennucci}, {Ransom}, {Ray}, {Spiewak}, {Stairs}, {Stovall}, {Swiggum}, \& {Zhu}}]{Madison2019}
{Madison}, D.~R., {Cordes}, J.~M., {Arzoumanian}, Z., {et~al.} 2019, \apj, 872, 150, \dodoi{10.3847/1538-4357/ab01fd}

\bibitem[{{Manchester} {et~al.}(2013){Manchester}, {Hobbs}, {Bailes}, {Coles}, {van Straten}, {Keith}, {Shannon}, {Bhat}, {Brown}, {Burke-Spolaor}, {Champion}, {Chaudhary}, {Edwards}, {Hampson}, {Hotan}, {Jameson}, {Jenet}, {Kesteven}, {Khoo}, {Kocz}, {Maciesiak}, {Oslowski}, {Ravi}, {Reynolds}, {Sarkissian}, {Verbiest}, {Wen}, {Wilson}, {Yardley}, {Yan}, \& {You}}]{2013PASA...30...17M}
{Manchester}, R.~N., {Hobbs}, G., {Bailes}, M., {et~al.} 2013, \pasa, 30, e017, \dodoi{10.1017/pasa.2012.017}

\bibitem[{{Mandow} {et~al.}(2025){Mandow}, {Zic}, {Dawson}, {Wang}, {Curylo}, {Dai}, {Di Marco}, {Hobbs}, {Gupta}, {Kapur}, {Kerr}, {Lower}, {Mishra}, {Reardon}, {Russell}, {Shannon}, {Zhang}, \& {Zhu}}]{2025arXiv250918972M}
{Mandow}, R.~F., {Zic}, A., {Dawson}, J.~R., {et~al.} 2025, arXiv e-prints, arXiv:2509.18972, \dodoi{10.48550/arXiv.2509.18972}

\bibitem[{{Miles} {et~al.}(2022){Miles}, {Shannon}, {Bailes}, {Reardon}, {Buchner}, {Middleton}, \& {Spiewak}}]{2022MNRAS.510.5908M}
{Miles}, M.~T., {Shannon}, R.~M., {Bailes}, M., {et~al.} 2022, \mnras, 510, 5908, \dodoi{10.1093/mnras/stab3549}

\bibitem[{{Miles} {et~al.}(2023){Miles}, {Shannon}, {Bailes}, {Reardon}, {Keith}, {Cameron}, {Parthasarathy}, {Shamohammadi}, {Spiewak}, {van Straten}, {Buchner}, {Camilo}, {Geyer}, {Karastergiou}, {Kramer}, {Serylak}, {Theureau}, \& {Venkatraman Krishnan}}]{2023MNRAS.519.3976M}
---. 2023, \mnras, 519, 3976, \dodoi{10.1093/mnras/stac3644}

\bibitem[{{Miles} {et~al.}(2025{\natexlab{a}}){Miles}, {Shannon}, {Reardon}, {Bailes}, {Champion}, {Geyer}, {Gitika}, {Grunthal}, {Keith}, {Kramer}, {Kulkarni}, {Nathan}, {Parthasarathy}, {Porayko}, {Singha}, {Theureau}, {Abbate}, {Buchner}, {Cameron}, {Camilo}, {Moreschi}, {Shaifullah}, {Shamohammadi}, \& {Krishnan}}]{2025MNRAS.536.1467M}
{Miles}, M.~T., {Shannon}, R.~M., {Reardon}, D.~J., {et~al.} 2025{\natexlab{a}}, \mnras, 536, 1467, \dodoi{10.1093/mnras/stae2572}

\bibitem[{{Miles} {et~al.}(2025{\natexlab{b}}){Miles}, {Shannon}, {Reardon}, {Bailes}, {Champion}, {Geyer}, {Gitika}, {Grunthal}, {Keith}, {Kramer}, {Kulkarni}, {Nathan}, {Parthasarathy}, {Singha}, {Theureau}, {Thrane}, {Abbate}, {Buchner}, {Cameron}, {Camilo}, {Moreschi}, {Shaifullah}, {Shamohammadi}, {Possenti}, \& {Krishnan}}]{2025MNRAS.536.1489M}
---. 2025{\natexlab{b}}, \mnras, 536, 1489, \dodoi{10.1093/mnras/stae2571}

\bibitem[{{Milosavljevi{\'c}} \& {Merritt}(2003)}]{2003AIPC..686..201M}
{Milosavljevi{\'c}}, M., \& {Merritt}, D. 2003, in American Institute of Physics Conference Series, Vol. 686, The Astrophysics of Gravitational Wave Sources, ed. J.~M. {Centrella}, 201--210, \dodoi{10.1063/1.1629432}

\bibitem[{{Mingarelli} {et~al.}(2012){Mingarelli}, {Grover}, {Sidery}, {Smith}, \& {Vecchio}}]{Mingarelli2012}
{Mingarelli}, C.~M.~F., {Grover}, K., {Sidery}, T., {Smith}, R.~J.~E., \& {Vecchio}, A. 2012, \prl, 109, 081104, \dodoi{10.1103/PhysRevLett.109.081104}

\bibitem[{{Mingarelli} {et~al.}(2013){Mingarelli}, {Sidery}, {Mandel}, \& {Vecchio}}]{Mingarellietal2013}
{Mingarelli}, C.~M.~F., {Sidery}, T., {Mandel}, I., \& {Vecchio}, A. 2013, \prd, 88, 062005, \dodoi{10.1103/PhysRevD.88.062005}

\bibitem[{{Mingarelli} {et~al.}(2025){Mingarelli}, {Blecha}, {Bogdanovi{\'c}}, {Charisi}, {Chen}, {Escala}, {Goncharov}, {Graham}, {Komossa}, {McWilliams}, {Schwartz}, \& {Zrake}}]{2025NatAs...9..183M}
{Mingarelli}, C.~M.~F., {Blecha}, L., {Bogdanovi{\'c}}, T., {et~al.} 2025, Nature Astronomy, 9, 183, \dodoi{10.1038/s41550-025-02482-1}

\bibitem[{{Moran} {et~al.}(2023){Moran}, {Mingarelli}, {Bedell}, {Good}, \& {Spergel}}]{MoranEtAl2023}
{Moran}, A., {Mingarelli}, C. M.~F., {Bedell}, M., {Good}, D., \& {Spergel}, D.~N. 2023, \apj, 954, 89, \dodoi{10.3847/1538-4357/acec75}

\bibitem[{{Moreschi} {et~al.}(2025){Moreschi}, {Valtolina}, {Sesana}, {Shaifullah}, {Falxa}, {Speri}, {Izquierdo-Villalba}, \& {Chalumeau}}]{Moreschi2025aniso}
{Moreschi}, B.~E., {Valtolina}, S., {Sesana}, A., {et~al.} 2025, arXiv e-prints, arXiv:2506.14882, \dodoi{10.48550/arXiv.2506.14882}

\bibitem[{{Morrison} {et~al.}(2023){Morrison}, {Crosse}, {Sleap}, {Wayth}, {Williams}, {Johnston-Hollitt}, {Jones}, {Tingay}, {Walker}, \& {Williams}}]{2023PASA...40...19M}
{Morrison}, I.~S., {Crosse}, B., {Sleap}, G., {et~al.} 2023, \pasa, 40, e019, \dodoi{10.1017/pasa.2023.15}

\bibitem[{{NANOGrav Collaboration} {et~al.}(2015){NANOGrav Collaboration}, {Arzoumanian}, {Brazier}, {Burke-Spolaor}, {Chamberlin}, {Chatterjee}, {Christy}, {Cordes}, {Cornish}, {Crowter}, {Demorest}, {Dolch}, {Ellis}, {Ferdman}, {Fonseca}, {Garver-Daniels}, {Gonzalez}, {Jenet}, {Jones}, {Jones}, {Kaspi}, {Koop}, {Lam}, {Lazio}, {Levin}, {Lommen}, {Lorimer}, {Luo}, {Lynch}, {Madison}, {McLaughlin}, {McWilliams}, {Nice}, {Palliyaguru}, {Pennucci}, {Ransom}, {Siemens}, {Stairs}, {Stinebring}, {Stovall}, {Swiggum}, {Vallisneri}, {van Haasteren}, {Wang}, \& {Zhu}}]{2015ApJ...813...65N}
{NANOGrav Collaboration}, {Arzoumanian}, Z., {Brazier}, A., {et~al.} 2015, \apj, 813, 65, \dodoi{10.1088/0004-637X/813/1/65}

\bibitem[{{Nathan} {et~al.}(2023){Nathan}, {Miles}, {Ashton}, {Lasky}, {Thrane}, {Reardon}, {Shannon}, \& {Cameron}}]{2023MNRAS.523.4405N}
{Nathan}, R.~S., {Miles}, M.~T., {Ashton}, G., {et~al.} 2023, \mnras, 523, 4405, \dodoi{10.1093/mnras/stad1660}

\bibitem[{{Ni{\c{t}}u} {et~al.}(2024){Ni{\c{t}}u}, {Keith}, {Tiburzi}, {Br{\"u}ggen}, {Champion}, {Chen}, {Cognard}, {Desvignes}, {Dettmar}, {Grie{\ss}meier}, {Guillemot}, {Guo}, {Hoeft}, {Hu}, {Jang}, {Janssen}, {Jawor}, {Karuppusamy}, {Keane}, {Kramer}, {K{\"u}nsem{\"o}ller}, {Lackeos}, {Liu}, {Main}, {McKee}, {Porayko}, {Shaifullah}, {Theureau}, \& {Vocks}}]{Nitu2024}
{Ni{\c{t}}u}, I.~C., {Keith}, M.~J., {Tiburzi}, C., {et~al.} 2024, \mnras, 528, 3304, \dodoi{10.1093/mnras/stae220}

\bibitem[{Nomura {et~al.}(2020)Nomura, Rodd, \& Terrano}]{Nomura:2020tpc}
Nomura, Y., Rodd, N.~L., \& Terrano, W.~A. 2020, \prl, 125, 251801, \dodoi{10.1103/PhysRevLett.125.251801}

\bibitem[{{Ocker} {et~al.}(2024{\natexlab{a}}){Ocker}, {Anderson}, {Lazio}, {Cordes}, \& {Ravi}}]{Ocker:2024ApJ_HII}
{Ocker}, S.~K., {Anderson}, L.~D., {Lazio}, T. J.~W., {Cordes}, J.~M., \& {Ravi}, V. 2024{\natexlab{a}}, \apj, 974, 10, \dodoi{10.3847/1538-4357/ad6a51}

\bibitem[{{Ocker} {et~al.}(2020){Ocker}, {Cordes}, \& {Chatterjee}}]{Ocker2020ApJ}
{Ocker}, S.~K., {Cordes}, J.~M., \& {Chatterjee}, S. 2020, \apj, 897, 124, \dodoi{10.3847/1538-4357/ab98f9}

\bibitem[{{Ocker} {et~al.}(2024{\natexlab{b}}){Ocker}, {Cordes}, {Chatterjee}, {Stinebring}, {Dolch}, {Giannakopoulos}, {Pelgrims}, {McKee}, \& {Reardon}}]{Ocker2024MNRAS}
{Ocker}, S.~K., {Cordes}, J.~M., {Chatterjee}, S., {et~al.} 2024{\natexlab{b}}, \mnras, 527, 7568, \dodoi{10.1093/mnras/stad3683}

\bibitem[{{Os{\l}owski} {et~al.}(2014){Os{\l}owski}, {van Straten}, {Bailes}, {Jameson}, \& {Hobbs}}]{ovb+14}
{Os{\l}owski}, S., {van Straten}, W., {Bailes}, M., {Jameson}, A., \& {Hobbs}, G. 2014, \mnras, 441, 3148, \dodoi{10.1093/mnras/stu804}

\bibitem[{{Os{\l}owski} {et~al.}(2011){Os{\l}owski}, {van Straten}, {Hobbs}, {Bailes}, \& {Demorest}}]{ovh+11}
{Os{\l}owski}, S., {van Straten}, W., {Hobbs}, G.~B., {Bailes}, M., \& {Demorest}, P. 2011, \mnras, 418, 1258, \dodoi{10.1111/j.1365-2966.2011.19578.x}

\bibitem[{{Oswald} {et~al.}(2025){Oswald}, {Basu}, {Chakraborty}, {Joshi}, \& et~al}]{Oswald2025_SKA_SKAPTA}
{Oswald}, L., {Basu}, A., {Chakraborty}, M., {Joshi}, B.~C., \& et~al. 2025

\bibitem[{{Paladi} {et~al.}(2024){Paladi}, {Dwivedi}, {Rana}, {Nobleson}, {Susobhanan}, {Joshi}, {Tarafdar}, {Deb}, {Arumugam}, {Gopakumar}, {Krishnakumar}, {Batra}, {Debnath}, {Kareem}, {Arumugam}, {Bagchi}, {Bathula}, {Dandapat}, {Desai}, {Gupta}, {Hisano}, {Kharbanda}, {Kikunaga}, {Kolhe}, {Maan}, {Manoharan}, {Singha}, {Srivastava}, {Surnis}, \& {Takahashi}}]{2024MNRAS.paladi.527..213P}
{Paladi}, A.~K., {Dwivedi}, C., {Rana}, P., {et~al.} 2024, \mnras, 527, 213, \dodoi{10.1093/mnras/stad3122}

\bibitem[{{Park} {et~al.}(2021){Park}, {Folkner}, {Williams}, \& {Boggs}}]{2021AJ....161..105P}
{Park}, R.~S., {Folkner}, W.~M., {Williams}, J.~G., \& {Boggs}, D.~H. 2021, \aj, 161, 105, \dodoi{10.3847/1538-3881/abd414}

\bibitem[{{Parthasarathy} {et~al.}(2021){Parthasarathy}, {Bailes}, {Shannon}, {van Straten}, {Os{\l}owski}, {Johnston}, {Spiewak}, {Reardon}, {Kramer}, {Venkatraman Krishnan}, {Pennucci}, {Abbate}, {Buchner}, {Camilo}, {Champion}, {Geyer}, {Hugo}, {Jameson}, {Karastergiou}, {Keith}, \& {Serylak}}]{2021MNRAS.502..407P}
{Parthasarathy}, A., {Bailes}, M., {Shannon}, R.~M., {et~al.} 2021, \mnras, 502, 407, \dodoi{10.1093/mnras/stab037}

\bibitem[{{Pennucci} {et~al.}(2014){Pennucci}, {Demorest}, \& {Ransom}}]{2014ApJ...790...93P}
{Pennucci}, T.~T., {Demorest}, P.~B., \& {Ransom}, S.~M. 2014, \apj, 790, 93, \dodoi{10.1088/0004-637X/790/2/93}

\bibitem[{{Perera} {et~al.}(2019){Perera}, {DeCesar}, {Demorest}, {Kerr}, {Lentati}, {Nice}, {Os{\l}owski}, {Ransom}, {Keith}, {Arzoumanian}, {Bailes}, {Baker}, {Bassa}, {Bhat}, {Brazier}, {Burgay}, {Burke-Spolaor}, {Caballero}, {Champion}, {Chatterjee}, {Chen}, {Cognard}, {Cordes}, {Crowter}, {Dai}, {Desvignes}, {Dolch}, {Ferdman}, {Ferrara}, {Fonseca}, {Goldstein}, {Graikou}, {Guillemot}, {Hazboun}, {Hobbs}, {Hu}, {Islo}, {Janssen}, {Karuppusamy}, {Kramer}, {Lam}, {Lee}, {Liu}, {Luo}, {Lyne}, {Manchester}, {McKee}, {McLaughlin}, {Mingarelli}, {Parthasarathy}, {Pennucci}, {Perrodin}, {Possenti}, {Reardon}, {Russell}, {Sanidas}, {Sesana}, {Shaifullah}, {Shannon}, {Siemens}, {Simon}, {Spiewak}, {Stairs}, {Stappers}, {Swiggum}, {Taylor}, {Theureau}, {Tiburzi}, {Vallisneri}, {Vecchio}, {Wang}, {Zhang}, {Zhang}, {Zhu}, \& {Zhu}}]{2019MNRAS.490.4666P}
{Perera}, B.~B.~P., {DeCesar}, M.~E., {Demorest}, P.~B., {et~al.} 2019, \mnras, 490, 4666, \dodoi{10.1093/mnras/stz2857}

\bibitem[{{Peters} \& {Mathews}(1963)}]{1963PhRv..131..435P}
{Peters}, P.~C., \& {Mathews}, J. 1963, Physical Review, 131, 435, \dodoi{10.1103/PhysRev.131.435}

\bibitem[{{Phinney}(2001)}]{2001astro.ph..8028P}
{Phinney}, E.~S. 2001, arXiv e-prints, astro, \dodoi{10.48550/arXiv.astro-ph/0108028}

\bibitem[{{Planck Collaboration} {et~al.}(2020){Planck Collaboration}, {Aghanim}, {Akrami}, {Arroja}, {Ashdown}, {Aumont}, {Baccigalupi}, {Ballardini}, {Banday}, {Barreiro}, {Bartolo}, {Basak}, {Battye}, {Benabed}, {Bernard}, {Bersanelli}, {Bielewicz}, {Bock}, {Bond}, {Borrill}, {Bouchet}, {Boulanger}, {Bucher}, {Burigana}, {Butler}, {Calabrese}, {Cardoso}, {Carron}, {Casaponsa}, {Challinor}, {Chiang}, {Colombo}, {Combet}, {Contreras}, {Crill}, {Cuttaia}, {de Bernardis}, {de Zotti}, {Delabrouille}, {Delouis}, {D{\'e}sert}, {Di Valentino}, {Dickinson}, {Diego}, {Donzelli}, {Dor{\'e}}, {Douspis}, {Ducout}, {Dupac}, {Efstathiou}, {Elsner}, {En{\ss}lin}, {Eriksen}, {Falgarone}, {Fantaye}, {Fergusson}, {Fernandez-Cobos}, {Finelli}, {Forastieri}, {Frailis}, {Franceschi}, {Frolov}, {Galeotta}, {Galli}, {Ganga}, {G{\'e}nova-Santos}, {Gerbino}, {Ghosh}, {Gonz{\'a}lez-Nuevo}, {G{\'o}rski}, {Gratton}, {Gruppuso}, {Gudmundsson}, {Hamann}, {Handley}, {Hansen}, {Helou}, {Herranz}, {Hildebrandt}, {Hivon}, {Huang}, {Jaffe},
  {Jones}, {Karakci}, {Keih{\"a}nen}, {Keskitalo}, {Kiiveri}, {Kim}, {Kisner}, {Knox}, {Krachmalnicoff}, {Kunz}, {Kurki-Suonio}, {Lagache}, {Lamarre}, {Langer}, {Lasenby}, {Lattanzi}, {Lawrence}, {Le Jeune}, {Leahy}, {Lesgourgues}, {Levrier}, {Lewis}, {Liguori}, {Lilje}, {Lilley}, {Lindholm}, {L{\'o}pez-Caniego}, {Lubin}, {Ma}, {Mac{\'\i}as-P{\'e}rez}, {Maggio}, {Maino}, {Mandolesi}, {Mangilli}, {Marcos-Caballero}, {Maris}, {Martin}, {Martinelli}, {Mart{\'\i}nez-Gonz{\'a}lez}, {Matarrese}, {Mauri}, {McEwen}, {Meerburg}, {Meinhold}, {Melchiorri}, {Mennella}, {Migliaccio}, {Millea}, {Mitra}, {Miville-Desch{\^e}nes}, {Molinari}, {Moneti}, {Montier}, {Morgante}, {Moss}, {Mottet}, {M{\"u}nchmeyer}, {Natoli}, {N{\o}rgaard-Nielsen}, {Oxborrow}, {Pagano}, {Paoletti}, {Partridge}, {Patanchon}, {Pearson}, {Peel}, {Peiris}, {Perrotta}, {Pettorino}, {Piacentini}, {Polastri}, {Polenta}, {Puget}, {Rachen}, {Reinecke}, {Remazeilles}, {Renault}, {Renzi}, {Rocha}, {Rosset}, {Roudier}, {Rubi{\~n}o-Mart{\'\i}n},
  {Ruiz-Granados}, {Salvati}, {Sandri}, {Savelainen}, {Scott}, {Shellard}, {Shiraishi}, {Sirignano}, {Sirri}, {Spencer}, {Sunyaev}, {Suur-Uski}, {Tauber}, {Tavagnacco}, {Tenti}, {Terenzi}, {Toffolatti}, {Tomasi}, {Trombetti}, {Valiviita}, {Van Tent}, {Vibert}, {Vielva}, {Villa}, {Vittorio}, {Wandelt}, {Wehus}, {White}, {White}, {Zacchei}, \& {Zonca}}]{PlanckCosmo2018}
{Planck Collaboration}, {Aghanim}, N., {Akrami}, Y., {et~al.} 2020, \aap, 641, A1, \dodoi{10.1051/0004-6361/201833880}

\bibitem[{{Pol} {et~al.}(2022){Pol}, {Taylor}, \& {Romano}}]{Pol_2022}
{Pol}, N., {Taylor}, S.~R., \& {Romano}, J.~D. 2022, \apj, 940, 173, \dodoi{10.3847/1538-4357/ac9836}

\bibitem[{Porayko \& Postnov(2014)}]{Porayko:2014rfa}
Porayko, N.~K., \& Postnov, K.~A. 2014, Phys. Rev. D, 90, 062008, \dodoi{10.1103/PhysRevD.90.062008}

\bibitem[{Porayko {et~al.}(2018)}]{Porayko:2018sfa}
Porayko, N.~K., {et~al.} 2018, Phys. Rev. D, 98, 102002, \dodoi{10.1103/PhysRevD.98.102002}

\bibitem[{{Porayko} {et~al.}(2019){Porayko}, {Noutsos}, {Tiburzi}, {Verbiest}, {Horneffer}, {K{\"u}nsem{\"o}ller}, {Os{\l}owski}, {Kramer}, {Schnitzeler}, {Anderson}, {Br{\"u}ggen}, {Grie{\ss}meier}, {Hoeft}, {Schwarz}, {Serylak}, \& {Wucknitz}}]{Porayko:2019MNRASsty3324}
{Porayko}, N.~K., {Noutsos}, A., {Tiburzi}, C., {et~al.} 2019, \mnras, 483, 4100, \dodoi{10.1093/mnras/sty3324}

\bibitem[{{Porayko} {et~al.}(2023){Porayko}, {Mevius}, {Hern{\'a}ndez-Pajares}, {Tiburzi}, {Olivares Pulido}, {Liu}, {Verbiest}, {K{\"u}nsem{\"o}ller}, {Krishnakumar}, {Bak Nielsen}, {Br{\"u}ggen}, {Graffigna}, {Dettmar}, {Kramer}, {Os{\l}owski}, {Schwarz}, {Shaifullah}, \& {Wucknitz}}]{Porayko:2023JGeodesy}
{Porayko}, N.~K., {Mevius}, M., {Hern{\'a}ndez-Pajares}, M., {et~al.} 2023, Journal of Geodesy, 97, 116, \dodoi{10.1007/s00190-023-01806-1}

\bibitem[{Porayko {et~al.}(2025)Porayko, Usynina, Terol-Calvo, Martin~Camalich, Shaifullah, Castillo, Blas, Guillemot, Peel, Tiburzi, Postnov, Kramer, Antoniadis, Babak, Bak~Nielsen, Barausse, Bassa, Blanchard, Bonetti, Bortolas, Brook, Burgay, Caballero, Chalumeau, Champion, Chanlaridis, Chen, Cognard, Desvignes, Falxa, Ferdman, Franchini, Gair, Golden, Goncharov, Graikou, Grie{\ss}meier, Guo, Hu, Iraci, Izquierdo-Villalba, Jang, Jawor, Janssen, Jessner, Karuppusamy, Keane, Keith, Krishnakumar, Lackeos, Lee, Liu, Liu, Lyne, McKee, Main, Mickaliger, Ni{\c{t}}u, Parthasarathy, Perera, Perrodin, Petiteau, Possenti, Quelquejay~Leclere, Samajdar, Sanidas, Sesana, Speri, Spiewak, Stappers, Susarla, Theureau, van~der Wateren, Vecchio, Venkatraman~Krishnan, Wang, Wang, Wu, \& {EPTA Collaboration}}]{Porayko:2025EPTA-Polarimetry-ALDM}
Porayko, N.~K., Usynina, P., Terol-Calvo, J., {et~al.} 2025, Phys. Rev. D, 111, 062005, \dodoi{10.1103/PhysRevD.111.062005}

\bibitem[{{Pshirkov} {et~al.}(2010){Pshirkov}, {Baskaran}, \& {Postnov}}]{2010MNRAS.402..417P}
{Pshirkov}, M.~S., {Baskaran}, D., \& {Postnov}, K.~A. 2010, \mnras, 402, 417, \dodoi{10.1111/j.1365-2966.2009.15887.x}

\bibitem[{{Punturo} {et~al.}(2010){Punturo}, {Abernathy}, {Acernese}, {Allen}, {Andersson}, {Arun}, {Barone}, {Barr}, {Barsuglia}, {Beker}, {Beveridge}, {Birindelli}, {Bose}, {Bosi}, {Braccini}, {Bradaschia}, {Bulik}, {Calloni}, {Cella}, {Chassande Mottin}, {Chelkowski}, {Chincarini}, {Clark}, {Coccia}, {Colacino}, {Colas}, {Cumming}, {Cunningham}, {Cuoco}, {Danilishin}, {Danzmann}, {De Luca}, {De Salvo}, {Dent}, {De Rosa}, {Di Fiore}, {Di Virgilio}, {Doets}, {Fafone}, {Falferi}, {Flaminio}, {Franc}, {Frasconi}, {Freise}, {Fulda}, {Gair}, {Gemme}, {Gennai}, {Giazotto}, {Glampedakis}, {Granata}, {Grote}, {Guidi}, {Hammond}, {Hannam}, {Harms}, {Heinert}, {Hendry}, {Heng}, {Hennes}, {Hild}, {Hough}, {Husa}, {Huttner}, {Jones}, {Khalili}, {Kokeyama}, {Kokkotas}, {Krishnan}, {Lorenzini}, {L{\"u}ck}, {Majorana}, {Mandel}, {Mandic}, {Martin}, {Michel}, {Minenkov}, {Morgado}, {Mosca}, {Mours}, {M{\"u}ller{\textendash}Ebhardt}, {Murray}, {Nawrodt}, {Nelson}, {Oshaughnessy}, {Ott}, {Palomba}, {Paoli}, {Parguez},
  {Pasqualetti}, {Passaquieti}, {Passuello}, {Pinard}, {Poggiani}, {Popolizio}, {Prato}, {Puppo}, {Rabeling}, {Rapagnani}, {Read}, {Regimbau}, {Rehbein}, {Reid}, {Rezzolla}, {Ricci}, {Richard}, {Rocchi}, {Rowan}, {R{\"u}diger}, {Sassolas}, {Sathyaprakash}, {Schnabel}, {Schwarz}, {Seidel}, {Sintes}, {Somiya}, {Speirits}, {Strain}, {Strigin}, {Sutton}, {Tarabrin}, {Th{\"u}ring}, {van den Brand}, {van Leewen}, {van Veggel}, {van den Broeck}, {Vecchio}, {Veitch}, {Vetrano}, {Vicere}, {Vyatchanin}, {Willke}, {Woan}, {Wolfango}, \& {Yamamoto}}]{2010CQGra..27s4002P}
{Punturo}, M., {Abernathy}, M., {Acernese}, F., {et~al.} 2010, Classical and Quantum Gravity, 27, 194002, \dodoi{10.1088/0264-9381/27/19/194002}

\bibitem[{Quelquejay~Leclere {et~al.}(2023)}]{EuropeanPulsarTimingArray:2023lqe}
Quelquejay~Leclere, H., {et~al.} 2023, Phys. Rev. D, 108, 123527, \dodoi{10.1103/PhysRevD.108.123527}

\bibitem[{{Rajagopal} \& {Romani}(1995)}]{1995ApJ...446..543R}
{Rajagopal}, M., \& {Romani}, R.~W. 1995, \apj, 446, 543, \dodoi{10.1086/175813}

\bibitem[{{Rana} {et~al.}(2025){Rana}, {Tarafdar}, {Nobleson}, {Dwivedi}, {Chandra Joshi}, {Deb}, {Mondal}, {Krishnakumar}, {Shukla}, {Singha}, {Grover}, {Tahbildar}, {Susobhanan}, {Surnis}, {Desai}, {Batra}, {Srivastava}, {Bharambe}, {Jose}, {Vyasraj}, {Jose Jacob}, {Amarnath}, {Singh}, {Zuraiq}, {Sengupta}, {Ogi}, {Kumar}, {Jagadeesh}, {Kareem}, {Maity}, {Rai}, {Vara}, {Chowdhury}, {Kato}, {Arumugam}, {Mamidipaka}, {Arul Pandian}, {Shaji}, {Thiagaraj}, {Arumugam}, {Bagchi}, {Chakraborty}, {Gopakumar}, {Gupta}, {Maan}, {Kumar Paladi}, \& {Takahashi}}]{2025PASA...42..108R}
{Rana}, P., {Tarafdar}, P., {Nobleson}, K., {et~al.} 2025, \pasa, 42, e108, \dodoi{10.1017/pasa.2025.10066}

\bibitem[{{Ravi} {et~al.}(2014){Ravi}, {Wyithe}, {Shannon}, {Hobbs}, \& {Manchester}}]{2014MNRAS.442...56R}
{Ravi}, V., {Wyithe}, J.~S.~B., {Shannon}, R.~M., {Hobbs}, G., \& {Manchester}, R.~N. 2014, \mnras, 442, 56, \dodoi{10.1093/mnras/stu779}

\bibitem[{{Reardon} {et~al.}(2023{\natexlab{a}}){Reardon}, {Zic}, {Shannon}, {Hobbs}, {Bailes}, {Di Marco}, {Kapur}, {Rogers}, {Thrane}, {Askew}, {Bhat}, {Cameron}, {Cury{\l}o}, {Coles}, {Dai}, {Goncharov}, {Kerr}, {Kulkarni}, {Levin}, {Lower}, {Manchester}, {Mandow}, {Miles}, {Nathan}, {Os{\l}owski}, {Russell}, {Spiewak}, {Zhang}, \& {Zhu}}]{2023ApJ...951L...6R}
{Reardon}, D.~J., {Zic}, A., {Shannon}, R.~M., {et~al.} 2023{\natexlab{a}}, \apjl, 951, L6, \dodoi{10.3847/2041-8213/acdd02}

\bibitem[{{Reardon} {et~al.}(2023{\natexlab{b}}){Reardon}, {Zic}, {Shannon}, {Di Marco}, {Hobbs}, {Kapur}, {Lower}, {Mandow}, {Middleton}, {Miles}, {Rogers}, {Askew}, {Bailes}, {Bhat}, {Cameron}, {Kerr}, {Kulkarni}, {Manchester}, {Nathan}, {Russell}, {Os{\l}owski}, \& {Zhu}}]{2023ApJ...951L...7R}
---. 2023{\natexlab{b}}, \apjl, 951, L7, \dodoi{10.3847/2041-8213/acdd03}

\bibitem[{{Rhook} \& {Wyithe}(2005)}]{RhookWhithe:2005}
{Rhook}, K.~J., \& {Wyithe}, J. S.~B. 2005, \mnras, 361, 1145, \dodoi{10.1111/j.1365-2966.2005.08987.x}

\bibitem[{{Rickett}(1990)}]{1990ARA&A..28..561R}
{Rickett}, B.~J. 1990, \araa, 28, 561, \dodoi{10.1146/annurev.aa.28.090190.003021}

\bibitem[{{Roebber}(2019)}]{Roebber2019}
{Roebber}, E. 2019, \apj, 876, 55, \dodoi{10.3847/1538-4357/ab100e}

\bibitem[{{Ruan} {et~al.}(2020){Ruan}, {Guo}, {Cai}, \& {Zhang}}]{Taiji:2020}
{Ruan}, W.-H., {Guo}, Z.-K., {Cai}, R.-G., \& {Zhang}, Y.-Z. 2020, International Journal of Modern Physics A, 35, 2050075, \dodoi{10.1142/S0217751X2050075X}

\bibitem[{Saito \& Yokoyama(2009)}]{Saito:2008jc}
Saito, R., \& Yokoyama, J. 2009, Phys. Rev. Lett., 102, 161101, \dodoi{10.1103/PhysRevLett.102.161101}

\bibitem[{{Sampson} {et~al.}(2015){Sampson}, {Cornish}, \& {McWilliams}}]{Sampson2015}
{Sampson}, L., {Cornish}, N.~J., \& {McWilliams}, S.~T. 2015, \prd, 91, 084055, \dodoi{10.1103/PhysRevD.91.084055}

\bibitem[{{Sazhin}(1978)}]{1978SvA....22...36S}
{Sazhin}, M.~V. 1978, \sovast, 22, 36

\bibitem[{Schwaller(2015)}]{Schwaller:2015tja}
Schwaller, P. 2015, Phys. Rev. Lett., 115, 181101, \dodoi{10.1103/PhysRevLett.115.181101}

\bibitem[{{Semenzato} {et~al.}(2025){Semenzato}, {Bellomo}, {Raccanelli}, \& {Mingarelli}}]{semenzato2025-leakage}
{Semenzato}, F., {Bellomo}, N., {Raccanelli}, A., \& {Mingarelli}, C. M.~F. 2025, arXiv e-prints, arXiv:2510.24857, \dodoi{10.48550/arXiv.2510.24857}

\bibitem[{{Semenzato} {et~al.}(2024){Semenzato}, {Casey-Clyde}, {Mingarelli}, {Raccanelli}, {Bellomo}, {Bartolo}, \& {Bertacca}}]{SemenzatoEtAl2025}
{Semenzato}, F., {Casey-Clyde}, J.~A., {Mingarelli}, C. M.~F., {et~al.} 2024, arXiv e-prints, arXiv:2411.00532, \dodoi{10.48550/arXiv.2411.00532}

\bibitem[{{Sesana}(2013)}]{2013CQGra..30v4014S}
{Sesana}, A. 2013, Classical and Quantum Gravity, 30, 224014, \dodoi{10.1088/0264-9381/30/22/224014}

\bibitem[{{Sesana} {et~al.}(2011){Sesana}, {Gair}, {Berti}, \& {Volonteri}}]{SesanaEtAl:2011}
{Sesana}, A., {Gair}, J., {Berti}, E., \& {Volonteri}, M. 2011, \prd, 83, 044036, \dodoi{10.1103/PhysRevD.83.044036}

\bibitem[{{Sesana} \& {Vecchio}(2010)}]{2010CQGra..27h4016S}
{Sesana}, A., \& {Vecchio}, A. 2010, Classical and Quantum Gravity, 27, 084016, \dodoi{10.1088/0264-9381/27/8/084016}

\bibitem[{{Sesana} {et~al.}(2008){Sesana}, {Vecchio}, \& {Colacino}}]{2008MNRAS.390..192S}
{Sesana}, A., {Vecchio}, A., \& {Colacino}, C.~N. 2008, \mnras, 390, 192, \dodoi{10.1111/j.1365-2966.2008.13682.x}

\bibitem[{{Shaifullah} {et~al.}(2020){Shaifullah}, {Tiburzi}, \& {Zucca}}]{Shaifullah2020}
{Shaifullah}, G., {Tiburzi}, C., \& {Zucca}, P. 2020, \solphys, 295, 136, \dodoi{10.1007/s11207-020-01705-0}

\bibitem[{{Shannon} \& {Cordes}(2010)}]{2010ApJ...725.1607S}
{Shannon}, R.~M., \& {Cordes}, J.~M. 2010, \apj, 725, 1607, \dodoi{10.1088/0004-637X/725/2/1607}

\bibitem[{{Shannon} \& {Cordes}(2017)}]{2017MNRAS.464.2075S}
---. 2017, \mnras, 464, 2075, \dodoi{10.1093/mnras/stw2449}

\bibitem[{{Shannon} {et~al.}(2014){Shannon}, {Os{\l}owski}, {Dai}, {Bailes}, {Hobbs}, {Manchester}, {van Straten}, {Raithel}, {Ravi}, {Toomey}, {Bhat}, {Burke-Spolaor}, {Coles}, {Keith}, {Kerr}, {Levin}, {Sarkissian}, {Wang}, {Wen}, \& {Zhu}}]{2014MNRAS.443.1463S}
{Shannon}, R.~M., {Os{\l}owski}, S., {Dai}, S., {et~al.} 2014, \mnras, 443, 1463, \dodoi{10.1093/mnras/stu1213}

\bibitem[{{Shannon} {et~al.}(2015){Shannon}, {Ravi}, {Lentati}, {Lasky}, {Hobbs}, {Kerr}, {Manchester}, {Coles}, {Levin}, {Bailes}, {Bhat}, {Burke-Spolaor}, {Dai}, {Keith}, {Os{\l}owski}, {Reardon}, {van Straten}, {Toomey}, {Wang}, {Wen}, {Wyithe}, \& {Zhu}}]{2015Sci...349.1522S}
{Shannon}, R.~M., {Ravi}, V., {Lentati}, L.~T., {et~al.} 2015, Science, 349, 1522, \dodoi{10.1126/science.aab1910}

\bibitem[{{Shannon} {et~al.}(2016){Shannon}, {Lentati}, {Kerr}, {Bailes}, {Bhat}, {Coles}, {Dai}, {Dempsey}, {Hobbs}, {Keith}, {Lasky}, {Levin}, {Manchester}, {Os{\l}owski}, {Ravi}, {Reardon}, {Rosado}, {Spiewak}, {van Straten}, {Toomey}, {Wang}, {Wen}, {You}, \& {Zhu}}]{2016ApJ...828L...1S}
{Shannon}, R.~M., {Lentati}, L.~T., {Kerr}, M., {et~al.} 2016, \apjl, 828, L1, \dodoi{10.3847/2041-8205/828/1/L1}

\bibitem[{Siegel {et~al.}(2007)Siegel, Hertzberg, \& Fry}]{Siegel:2007fz}
Siegel, E.~R., Hertzberg, M.~P., \& Fry, J.~N. 2007, \mnras, 382, 879, \dodoi{10.1111/j.1365-2966.2007.12435.x}

\bibitem[{{Siemens} {et~al.}(2013){Siemens}, {Ellis}, {Jenet}, \& {Romano}}]{2013CQGra..30v4015S}
{Siemens}, X., {Ellis}, J., {Jenet}, F., \& {Romano}, J.~D. 2013, Classical and Quantum Gravity, 30, 224015, \dodoi{10.1088/0264-9381/30/22/224015}

\bibitem[{Sikivie(1983)}]{Sikivie:1983ip}
Sikivie, P. 1983, Phys. Rev. Lett., 51, 1415, \dodoi{10.1103/PhysRevLett.51.1415}

\bibitem[{{Singha} {et~al.}(2021){Singha}, {Surnis}, {Joshi}, {Tarafdar}, {Rana}, {Susobhanan}, {Girgaonkar}, {Kolhe}, {Agarwal}, {Desai}, {Prabu}, {Bathula}, {Dandapat}, {Dey}, {Hisano}, {Kato}, {Kharbanda}, {Kikunaga}, {Marmat}, {Susarla}, {Bagchi}, {Dhanda Batra}, {Choudhury}, {Gopakumar}, {Gupta}, {Krishnakumar}, {Maan}, {Manoharan}, {Nobleson}, {Pandian}, {Pathak}, \& {Takahashi}}]{2021MNRAS.507L..57S}
{Singha}, J., {Surnis}, M.~P., {Joshi}, B.~C., {et~al.} 2021, \mnras, 507, L57, \dodoi{10.1093/mnrasl/slab098}

\bibitem[{{Singha} {et~al.}(2024){Singha}, {Joshi}, {Krishnakumar}, {Kareem}, {Bathula}, {Dwivedi}, {Jacob}, {Desai}, {Tarafdar}, {Arumugam}, {Arumugam}, {Bagchi}, {Batra}, {Dandapat}, {Deb}, {Debnath}, {Gopakumar}, {Gupta}, {Hisano}, {Kato}, {Kikunaga}, {Marmat}, {Nobleson}, {Paladi}, {Arul Pandian}, {Prabu}, {Rana}, {Srivastava}, {Surnis}, {Susobhanan}, \& {Takahashi}}]{2024MNRAS.535.1184S}
{Singha}, J., {Joshi}, B.~C., {Krishnakumar}, M.~A., {et~al.} 2024, \mnras, 535, 1184, \dodoi{10.1093/mnras/stae2405}

\bibitem[{Smarra {et~al.}(2023)Smarra, Goncharov, Barausse, Antoniadis, Babak, Nielsen, Bassa, Berthereau, Bonetti, Bortolas, Brook, Burgay, Caballero, Chalumeau, Champion, Chanlaridis, Chen, Cognard, Desvignes, Falxa, Ferdman, Franchini, Gair, Graikou, Grie{\ss}meier, Guillemot, Guo, Hu, Iraci, Izquierdo-Villalba, Jang, Jawor, Janssen, Jessner, Karuppusamy, Keane, Keith, Kramer, Krishnakumar, Lackeos, Lee, Liu, Liu, Lyne, McKee, Main, Mickaliger, Ni{\c{t}}u, Parthasarathy, Perera, Perrodin, Petiteau, Porayko, Possenti, Leclere, Samajdar, Sanidas, Sesana, Shaifullah, Speri, Spiewak, Stappers, Susarla, Theureau, Tiburzi, van~der Wateren, Vecchio, Venkatraman~Krishnan, Wang, Wang, Wu, \& Array}]{EPTA:2023uldm}
Smarra, C., Goncharov, B., Barausse, E., {et~al.} 2023, Phys. Rev. Lett., 131, 171001, \dodoi{10.1103/PhysRevLett.131.171001}

\bibitem[{Smarra {et~al.}(2024)Smarra, Kuntz, Barausse, Goncharov, L{\'o}pez~Nacir, Blas, Shao, Antoniadis, Champion, Cognard, Guillemot, Hu, Keith, Kramer, Liu, Perrodin, Sanidas, \& Theureau}]{Smarra:2024ConformalULDM}
Smarra, C., Kuntz, A., Barausse, E., {et~al.} 2024, Phys. Rev. D, 110, 043033, \dodoi{10.1103/PhysRevD.110.043033}

\bibitem[{{Smith} {et~al.}(2023){Smith}, {Abdollahi}, {Ajello}, {Bailes}, {Baldini}, {Ballet}, {Baring}, {Bassa}, {Gonzalez}, {Bellazzini}, {Berretta}, {Bhattacharyya}, {Bissaldi}, {Bonino}, {Bottacini}, {Bregeon}, {Bruel}, {Burgay}, {Burnett}, {Cameron}, {Camilo}, {Caputo}, {Caraveo}, {Cavazzuti}, {Chiaro}, {Ciprini}, {Clark}, {Cognard}, {Corongiu}, {Orestano}, {Crnogorcevic}, {Cuoco}, {Cutini}, {D'Ammando}, {de Angelis}, {DeCesar}, {De Gaetano}, {de Menezes}, {Deneva}, {de Palma}, {Di Lalla}, {Dirirsa}, {Di Venere}, {Dom{\'\i}nguez}, {Dumora}, {Fegan}, {Ferrara}, {Fiori}, {Fleischhack}, {Flynn}, {Franckowiak}, {Freire}, {Fukazawa}, {Fusco}, {Galanti}, {Gammaldi}, {Gargano}, {Gasparrini}, {Giacchino}, {Giglietto}, {Giordano}, {Giroletti}, {Green}, {Grenier}, {Guillemot}, {Guiriec}, {Gustafsson}, {Harding}, {Hays}, {Hewitt}, {Horan}, {Hou}, {Jankowski}, {Johnson}, {Johnson}, {Johnston}, {Kataoka}, {Keith}, {Kerr}, {Kramer}, {Kuss}, {Latronico}, {Lee}, {Li}, {Li}, {Limyansky}, {Longo}, {Loparco}, {Lorusso},
  {Lovellette}, {Lower}, {Lubrano}, {Lyne}, {Maan}, {Maldera}, {Manchester}, {Manfreda}, {Marelli}, {Mart{\'\i}-Devesa}, {Mazziotta}, {McEnery}, {Mereu}, {Michelson}, {Mickaliger}, {Mitthumsiri}, {Mizuno}, {Moiseev}, {Monzani}, {Morselli}, {Negro}, {Nemmen}, {Nieder}, {Nuss}, {Omodei}, {Orienti}, {Orlando}, {Ormes}, {Palatiello}, {Paneque}, {Panzarini}, {Parthasarathy}, {Persic}, {Pesce-Rollins}, {Pillera}, {Poon}, {Porter}, {Possenti}, {Principe}, {Rain{\`o}}, {Rando}, {Ransom}, {Ray}, {Razzano}, {Razzaque}, {Reimer}, {Reimer}, {Renault-Tinacci}, {Romani}, {S{\'a}nchez-Conde}, {Parkinson}, {Scotton}, {Serini}, {Sgr{\`o}}, {Shannon}, {Sharma}, {Shen}, {Siskind}, {Spandre}, {Spinelli}, {Stappers}, {Stephens}, {Suson}, {Tabassum}, {Tajima}, {Tak}, {Theureau}, {Thompson}, {Tibolla}, {Torres}, {Valverde}, {Venter}, {Wadiasingh}, {Wang}, {Wang}, {Wang}, {Weltevrede}, {Wood}, {Yan}, {Zaharijas}, {Zhang}, \& {Zhu}}]{2023ApJ...958..191S}
{Smith}, D.~A., {Abdollahi}, S., {Ajello}, M., {et~al.} 2023, \apj, 958, 191, \dodoi{10.3847/1538-4357/acee67}

\bibitem[{{Spiewak} {et~al.}(2022){Spiewak}, {Bailes}, {Miles}, {Parthasarathy}, {Reardon}, {Shamohammadi}, {Shannon}, {Bhat}, {Buchner}, {Cameron}, {Camilo}, {Geyer}, {Johnston}, {Karastergiou}, {Keith}, {Kramer}, {Serylak}, {van Straten}, {Theureau}, \& {Venkatraman Krishnan}}]{2022PASA...39...27S}
{Spiewak}, R., {Bailes}, M., {Miles}, M.~T., {et~al.} 2022, \pasa, 39, e027, \dodoi{10.1017/pasa.2022.19}

\bibitem[{{Springel} {et~al.}(2005){Springel}, {White}, {Jenkins}, {Frenk}, {Yoshida}, {Gao}, {Navarro}, {Thacker}, {Croton}, {Helly}, {Peacock}, {Cole}, {Thomas}, {Couchman}, {Evrard}, {Colberg}, \& {Pearce}}]{2005Natur.435..629S}
{Springel}, V., {White}, S. D.~M., {Jenkins}, A., {et~al.} 2005, \nat, 435, 629, \dodoi{10.1038/nature03597}

\bibitem[{Starobinsky(1979)}]{Starobinsky:1979ty}
Starobinsky, A.~A. 1979, JETP Lett., 30, 682

\bibitem[{{Steinle} {et~al.}(2023){Steinle}, {Middleton}, {Moore}, {Chen}, {Klein}, {Pratten}, {Buscicchio}, {Finch}, \& {Vecchio}}]{SteinleEtAl:2023}
{Steinle}, N., {Middleton}, H., {Moore}, C.~J., {et~al.} 2023, \mnras, 525, 2851, \dodoi{10.1093/mnras/stad2408}

\bibitem[{{Susarla} {et~al.}(2024){Susarla}, {Chalumeau}, {Tiburzi}, {Keane}, {Verbiest}, {Hazboun}, {Krishnakumar}, {Iraci}, {Shaifullah}, {Golden}, {Bak Nielsen}, {Donner}, {Grie{\ss}meier}, {Keith}, {Os{\l}owski}, {Porayko}, {Serylak}, {Anderson}, {Br{\"u}ggen}, {Ciardi}, {Dettmar}, {Hoeft}, {K{\"u}nsem{\"o}ller}, {Schwarz}, \& {Vocks}}]{Susarla:2024AandA}
{Susarla}, S.~C., {Chalumeau}, A., {Tiburzi}, C., {et~al.} 2024, \aap, 692, A18, \dodoi{10.1051/0004-6361/202450680}

\bibitem[{{Susobhanan}(2023)}]{AS23}
{Susobhanan}, A. 2023, Classical and Quantum Gravity, 40, 155014, \dodoi{10.1088/1361-6382/ace234}

\bibitem[{{Susobhanan} {et~al.}(2020){Susobhanan}, {Gopakumar}, {Hobbs}, \& {Taylor}}]{AS20}
{Susobhanan}, A., {Gopakumar}, A., {Hobbs}, G., \& {Taylor}, S.~R. 2020, \prd, 101, 043022, \dodoi{10.1103/PhysRevD.101.043022}

\bibitem[{{Susobhanan} {et~al.}(2021){Susobhanan}, {Maan}, {Joshi}, {Prabu}, {Desai}, {Nobleson}, {Susarla}, {Girgaonkar}, {Dey}, {Batra}, {Gupta}, {Gopakumar}, {Bagchi}, {Basu}, {Bethapudi}, {Choudhary}, {De}, {Krishnakumar}, {Manoharan}, {Naidu}, {Pathak}, {Singha}, \& {Surnis}}]{2021PASA...38...17S}
{Susobhanan}, A., {Maan}, Y., {Joshi}, B.~C., {et~al.} 2021, \pasa, 38, e017, \dodoi{10.1017/pasa.2021.12}

\bibitem[{{Tarafdar} {et~al.}(2022){Tarafdar}, {Nobleson}, {Rana}, {Singha}, {Krishnakumar}, {Joshi}, {Paladi}, {Kolhe}, {Batra}, {Agarwal}, {Bathula}, {Dandapat}, {Desai}, {Dey}, {Hisano}, {Ingale}, {Kato}, {Kharbanda}, {Kikunaga}, {Marmat}, {Pandian}, {Prabu}, {Srivastava}, {Surnis}, {Susarla}, {Susobhanan}, {Takahashi}, {Arumugam}, {Bagchi}, {Banik}, {De}, {Girgaonkar}, {Gopakumar}, {Gupta}, {Maan}, {Manoharan}, {Naidu}, \& {Pathak}}]{2022PASA...39...53T}
{Tarafdar}, P., {Nobleson}, K., {Rana}, P., {et~al.} 2022, \pasa, 39, e053, \dodoi{10.1017/pasa.2022.46}

\bibitem[{{Taylor} \& {Weisberg}(1982)}]{1982ApJ...253..908T}
{Taylor}, J.~H., \& {Weisberg}, J.~M. 1982, \apj, 253, 908, \dodoi{10.1086/159690}

\bibitem[{{Thorne} \& {Braginskii}(1976)}]{1976ApJ...204L...1T}
{Thorne}, K.~S., \& {Braginskii}, V.~B. 1976, \apjl, 204, L1, \dodoi{10.1086/182042}

\bibitem[{{Tiburzi} {et~al.}(2025){Tiburzi}, {Lam}, {Reardon}, {Porayko}, {Koch}, \& {{The SKA Pulsar Science Working Group}}}]{Tiburzi2025_SKA_SKAPTA}
{Tiburzi}, C., {Lam}, M., {Reardon}, D., {et~al.} 2025

\bibitem[{{Tiburzi} {et~al.}(2016){Tiburzi}, {Hobbs}, {Kerr}, {Coles}, {Dai}, {Manchester}, {Possenti}, {Shannon}, \& {You}}]{2016MNRAS.455.4339T}
{Tiburzi}, C., {Hobbs}, G., {Kerr}, M., {et~al.} 2016, \mnras, 455, 4339, \dodoi{10.1093/mnras/stv2143}

\bibitem[{{Tiburzi} {et~al.}(2019){Tiburzi}, {Verbiest}, {Shaifullah}, {Janssen}, {Anderson}, {Horneffer}, {K{\"u}nsem{\"o}ller}, {Os{\l}owski}, {Donner}, {Kramer}, {Kumari}, {Porayko}, {Zucca}, {Ciardi}, {Dettmar}, {Grie{\ss}meier}, {Hoeft}, \& {Serylak}}]{Tiburzi2019}
{Tiburzi}, C., {Verbiest}, J.~P.~W., {Shaifullah}, G.~M., {et~al.} 2019, \mnras, 487, 394, \dodoi{10.1093/mnras/stz1278}

\bibitem[{{Tiburzi} {et~al.}(2021){Tiburzi}, {Shaifullah}, {Bassa}, {Zucca}, {Verbiest}, {Porayko}, {van der Wateren}, {Fallows}, {Main}, {Janssen}, {Anderson}, {Bak Nielsen}, {Donner}, {Keane}, {K{\"u}nsem{\"o}ller}, {Os{\l}owski}, {Grie{\ss}meier}, {Serylak}, {Br{\"u}ggen}, {Ciardi}, {Dettmar}, {Hoeft}, {Kramer}, {Mann}, \& {Vocks}}]{Tiburzi2021}
{Tiburzi}, C., {Shaifullah}, G.~M., {Bassa}, C.~G., {et~al.} 2021, \aap, 647, A84, \dodoi{10.1051/0004-6361/202039846}

\bibitem[{{Toubiana} {et~al.}(2025){Toubiana}, {Sberna}, {Volonteri}, {Barausse}, {Babak}, {Enficiaud}, {Izquierdo{\textendash}Villalba}, {Gair}, {Greene}, \& {Quelquejay Leclere}}]{ToubianaEtAl:2025}
{Toubiana}, A., {Sberna}, L., {Volonteri}, M., {et~al.} 2025, \aap, 700, A135, \dodoi{10.1051/0004-6361/202453027}

\bibitem[{{Traianou} {et~al.}(2025){Traianou}, {G{\'o}mez}, {Cho}, {Chael}, {Fuentes}, {Myserlis}, {Wielgus}, {Zhao}, {Lico}, {Moriyama}, {Dey}, {Bruni}, {Dahale}, {Toscano}, {Gurvits}, {Lisakov}, {Kovalev}, {Lobanov}, {Pushkarev}, \& {Sokolovsky}}]{RadAstOJ287}
{Traianou}, E., {G{\'o}mez}, J.~L., {Cho}, I., {et~al.} 2025, \aap, 700, A16, \dodoi{10.1051/0004-6361/202554929}

\bibitem[{{Tristram} {et~al.}(2022){Tristram}, {Banday}, {G{\'o}rski}, {Keskitalo}, {Lawrence}, {Andersen}, {Barreiro}, {Borrill}, {Colombo}, {Eriksen}, {Fernandez-Cobos}, {Kisner}, {Mart{\'\i}nez-Gonz{\'a}lez}, {Partridge}, {Scott}, {Svalheim}, \& {Wehus}}]{Tristram2022}
{Tristram}, M., {Banday}, A.~J., {G{\'o}rski}, K.~M., {et~al.} 2022, \prd, 105, 083524, \dodoi{10.1103/PhysRevD.105.083524}

\bibitem[{{Truant} {et~al.}(2025){Truant}, {Izquierdo-Villalba}, {Sesana}, {Shaifullah}, \& {Bonetti}}]{2025A&A...694A.282T}
{Truant}, R.~J., {Izquierdo-Villalba}, D., {Sesana}, A., {Shaifullah}, G.~M., \& {Bonetti}, M. 2025, \aap, 694, A282, \dodoi{10.1051/0004-6361/202451556}

\bibitem[{Vachaspati \& Vilenkin(1985)}]{Vachaspati:1984gt}
Vachaspati, T., \& Vilenkin, A. 1985, Phys. Rev. D, 31, 3052, \dodoi{10.1103/PhysRevD.31.3052}

\bibitem[{{Vallisneri} {et~al.}(2020){Vallisneri}, {Taylor}, {Simon}, {Folkner}, {Park}, {Cutler}, {Ellis}, {Lazio}, {Vigeland}, {Aggarwal}, {Arzoumanian}, {Baker}, {Brazier}, {Brook}, {Burke-Spolaor}, {Chatterjee}, {Cordes}, {Cornish}, {Crawford}, {Cromartie}, {Crowter}, {DeCesar}, {Demorest}, {Dolch}, {Ferdman}, {Ferrara}, {Fonseca}, {Garver-Daniels}, {Gentile}, {Good}, {Hazboun}, {Holgado}, {Huerta}, {Islo}, {Jennings}, {Jones}, {Jones}, {Kaplan}, {Kelley}, {Key}, {Lam}, {Levin}, {Lorimer}, {Luo}, {Lynch}, {Madison}, {McLaughlin}, {McWilliams}, {Mingarelli}, {Ng}, {Nice}, {Pennucci}, {Pol}, {Ransom}, {Ray}, {Siemens}, {Spiewak}, {Stairs}, {Stinebring}, {Stovall}, {Swiggum}, {van Haasteren}, {Witt}, \& {Zhu}}]{2020ApJ...893..112V}
{Vallisneri}, M., {Taylor}, S.~R., {Simon}, J., {et~al.} 2020, \apj, 893, 112, \dodoi{10.3847/1538-4357/ab7b67}

\bibitem[{{van Haasteren}(2024)}]{2024ApJS..273...23V}
{van Haasteren}, R. 2024, \apjs, 273, 23, \dodoi{10.3847/1538-4365/ad530f}

\bibitem[{{van Haasteren} \& {Levin}(2010)}]{2010MNRAS.401.2372V}
{van Haasteren}, R., \& {Levin}, Y. 2010, \mnras, 401, 2372, \dodoi{10.1111/j.1365-2966.2009.15885.x}

\bibitem[{{van Straten}(2006)}]{2006ApJ...642.1004V}
{van Straten}, W. 2006, \apj, 642, 1004, \dodoi{10.1086/501001}

\bibitem[{{van Straten}(2013)}]{2013ApJS..204...13V}
---. 2013, \apjs, 204, 13, \dodoi{10.1088/0067-0049/204/1/13}

\bibitem[{{Venkatraman Krishnan} {et~al.}(2025){Venkatraman Krishnan}, {Shao}, {Balakrishnan}, {Colom i Bernadichand}, \& et~al}]{Krishnan2025_SKA_SKAPTA}
{Venkatraman Krishnan}, V.~V., {Shao}, L., {Balakrishnan}, V., {Colom i Bernadichand}, M., \& et~al. 2025

\bibitem[{{Verbiest} \& {Shaifullah}(2018)}]{2018CQGra..35m3001V}
{Verbiest}, J. P.~W., \& {Shaifullah}, G.~M. 2018, Classical and Quantum Gravity, 35, 133001, \dodoi{10.1088/1361-6382/aac412}

\bibitem[{{Verbiest} {et~al.}(2024){Verbiest}, {Vigeland}, {Porayko}, {Chen}, \& {Reardon}}]{Verbiest:2024RINP}
{Verbiest}, J. P.~W., {Vigeland}, S.~J., {Porayko}, N.~K., {Chen}, S., \& {Reardon}, D.~J. 2024, Results in Physics, 61, 107719, \dodoi{10.1016/j.rinp.2024.107719}

\bibitem[{{Verbiest} {et~al.}(2016){Verbiest}, {Lentati}, {Hobbs}, {van Haasteren}, {Demorest}, {Janssen}, {Wang}, {Desvignes}, {Caballero}, {Keith}, {Champion}, {Arzoumanian}, {Babak}, {Bassa}, {Bhat}, {Brazier}, {Brem}, {Burgay}, {Burke-Spolaor}, {Chamberlin}, {Chatterjee}, {Christy}, {Cognard}, {Cordes}, {Dai}, {Dolch}, {Ellis}, {Ferdman}, {Fonseca}, {Gair}, {Garver-Daniels}, {Gentile}, {Gonzalez}, {Graikou}, {Guillemot}, {Hessels}, {Jones}, {Karuppusamy}, {Kerr}, {Kramer}, {Lam}, {Lasky}, {Lassus}, {Lazarus}, {Lazio}, {Lee}, {Levin}, {Liu}, {Lynch}, {Lyne}, {Mckee}, {McLaughlin}, {McWilliams}, {Madison}, {Manchester}, {Mingarelli}, {Nice}, {Os{\l}owski}, {Palliyaguru}, {Pennucci}, {Perera}, {Perrodin}, {Possenti}, {Petiteau}, {Ransom}, {Reardon}, {Rosado}, {Sanidas}, {Sesana}, {Shaifullah}, {Shannon}, {Siemens}, {Simon}, {Smits}, {Spiewak}, {Stairs}, {Stappers}, {Stinebring}, {Stovall}, {Swiggum}, {Taylor}, {Theureau}, {Tiburzi}, {Toomey}, {Vallisneri}, {van Straten}, {Vecchio}, {Wang}, {Wen}, {You},
  {Zhu}, \& {Zhu}}]{2016MNRAS.458.1267V}
{Verbiest}, J.~P.~W., {Lentati}, L., {Hobbs}, G., {et~al.} 2016, \mnras, 458, 1267, \dodoi{10.1093/mnras/stw347}

\bibitem[{Vilenkin(1981)}]{Vilenkin:1981zs}
Vilenkin, A. 1981, Phys. Rev. D, 23, 852, \dodoi{10.1103/PhysRevD.23.852}

\bibitem[{{Volonteri} {et~al.}(2003){Volonteri}, {Haardt}, \& {Madau}}]{2003ApJ...582..559V}
{Volonteri}, M., {Haardt}, F., \& {Madau}, P. 2003, \apj, 582, 559, \dodoi{10.1086/344675}

\bibitem[{{Wang} {et~al.}(2015){Wang}, {Hobbs}, {Coles}, {Shannon}, {Zhu}, {Madison}, {Kerr}, {Ravi}, {Keith}, {Manchester}, {Levin}, {Bailes}, {Bhat}, {Burke-Spolaor}, {Dai}, {Os{\l}owski}, {van Straten}, {Toomey}, {Wang}, \& {Wen}}]{2015MNRAS.446.1657W}
{Wang}, J.~B., {Hobbs}, G., {Coles}, W., {et~al.} 2015, \mnras, 446, 1657, \dodoi{10.1093/mnras/stu2137}

\bibitem[{Weinberg(2004)}]{Weinberg:2004kr}
Weinberg, S. 2004, \prd, 69, 023503, \dodoi{10.1103/PhysRevD.69.023503}

\bibitem[{{Weltevrede} {et~al.}(2006){Weltevrede}, {Edwards}, \& {Stappers}}]{wes06}
{Weltevrede}, P., {Edwards}, R.~T., \& {Stappers}, B.~W. 2006, \aap, 445, 243, \dodoi{10.1051/0004-6361:20053088}

\bibitem[{{Witten}(1984)}]{Witten:1984rs}
{Witten}, E. 1984, \prd, 30, 272, \dodoi{10.1103/PhysRevD.30.272}

\bibitem[{{Wood} {et~al.}(2020){Wood}, {Tun-Beltran}, {Kooi}, {Polisensky}, \& {Nieves-Chinchilla}}]{Wood_2020}
{Wood}, B.~E., {Tun-Beltran}, S., {Kooi}, J.~E., {Polisensky}, E.~J., \& {Nieves-Chinchilla}, T. 2020, \apj, 896, 99, \dodoi{10.3847/1538-4357/ab93b8}

\bibitem[{Wu {et~al.}(2022)Wu, Chen, Huang, Zhu, Bhat, Feng, Hobbs, Manchester, Russell, \& Shannon}]{PPTA:2022eul}
Wu, Y.-M., Chen, Z.-C., Huang, Q.-G., {et~al.} 2022, \prd, 106, L081101, \dodoi{10.1103/PhysRevD.106.L081101}

\bibitem[{{Wu} {et~al.}(2022){Wu}, {Verbiest}, {Main}, {Grie{\ss}meier}, {Liu}, {Os{\l}owski}, {Moochickal Ambalappat}, {Nielsen}, {K{\"u}nsem{\"o}ller}, {Donner}, {Tiburzi}, {Porayko}, {Serylak}, {K{\"u}nkel}, {Br{\"u}ggen}, \& {Vocks}}]{Wu:2022AandA657A98}
{Wu}, Z., {Verbiest}, J. P.~W., {Main}, R.~A., {et~al.} 2022, \aap, 663, A116, \dodoi{10.1051/0004-6361/202142980}

\bibitem[{{Wu} {et~al.}(2023){Wu}, {Coles}, {Verbiest}, {Ambalappat}, {Tiburzi}, {Grie{\ss}meier}, {Main}, {Liu}, {Kramer}, {Wucknitz}, {Porayko}, {Os{\l}owski}, {Nielsen}, {Donner}, {Hoeft}, {Br{\"u}ggen}, {Vocks}, {Dettmar}, {Theureau}, {Serylak}, {Kondratiev}, {McKee}, {Shaifullah}, {Kravtsov}, {Zakharenko}, {Ulyanov}, {Konovalenko}, {Zarka}, {Cecconi}, {Koopmans}, \& {Corbel}}]{wcv+23}
{Wu}, Z., {Coles}, W.~A., {Verbiest}, J. P.~W., {et~al.} 2023, \mnras, 520, 5536, \dodoi{10.1093/mnras/stad429}

\bibitem[{{Wyithe} \& {Loeb}(2003)}]{2003ApJ...590..691W}
{Wyithe}, J. S.~B., \& {Loeb}, A. 2003, \apj, 590, 691, \dodoi{10.1086/375187}

\bibitem[{Xia {et~al.}(2023)Xia, Tang, Huang, Yuan, \& Fan}]{Xia:2023hov}
Xia, Z.-Q., Tang, T.-P., Huang, X., Yuan, Q., \& Fan, Y.-Z. 2023, Phys. Rev. D, 107, L121302, \dodoi{10.1103/PhysRevD.107.L121302}

\bibitem[{{Xin} {et~al.}(2021){Xin}, {Mingarelli}, \& {Hazboun}}]{XinMingarelliHazboun2021}
{Xin}, C., {Mingarelli}, C. M.~F., \& {Hazboun}, J.~S. 2021, \apj, 915, 97, \dodoi{10.3847/1538-4357/ac01c5}

\bibitem[{{Xu} {et~al.}(2023){Xu}, {Chen}, {Guo}, {Jiang}, {Wang}, {Xu}, {Xue}, {Nicolas Caballero}, {Yuan}, {Xu}, {Wang}, {Hao}, {Luo}, {Lee}, {Han}, {Jiang}, {Shen}, {Wang}, {Wang}, {Xu}, {Wu}, {Manchester}, {Qian}, {Guan}, {Huang}, {Sun}, \& {Zhu}}]{2023RAA....23g5024X}
{Xu}, H., {Chen}, S., {Guo}, Y., {et~al.} 2023, Research in Astronomy and Astrophysics, 23, 075024, \dodoi{10.1088/1674-4527/acdfa5}

\bibitem[{Xue {et~al.}(2021)}]{Xue:2021gyq}
Xue, X., {et~al.} 2021, Phys. Rev. Lett., 127, 251303, \dodoi{10.1103/PhysRevLett.127.251303}

\bibitem[{Xue {et~al.}(2022)}]{PPTA:2021uzb}
---. 2022, Phys. Rev. Res., 4, L012022, \dodoi{10.1103/PhysRevResearch.4.L012022}

\bibitem[{{Xue} {et~al.}(2024){Xue}, {Dai}, {Nhan Luu}, {Liu}, {Ren}, {Shu}, {Zhao}, {Zic}, {Bhat}, {Chen}, {Feng}, {Hobbs}, {Kapur}, {Manchester}, {Mandow}, {Mishra}, {Reardon}, {Russell}, {Shannon}, {Wang}, {Zhang}, {Zhang}, \& {Zhu}}]{Xue:2024PPA-ALDM}
{Xue}, X., {Dai}, S., {Nhan Luu}, H., {et~al.} 2024, arXiv e-prints, arXiv:2412.02229, \dodoi{10.48550/arXiv.2412.02229}

\bibitem[{Yonemaru {et~al.}(2021)}]{Yonemaru:2020bmr}
Yonemaru, N., {et~al.} 2021, \mnras, 501, 701, \dodoi{10.1093/mnras/staa3721}

\bibitem[{{You} {et~al.}(2007){You}, {Hobbs}, {Coles}, {Manchester}, {Edwards}, {Bailes}, {Sarkissian}, {Verbiest}, {van Straten}, {Hotan}, {Ord}, {Jenet}, {Bhat}, \& {Teoh}}]{2007MNRAS.378..493Y}
{You}, X.~P., {Hobbs}, G., {Coles}, W.~A., {et~al.} 2007, \mnras, 378, 493, \dodoi{10.1111/j.1365-2966.2007.11617.x}

\bibitem[{{Zeng} {et~al.}(2025){Zeng}, {Ning}, {Yuwen}, {Wang}, {Deng}, \& {Cai}}]{Zeng:2025law}
{Zeng}, X.-X., {Ning}, Z., {Yuwen}, Z.-Y., {et~al.} 2025, arXiv e-prints, arXiv:2504.11275, \dodoi{10.48550/arXiv.2504.11275}

\bibitem[{{Zhao} {et~al.}(2025){Zhao}, {Chen}, {Cardinal Tremblay}, {Goncharov}, {Zhu}, {Bhat}, {Cury{\l}o}, {Dai}, {Di Marco}, {Ding}, {Hobbs}, {Kapur}, {Ling}, {Liu}, {Mandow}, {Mishra}, {Reardon}, {Russell}, {Shannon}, {Wang}, {Zhang}, \& {Zic}}]{2025arXiv250813944Z}
{Zhao}, S.-Y., {Chen}, Z.-C., {Cardinal Tremblay}, J., {et~al.} 2025, \apj, 992, 181, \dodoi{10.3847/1538-4357/ae0719}

\bibitem[{Zhao {et~al.}(2013)Zhao, Zhang, You, \& Zhu}]{Zhao:2013bba}
Zhao, W., Zhang, Y., You, X.-P., \& Zhu, Z.-H. 2013, Phys. Rev. D, 87, 124012, \dodoi{10.1103/PhysRevD.87.124012}

\bibitem[{{Zheng} {et~al.}(2025){Zheng}, {Mingarelli}, {DeRocco}, {Nay}, {Boddy}, \& {Dror}}]{ZhengEtAl2025}
{Zheng}, Q., {Mingarelli}, C. M.~F., {DeRocco}, W., {et~al.} 2025, arXiv e-prints, arXiv:2508.21582, \dodoi{10.48550/arXiv.2508.21582}

\bibitem[{{Zic} {et~al.}(2022){Zic}, {Hobbs}, {Shannon}, {Reardon}, {Goncharov}, {Bhat}, {Cameron}, {Dai}, {Dawson}, {Kerr}, {Manchester}, {Mandow}, {Marshman}, {Russell}, {Thyagarajan}, \& {Zhu}}]{2022MNRAS.516..410Z}
{Zic}, A., {Hobbs}, G., {Shannon}, R.~M., {et~al.} 2022, \mnras, 516, 410, \dodoi{10.1093/mnras/stac2100}

\bibitem[{{Zic} {et~al.}(2023){Zic}, {Reardon}, {Kapur}, {Hobbs}, {Mandow}, {Cury{\l}o}, {Shannon}, {Askew}, {Bailes}, {Bhat}, {Cameron}, {Chen}, {Dai}, {Di Marco}, {Feng}, {Kerr}, {Kulkarni}, {Lower}, {Luo}, {Manchester}, {Miles}, {Nathan}, {Os{\l}owski}, {Rogers}, {Russell}, {Sarkissian}, {Shamohammadi}, {Spiewak}, {Thyagarajan}, {Toomey}, {Wang}, {Zhang}, {Zhang}, \& {Zhu}}]{2023PASA...40...49Z}
{Zic}, A., {Reardon}, D.~J., {Kapur}, A., {et~al.} 2023, \pasa, 40, e049, \dodoi{10.1017/pasa.2023.36}

\end{thebibliography}




\end{document}